\documentclass[useAMS,usenatbib]{mn2e}

\usepackage{graphicx}
\usepackage{amssymb}
\usepackage{multirow}
\usepackage{pifont}
\usepackage{longtable}
\usepackage{bbold}
\usepackage{color}

\makeatletter
    \newcommand{\rmnum}[1]{\romannumeral #1}
    \newcommand{\Rmnum}[1]{\expandafter\@slowromancap\romannumeral #1@}
\makeatother

\title[Inclined KBOs in the resonances beyond 50 AU]
{A study of the high-inclination population in the Kuiper belt -- \Rmnum5. Mean motion resonances beyond 50 AU}
\author[Jian Li]
{Jian Li$^{1,2}$\thanks{E-mail: ljian@nju.edu.cn}\\
$^1$School of Astronomy and Space Science, Nanjing University, 163 Xianlin Avenue, Nanjing 210023, PR China\\
$^2$ Key Laboratory of Modern Astronomy and Astrophysics in Ministry of Education, Nanjing University, Nanjing 210023, PR China}
\begin{document}

\date{Accepted 1988 December 15. Received 1988 December 14; in original form 1988 October 11}

\pagerange{\pageref{firstpage}--\pageref{lastpage}} \pubyear{2002}

\maketitle

\label{firstpage}

\begin{abstract}

In this paper, we present the most comprehensive study to date on Neptune's mean-motion resonances (MMRs) in the distant Kuiper belt from 50 to 100 AU. Over 200 resonant Kuiper belt objects (KBOs) have been identified in this region, spanning resonances from the 2nd-order 1:3 MMR to the 22nd-order 7:29 MMR, with inclinations $i<40^\circ$. Building on these diverse distributions, we first analyse the dynamical features of numerous $m$:$n$ MMRs, providing an informative database that includes the possible eccentricity ($e$) range, resonance widths, resonance centres, and permissible $(e,i)$ distributions. We then conduct numerical simulations to explore the long-term stability of these MMRs. Our results show that: (1) resonators can occupy all 1:$n$ to 7:$n$ MMRs, with nearly any $n$ corresponding to the 50-100 AU region, including the farthest-out MMRs of 5:29 (24th-order), 6:35 (29th-order), and 7:40 (33rd-order). This suggests that KBOs could potentially exist in even higher-order MMRs than those currently observed. (2) For each set of $m$:$n$ resonances with the same $m$, resonators consistently exhibit inclinations up to $40^\circ$, while eccentricities remain strictly restricted below 0.7. (3) For the 1:3 and 1:4 MMRs, the leading population is less stable than the trailing population. Most interestingly, we discover a novel phenomenon of number reversal, where the higher-order, weaker 3:8 MMR (at semimajor axis $a\approx57.9$ AU) hosts more resonators, rather than fewer as expected, compared to the lower-order, stronger 3:7 MMR (at $a\approx53.0$ AU). Future observations, whether confirming or challenging this phenomenon, will offer valuable insight into the eccentricity and inclination distributions of primordial KBOs.

 \end{abstract}


\section{Introduction}

In our previous study, we provided a comprehensive overview of the dynamics of high-inclination objects in Neptune's mean motion resonances (MMRs) within the Kuiper belt. The first two papers examined the 1st-order 2:3 ($\sim39.4$ AU) and 1:2 ($\sim47.8$ AU) MMRs \citep{Li2014a, Li2014b}, while the subsequent papers considered higher-order MMRs between roughly 42 and 47 AU \citep{Li2020, Li2023}. For reference, we note that for an $m$:$n$ MMR ($m$ and $n$ are integers), the absolute difference $|n-m|$ is called the order of the resonance, and its increase leads to a decrease in the resonance's strength. All of the MMRs we have considered before are located in a region extending from beyond the orbit of Neptune to about 50 AU, commonly referred to as the main Kuiper belt \citep{Glad2008,Peti2011}. In this paper, we expand this series of works by investigating resonant Kuiper belt objects (RKBOs) with semimajor axes $a>50$ AU. We will explore the effects of their high-inclination orbits on the libration centre, resonant amplitude, permissible region, and other dynamical features. Notably, in the distant Kuiper belt beyond 50 AU, resonances could be more heavily populated than current Neptune migration models predict \citep{Bann2018}. For example, the 2:5 RKBOs ($a\approx55.4$ AU) could be as numerous as the largest resonant population, the Plutinos, in the 2:3 MMR \citep{volk2016}.

The distant Kuiper belt is continuously extending outward to connect with the inner Oort cloud, which is located as far as $\gtrsim1000$-2000 AU \citep{Peti2017,Shan19,Khai20}. To avoid complications arising from the potential existence of Planet 9, which may reside in a wide orbit with $a=400$-800 AU, we need to set an outer boundary for the considered distant Kuiper belt. The Planet 9 hypothesis was initially proposed to explain the clustering in the argument of perihelion for extremely distant objects with $a>250$ AU, including Sedna and 2012 VP113 \citep{Truj14, Baty16, Brow16, Baty19}. In the context of this proposed Planet 9, we measured the mean plane of the Kuiper belt and found that it deviates noticeably from the invariable plane of the Solar system at $a>100$ AU \citep{LiX2020b}. This suggests that the gravitational influence of Planet 9 could be detectable much closer to the Sun than the extremely distant objects. Therefore, in this paper, we will focus solely on the RKBOs in Neptune's MMRs inside 100 AU, as the possible influence of Planet 9 on these objects is minimised.

Many published works have already been carried out on the study of RKBOs beyond 50 AU, particularly for the lowest-order 1:3 and 2:5 MMRs \citep{Beau94, Gomes97, Chia2003, Lyka2007, Glad2012, shep2012, Malh2018, namo2020}. This is likely because these two resonances are stronger and closer to the Sun, and thus currently host more observed RKBOs than other resonances. In addition, previous works have also considered some higher-order resonances, such as the 5:12, 4:15, and 5:18 MMRs in the Canada–France Ecliptic Plane Survey (CFEPS) \citep{Peti2011,Peti2017}, and the 2:9, 4:11, 5:11, 5:13, and 6:13 MMRs in the Outer Solar System Origins Survey (OSSOS) \citep{volk2016,Bann2018,Khai20}; and by combining the CFEPS, OSSOS and several other surveys, \citet{Crom22} studied more resonances beyond 50 AU, which are of even higher orders, such as 3:10, 4:17, 5:24, and 6:23. In this work, we extend the investigation to include all MMRs between 50 and 100 AU, with the aim of identifying all high-order resonances where the RKBOs could potentially be discovered. In fact, the recent study by \citet{Lyka2023} explored a similar issue, but was limited to only the 1:$n$ type MMRs. As we will demonstrate later, comparisons of the number of RKBOs in specific MMRs, other than the 1:$n$ type, may place strong constraints on the distribution of primordial objects in the Kuiper belt and perhaps on the planetary migration model.

Theoretical studies on the orbital distribution of RKBOs predicted that they could exist in extremely high-order resonances, reaching up to the 20th-order (e.g. the 3:23 MMR) \citep{Pike2017, PL2017}. Furthermore, \citet{yu2018} noted that there could be as many as 111 resonances inside 100 AU that small bodies may occupy. To gain a complete panorama of the resonances embedded in the distant Kuiper belt between 50 and 100 AU, we will first identify all observed RKBOs in this region, and then refine their orbital characteristics through theoretical and numerical studies, addressing the following issues that may be important for understanding the structure of the Kuiper belt: (1) the highest-order resonance(s) in which the RKBOs could potentially reside; (2) the distributions of eccentricities ($e$) and inclinations ($i$) of RKBOs as the order of the resonance increases; (3) the relative number of RKBOs in different resonances. Especially, for 1:$n$ type resonances, we will explore whether there is a number asymmetry between the leading and trailing resonators, similar to the 1:2 resonance within 50 AU \citep{Li2014b, Lih23, Lih24}. In addition, an important consideration of this study is the unique dynamics of RKBOs with high inclinations. Regarding the numerous resonances we will investigate, none of them is of the 1st-order, meaning that all the non-1:$n$ type resonances are constrained by the limiting curve theory that we developed in \citet{Li2020}.

The objective of the present paper is to provide a comprehensive analysis of the dynamical features (e.g. resonance centre, width, stability, and occupancy) related to the MMRs beyond 50 AU, along with a systematic database that can be directly referenced in the future. The rest of this paper is organised as follows: in Section 2, we identify the observed RKBOs with semimajor axes $a\in(50\mbox{AU}, 100\mbox{AU})$ and analyse their distribution properties. In Section 3, we examine the dynamical properties of individual resonances, ranging from 2nd- to 22nd-order, for eccentricities as large as $e=0.7$ and inclinations up to $i=40^{\circ}$. We also theoretically predict the possible ranges of $e$ and $i$ based on the permissible regions of the resonances. In Section 4, we conduct a 4-Gyr evolution study of objects originally occupying all the resonances within the 50-100 AU range, in order to inform future surveys on the orbital spaces where RKBOs beyond 50 AU may be discovered. Finally, the conclusions and discussion are given in Section 5.


\section{Observed RKBOs beyond 50 AU}

A generic $m$:$n$ resonance between the mean motions of Neptune and a Kuiper belt object (KBO) is characterised by the resonant angle:
\begin{equation}
\sigma^{(j)}=n\lambda-m\lambda_N+(m-n-2j)\varpi+2j\Omega,
\label{angle2j}
\end{equation}
where $\lambda$, $\varpi$ and $\Omega$ are the KBO's mean longitude, longitude of perihelion, and longitude of ascending node, respectively; $\lambda_N$ denotes the mean longitude of Neptune; $m(>0)$, $n(>0)$, and $j(\le0)$ are the integers, with the condition $m<n$ corresponding to external resonances. As mentioned above, the quantity $|n-m|$ is referred to as the order of the resonance. The strength of the resonance is proportional to $e^{|m-n-2j|}i^{|2j|}$, with the strongest case occurring when $j=0$, which corresponds to the resonant angle: 
\begin{equation}
\sigma_{m:n}=n\lambda-m\lambda_N+(m-n)\varpi.
\label{anglemn}
\end{equation}
This form represents the most common eccentricity-type resonance, which is generally stronger than the inclination-related resonance with $j=-1$, as seen in our previous studies of the 4:7 and 3:5 resonances within 50 AU \citep{Li2020, Li2023}. 
However, since a number of highly inclined KBOs--and even retrograde ones--have been detected \citep{Glad2009,Chen16,Peti2017}, it may be necessary to account for different values of $j\le-1$, introducing both pure $i$-type and mixed-$(e, i)$-type resonances. 
To obtain a comprehensive understanding of the $m$:$n$ resonances embedded in the distant Kuiper belt, we set the $m$ values from 1 to 20, and choose all possible $n$ values such that the resonances are consistent with semimajor axes $a=50$-100 AU. For example, regarding the 1:$n$ resonances, we explore from the 1:3 resonance (at $a=62.6$ AU) to the 1:6 resonance (at $a=99.4$ AU), and for the 2:$n$ resonances, they range from the 2:5 resonance (at $a=55.4$ AU) to the 2:11 resonance (at $a=93.8$ AU).

Then, for a given pair of $m$ and $n$, we identify the RKBOs by examining the resonant angle $\sigma^{(j)}$ defined in equation (\ref{angle2j}) for various values of $j$. As is usually done, we define the resonance centre of a resonance as $(\max{\sigma^{(j)}}-\min{\sigma^{(j)}})/2$, and the resonant amplitude $A$ as $(\max{\sigma^{(j)}}+\min{\sigma^{(j)}})/2$. The maximum and minimum values of $\sigma^{(j)}$, i.e. $\max{\sigma^{(j)}}$ and $\min{\sigma^{(j)}}$, are computed numerically over appropriate time spans. In this way, an object is regarded as an RKBO if it exhibits libration of the resonant angle $\sigma^{(j)}$, i.e. the resonant amplitude $A<180^{\circ}$, for any combination of $m$, $n$ and $j$.

\subsection{Candidate RKBO selection}

We select the KBOs with semimajor axes beyond 49 AU but within 102 AU registered in the Minor Planet Center (MPC) database\footnote{https://minorplanetcenter.net/iau/lists/MPLists.html}, as of April 29, 2024. Only samples with observations at two or more oppositions are considered. The chosen semimajor axis range is slightly wider than the $a=50$-100 AU range on which we focus. This is because the osculating $a$ of an RKBO can deviate from the nominal resonant
value:
\begin{equation}
a_{res}=(n/m)^{2/3}\cdot a_N,
\label{ares}
\end{equation}
which is calculated according to Kepler's third law, and $a_N$ is Neptune's semimajor axis. 
This choice of $a$ results in a total of 771 KBOs being considered. Note that the orbital elements of these KBOs, obtained from the MPC database, are best-fit values and do not include uncertainties. Orbital uncertainties will be taken into account when identifying an RKBO, as discussed later.

In order to explore a large number of combinations of $m$, $n$, and $j$ for each KBO, considering computational and analytical costs, we conduct short-term runs to select the candidate RKBOs. Specifically, we first integrate the orbits of the selected KBOs over a 1 Myr time span, under the gravitational perturbations of the four Jovian planets. This timescale is generally sufficient, as it exceeds the libration periods of Neptune's exterior resonances within 100 AU, such as the 1:6 resonance at $\sim99.4$ AU, which has a libration period of about 0.8 Myr \citep{Malh2019,Gall2020}. In this paper, we employ the SWIFT\_RMVS3 symplectic integrator \citep{Levi1994} for all numerical calculations, with a time-step of 0.5 yr, which is about 1/12 of the shortest orbital period (Jupiter’s); and, in the integration, the KBOs are treated as massless particles, i.e. they are affected by the planets but do not affect the motion of the other bodies.

\begin{figure}
 \hspace{0cm}
  \centering
  \includegraphics[width=9cm]{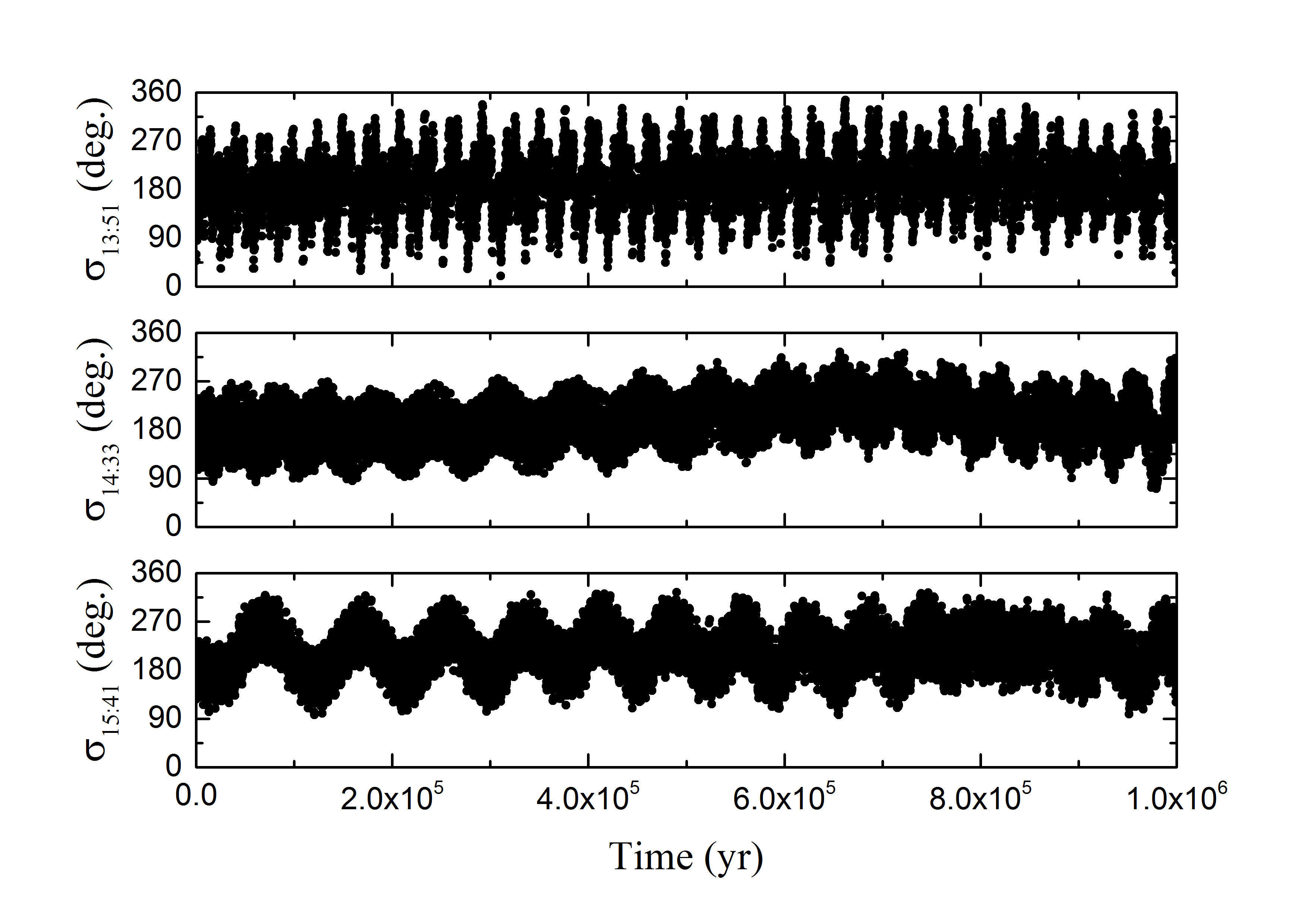}
  \caption{Time evolution of the resonant angle for candidate RKBOs in extremely high-order resonances from the 1 Myr integration. The upper, middle, and lower panels correspond to the 13:51, 14:33, and 15:41 resonances, respectively. All three cases are associated with eccentricity-type resonances, with resonant angles $\sigma_{m:n}$ as defined in equation (\ref{anglemn}).}
  \label{ResCandidates}
\end{figure}

At the end of the integration, we examine the time evolution of the resonant angles $\sigma^{(j)}$ for as many combinations of $m$, $n$, and $j$ as possible. An object is regarded as a candidate RKBO if a particular resonant angle $\sigma^{(j)}$ exhibits libration throughout the 1 Myr time span. 
The variation of each $\sigma^{(j)}$ was processed through a classification algorithm to diagnose its libration behaviour, determined by the resonant amplitude $A<180^{\circ}$. To ensure accurate identification, for a small subset of candidate RKBOs with the largest $A$ values of $160^{\circ}$-$180^{\circ}$, the time evolution of $\sigma^{(j)}$ was further examined by eye.
As a result, we found 261 candidate RKBOs beyond 50 AU for $m=1$ to 15, within the range of $m\le20$ that we consider. This indicates that, out of the 771 objects currently observed in the distant Kuiper belt, more than a third may be experiencing resonances. More interestingly, there are several KBOs located inside extremely high-order resonances that have not been noticed before, such as the 13:51, 14:33, and 15:41 eccentricity-type resonances (i.e. $j=0$). The time evolution of these candidate RKBOs is displayed in Fig. \ref{ResCandidates}. During the 1 Myr time span, all their resonant angles $\sigma_{m:n}$, as defined in equation (\ref{anglemn}), are consistently librating. As a matter of fact, under a longer integration time, as we will discuss below, these highest-order resonators will gradually diffuse out of their respective resonances, transitioning from libration to circulation. Nevertheless, their existence suggests that the potential objects in the distant Kuiper belt may also temporarily inhabit similar weak resonances, even for timescales shorter than a full libration period, similar to the phenomenon of resonance sticking \citep{Lyka2007b,yu2018}. For instance, \citet{Gall2006a} proposed that an object with $a>50$ AU may experience temporary captures in Neptune's MMRs and eventually diffuse into the Oort cloud.

\begin{figure}
 \hspace{0cm}
  \centering
  \includegraphics[width=9cm]{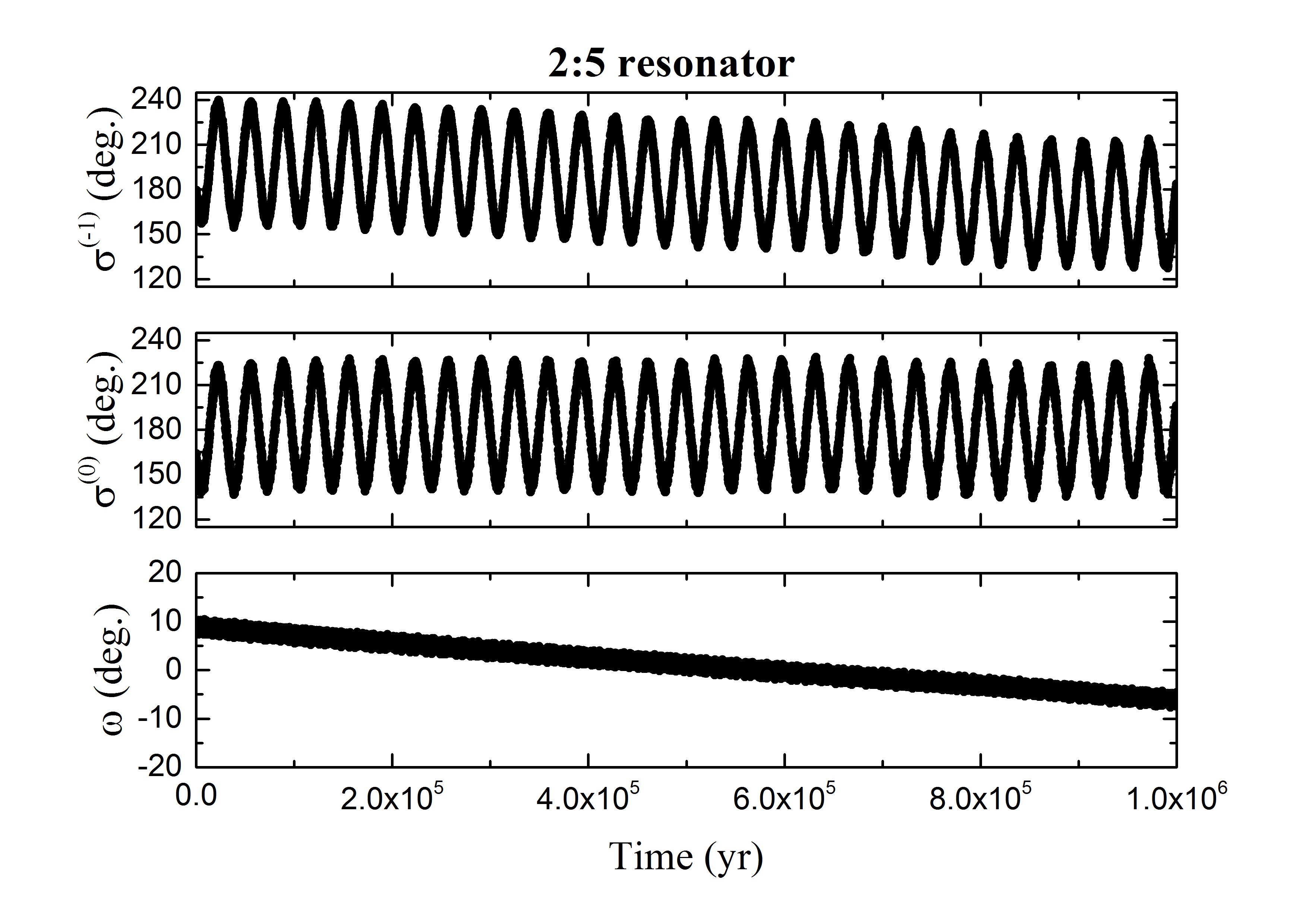}
   \caption{Time evolution of a candidate RKBO associated with the 2:5 resonance: (upper panel) the inclination-related resonant angle $\sigma^{(-1)}=5\lambda-2\lambda_N-\varpi-2\Omega$; (middle panel) the eccentricity-type resonant angle $\sigma^{(0)}=5\lambda-2\lambda_N-3\varpi$; (lower panel) the argument of perihelion $\omega$.}
  \label{ResCandidates25}
\end{figure}

For the resonant angle $\sigma^{(j)}$ with $j\le-1$, the corresponding resonance is related to the inclination and is referred to as either a pure $i$-type or a mixed-$(e, i)$-type resonance. For the low-inclination population, the resonance's strength, which is proportional to $i^{|2j|}$, could be too weak to sustain these resonances. In contrast, for the high-inclination population, it is theoretically possible for inclination-related resonances to arise, characterised by the libration of $\sigma^{(j)}$. In \citet{Li2020}, we identified an independent 4:7 resonator ($a\approx43.7$ AU) with $\sigma^{(-1)}$ librating while $\sigma^{(0)}$ simultaneously circulates. However, among all candidate RKBOs beyond 50 AU obtained here, if their $\sigma^{(j)}$ librate for $j \geq 1$, then $\sigma^{(0)}$ must also exhibit libration. In other words, no independent pure $i$-type or mixed-$(e, i)$-type resonances have been detected. A representative example of a 2:5 candidate RKBO, associated with the pseudo mixed-$(e, i)$-type, is given in Fig. \ref{ResCandidates25}. During the 1 Myr evolution, as seen in the upper panel, this object appears to be inside a mixed-$(e, i)$-type resonance, diagnosed by the libration of the resonant angle $\sigma^{(-1)}=5\lambda-2\lambda_N-\varpi-2\Omega$. But it is actually experiencing the eccentricity-type MMR coupled with the Kozai mechanism, as indicated by the libration of both another resonant angle $\sigma^{(0)}=5\lambda-2\lambda_N-3\varpi$ (i.e. $\sigma_{2:5}$) and the argument of perihelion $\omega$, shown in the middle and lower panels, respectively. We note that the two resonant angles are related via $\sigma^{(-1)}=\sigma^{(0)}+2\omega$. Thus, when $\sigma^{(0)}$ librates about the resonance centre at $180^{\circ}$, the libration of $\omega$ around $0^{\circ}$ (or $180^{\circ}$) leads to the pseudo-libration of $\sigma^{(-1)}$ around $180^{\circ}$. As we argued in \citet{Li2020}, this type of coupled resonance should be classified as an eccentricity-type resonance. Given that no independent inclination-related resonances are found among the observed KBOs beyond 50 AU, this suggests that the eccentricity-type resonance is the dominant mechanism. Therefore, in the longer-term evolution for identifying the RKBOs considered later, only the eccentricity-type resonance associated with the resonant angle $\sigma_{m:n}$ will be taken into account.

The resonance centres for some RKBOs, taking the non-1:$n$ type resonators as an example, may deviate from the nominal value of $180^{\circ}$ As illustrated in \citet{Li2014a}, there are two kinds of libration centres for the resonant angle: the special libration centre (SLC) and the general libration centre (GLC). The SLC corresponds to the stable equilibrium point where the time derivative of the semimajor axis ${\rm d}a/{\rm d}t=0$; whereas the GLC is the mean value of the SLC during time evolution, and it is actually what is usually called the `resonance centre'. For a visual illustration of these two kinds of libration centres, the readers are referred to Fig. 2 in \citet{Li2014a}. At low inclinations, the SLC and the GLC are identical and both are fixed at $180^{\circ}$. However, at high inclinations, the SLC would oscillate as the argument of perihelion $\omega$ varies. Since the SLC reflects the secular variation of the resonant angle, using $(\max{\sigma^{(j)}}-\min{\sigma^{(j)}})/2$ to accurately determine the GLC requires a timescale longer than the full variation period of the SLC. Below this timescale, we can acquire merely a partial variation of the SLC, rather than the full variation, resulting in a deviation of the GLC from the true value of $180^{\circ}$. Therefore, the timescale for evolution should at least be comparable to the period of $\omega$-change between $0^{\circ}$ and $360^{\circ}$, which is on the order of 10 Myr. Related to that, there is a special case in which the SLC exhibits an oscillation within a small subset of $0^{\circ}$-$360^{\circ}$ during the short 1 Myr integration performed above, but over a longer 10 Myr timescale, it indeed circulates. This can result in local libration but global circulation in the evolution of the resonant angle, which would affect the identification of intrinsic RKBOs if short integration is used. This special case, observed only in the high-inclination population, led to the development of our limiting curve theory, which defines the permissible region of libration in the $(e, i)$ space \citep{Li2020}. Based on the above arguments, after selecting the candidate RKBOs, we will examine their 10 Myr evolution in the next subsection to more securely identify the RKBOs with high inclinations.


Before proceeding further, there is a dynamical feature of Neptune's resonances in the Kuiper belt that we wish to briefly mention. \citet{Malh2019} pointed out that when the eccentricity of a KBO exceeds the critical value for Neptune-crossing, i.e. when its perihelion lies interior to Neptune’s orbit, a new resonance zone emerges with the stable libration centre at $0^{\circ}$. However, among the 771 observed KBOs with semimajor axes in the range of $a=50$-100 AU, none have been found to reside in this new resonance zone.

\subsection{Resonant population identification}

As discussed in Section 2.1, to account for the influence of the SLC variation observed in the high-inclination RKBOs, we need to perform the 10 Myr integration to confirm the resonant states of the 261 candidate RKBOs. Moreover, because some of these candidates reside in high-order resonances, the resonance widths in the semimajor axis space can be quite narrow. In this context, the orbital uncertainty may have a non-negligible impact on the identification of RKBOs. 
Therefore, in addition to using the best-fit orbital elements obtained from the MPC database, we introduce uncertainty for each candidate RKBO by generating 10 clones that differ only in their semimajor axes, sampled from a uniform distribution around the best-fit orbit \citep{Lyka2007}.

For the KBOs we considered, they have multiple-opposition observation arcs. According to data from the Asteroids Dynamic Site (AstDyS)\footnote{https://newton.spacedys.com/astdys/}, the 1-$\sigma$ uncertainties in semimajor axis, measured by $\Delta a / a$, are generally at the level of $\lesssim 0.05\%$ \citep{Li2020}. Within this uncertainty range, we generate clones with semimajor axes that differ marginally from the best-fit value. In fact, in \citet{Lyka2023}, the orbital uncertainties in $a$ for KBOs beyond 50 AU were assumed to be $\Delta a / a=0.025\%$. Therefore, we believe that the uncertainty range we have chosen for $a$ is reasonable.

The candidate RKBOs are then integrated, along with 10 clones for each, over a 10 Myr timescale. Since we are only concerned with eccentricity-type resonances, the focus is on determining whether the resonant angle $\sigma_{m:n}$, as defined in equation (\ref{anglemn}), undergoes libration or circulation. We remind that, as stated in Section 2.1, the integrator SWIFT\_RMVS3 is adopted to perform the numerical calculations with a time-step of 0.5 yr, and the candidate RKBOs are treated as massless particles in the integration; in addition, a classification algorithm is used to diagnose the resonant behaviour characterised by the libration of $\sigma_{m:n}$, i.e. with a resonant amplitude $A<180^{\circ}$, while an eye examination is further conducted if $A$ is as large as $160^{\circ}$-$180^{\circ}$.
We begin by identifying a candidate RKBO as an RKBO if the object with its best-fit orbit can consistently remain in resonance. Based on this criterion, we further classify such an RKBO as follows: (1) if all of its 10 clones also remain in the same resonance, it is classified as a secure RKBO; (2) if 1 to 9 clones remain in the same resonance, it is classified as a probable RKBO; (3) if none of the clones do, it is classified as an insecure RKBO.

During the revision of this manuscript, we became aware that the MPC Explorer\footnote{https://data.minorplanetcenter.net/explorer/} now provides orbital uncertainties, which were not available at the start of our work but can be used at present. Given the reasonableness of the assumed uncertainty $\Delta a / a=0.05\%$, we additionally examined the actual $\Delta a / a$ values for particular objects of interest and updated the resonant status of their respective clones accordingly. By `particular objects of interest', we refer to RKBOs residing in sparsely populated resonances that host only 1-3 samples, all of which have $\Delta a / a$ values significantly larger than 0.05\%. In total, there are four such objects, listed in Table \ref{uncommonRKBOs} (see Appendix \ref{sec:uncommon RKBOs} for details). From this table, we further note that the accuracy of the orbital determination is primarily governed by the number of oppositions.

\subsubsection{Resonance occupancy}

\begin{figure}
 \centering
  \begin{minipage}[c]{0.5\textwidth}
  \hspace{-1 cm}
  \centering
  \includegraphics[width=9.5cm]{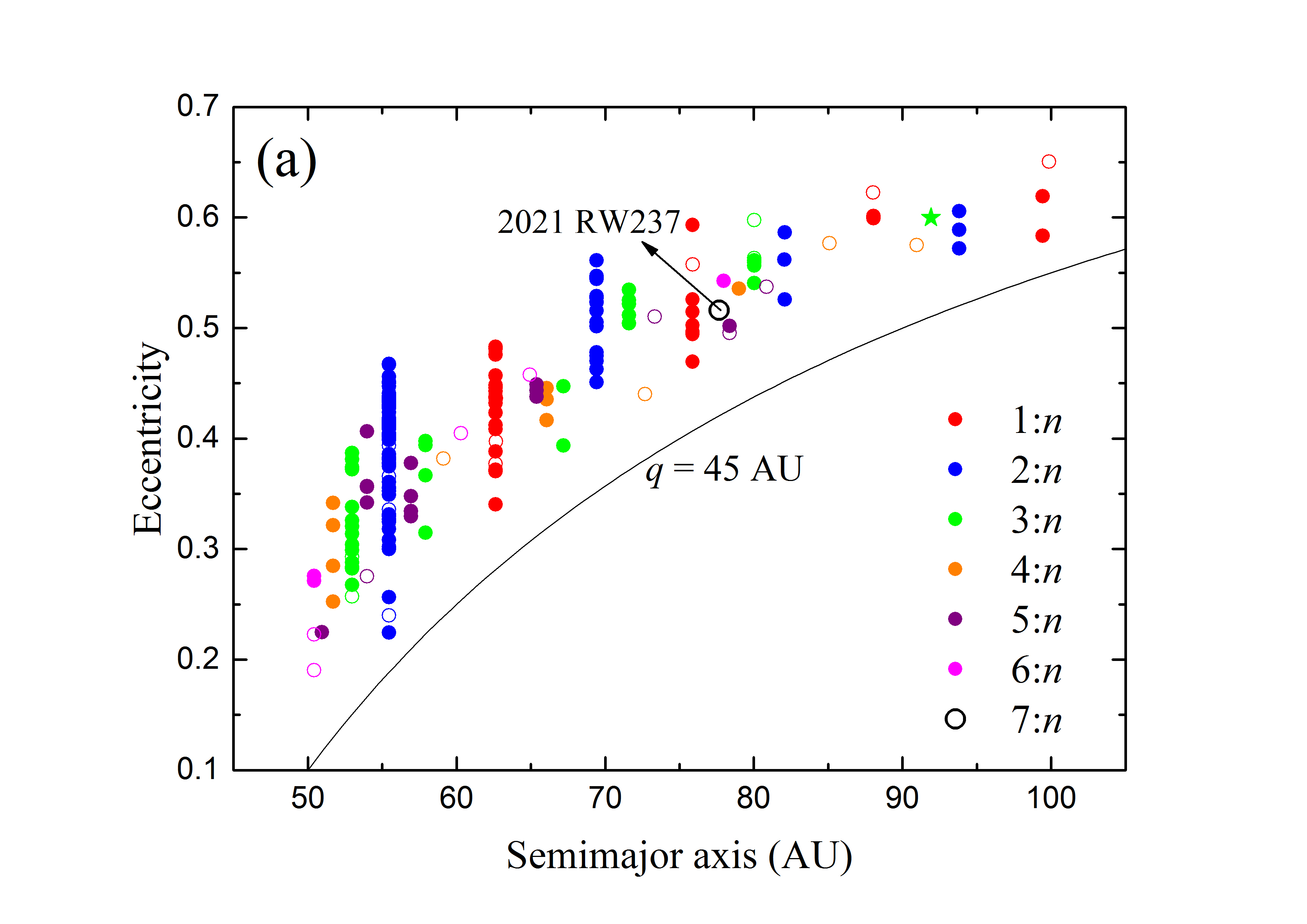}
  \end{minipage}
  \begin{minipage}[c]{0.5\textwidth}
  \hspace{-1 cm}
  \centering
  \includegraphics[width=9.5cm]{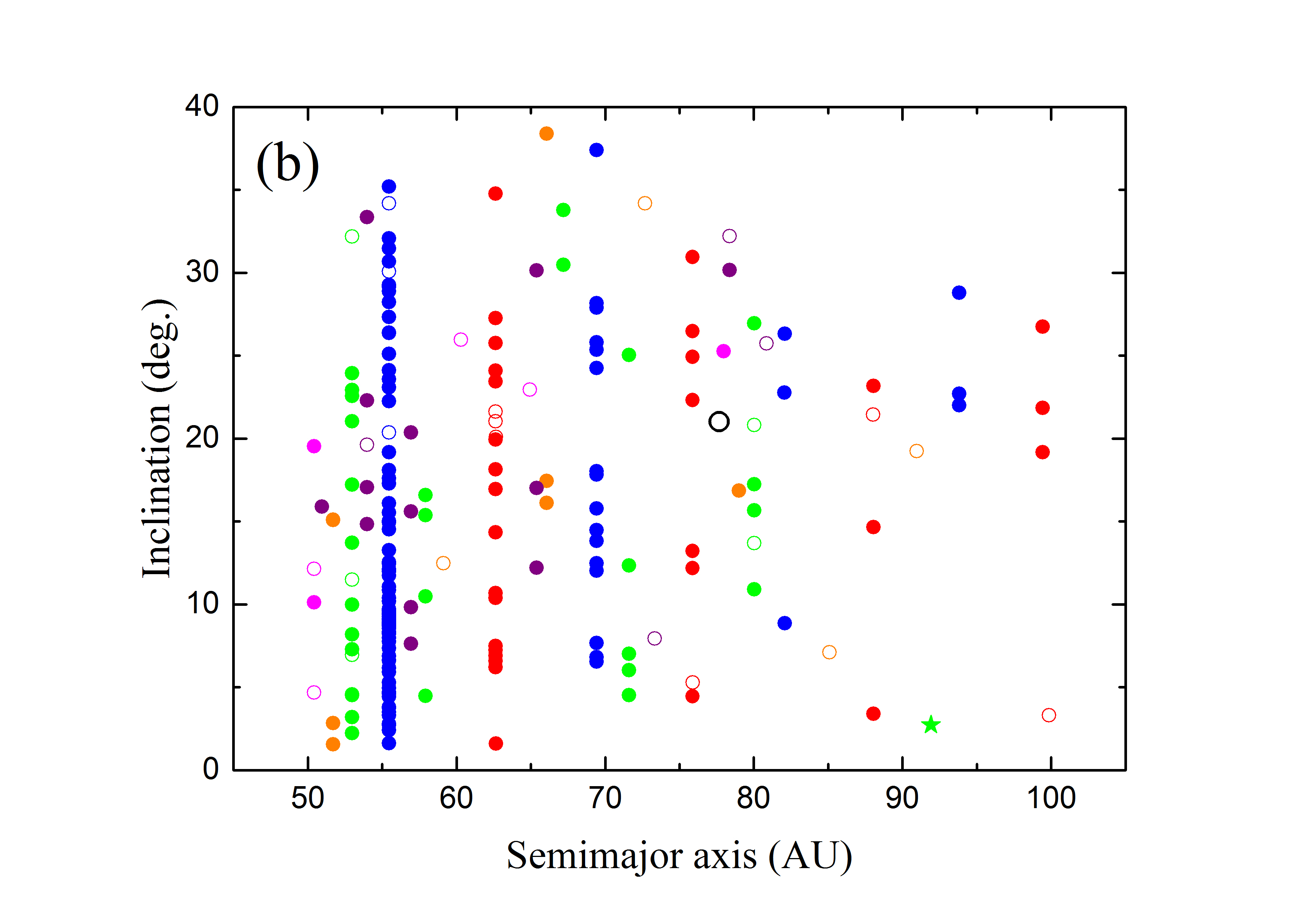}
  \end{minipage}
  \caption{Orbital distributions of the currently observed RKBOs beyond 50 AU. The original data is obtained from the MPC as of April 29, 2024, including only objects with multiple-opposition observation arcs. These RKBOs are distributed across the 1:$n$ (red), 2:$n$ (green), 3:$n$ (green), 4:$n$ (orange), 5:$n$ (purple), 6:$n$ (Magenta), and 7:$n$ (black) resonances; and the sole 7:$n$ RKBO, 2021 RW237 in the 7:29 resonance, is specifically marked in panel (a). For clarity, the semimajor axes, eccentricities and inclinations are taken to be the average values over the 10 Myr integration. The filled circles stand for secure RKBOs, the open circles denote probable RKBOs, while the star indicates an insecure RKBO (2021 LS43 in the 3:16 resonance). For reference, a solid line is plotted at a perihelion distance of $q = 45$ AU, above which all the MPC RKBOs reside (i.e. with $q > 45$ AU).}
  \label{real}
\end{figure}

Using the identification scheme described above, we find a total of 207 RKBOs in the distant Kuiper belt, extending from 50 to 100 AU. Of these, 177 are classified as secure RKBOs, 29 as probable RKBOs with at least one resonant clone, and 1 as an insecure RKBO (2021 LS43 in the 3:16 resonance). Fig. \ref{real} displays the orbital distributions of these RKBOs, with the secure, probable and insecure objects indicated by filled circles, open circles, and stars, respectively. In order to position each RKBO approximately at the nominal resonance location, as done by \citet{Lyka2005}, the figure uses numerical proper orbital elements (semimajor axis, eccentricity, and inclination), computed as the average values over the 10 Myr integration period. We note that a similar dynamical classification of multi-opposition KBOs was performed by \citet{volk2024}, and both their and our studies use nearly the same procedure to identify resonant objects (for the best-fit orbits), namely determining the entire resonant population regardless of whether the objects are classified as secure resonators or not. However, in \citet{volk2024}, the list of MPC KBOs is from December 2023 and contains 3357 objects in total, while our list, from April 2024, includes 4883 objects. Naturally, the inclusion of more observed KBOs leads to a larger population of identified resonators. 
For example, as found by Volk and Van Laerhoven, there are 56 objects in the 2:5 resonance, whereas we have identified 77. Among their 56 2:5 resonators, 52 are also included in our sample. The absence of the remaining four is due to our stricter criterion: we require an object to librate throughout the entire 10 Myr integration, whereas they only require libration for half of that duration.
We do not explore the classification difference in depth, as, in addition to our larger sample size, the orbital elements have also been updated based on a more recent MPC list.

As shown in Fig. \ref{real}, the RKBOs are distributed across numerous $m$:$n$ resonances, ranging from the 2nd-order 1:3 MMR to the 22nd-order 7:29 MMR. For candidate RKBOs in higher-order resonances (i.e. with a larger difference of $m-n$), or for those associated with $m\ge8$, they cannot maintain stable libration. Notably, the 29 probable RKBOs do not significantly affect the overall distribution of RKBOs among resonances of various orders, and some details will be discussed below. We point out that the number ratio of RKBOs to candidate RKBOs is found to be $207/261=79\%$, suggesting the presence of some high-order, weak resonances in the considered region beyond 50 AU that are unable to effectively trap KBOs over long timescales. This highlights the importance of using the 10 Myr integration timescale, rather than 1 Myr, to more securely identify the RKBOs. We note that a 10 Myr integration is the standard proposed by \citet{Glad2008} for resonant classification, and it is also used in our previous works \citep{Li2014a, Li2014b,Li2020, Li2023} as well as in \citet{volk2024}.

The details of how the RKBO occupancy depends on the resonance order are provided below:\\

(1) \underline{Resonance order $<10$:} We find that most RKBOs are associated with resonance orders lower than 10, accounting for 185 of the 207 objects (approximately 90\%). This result is unsurprising for two reasons. First, within the 50–100 AU region, all 1:$n$-type resonances (from the 1:3 to the 1:6 MMR) have orders of $|n-1|\le5$, and all 2:$n$-type resonances (from the 2:5 to the 2:11 MMR) have orders of $|n-2|\le9$. These two types of resonances are theoretically the most stable. Furthermore, the identified RKBOs occupy all individual $m$:$n$ resonances within these two categories, meaning that for $m=1$ and $m=2$, $n$ spans all potential values. Second, for $m=3$ to $m=6$, RKBOs inhabit nearly all $m$:$n$ resonances of orders $<10$, with the exception of the 5:14 MMR, which has a resonance order of 9 and is empty. Nevertheless, we observe that the higher-order 5:16 MMR is occupied. We believe that the absence of RKBOs in the 5:14 MMR is likely due to incomplete observations, and this argument will be supported by the stability analysis conducted subsequently.

(2) \underline{Resonance order of 10-20:} 
For the $m$:$n$ resonances ranging from the 10th-order to the 20th-order, only those with $m\ge3$ fall within the considered $a$-range of 50-100 AU.
A total of 21 RKBOs are identified in these resonances, with more than 50\% classified as secure RKBOs that span $m$ values from 3 to 6. We note that in \citet{volk2024}, their classification shows additional resonances with $m\ge7$, such as 2008 OG19 in the 7:23 resonance and 2010 TB192 in the 9:22 resonance. This difference arises because, in their classification scheme, an object is considered resonant if it librates for more than half of a 10 Myr integration, whereas we adopt a stricter criterion requiring libration throughout the entire 10 Myr integration.
Furthermore, we confirm that, associated with such high-order resonances, all RKBOs are in eccentricity-type resonances, characterised by the resonant angle $\sigma_{m:n}$. 
In Fig. \ref{real}(a), these RKBOs are located at semimajor axes greater than 60 AU, i.e. beyond the 6:17 MMR ($a=60.3$ AU), which is the innermost resonance within the 10th-order to the 20th-order range. By examining the RKBOs corresponding to the 3:$n$ (green), 4:$n$ (orange), 5:$n$ (purple), and 6:$n$ (Magenta) resonances in the region where $a>60$ AU, we observe that their eccentricities range from about 0.4 to 0.6, which are relatively large values. First of all, the large $e$-values could be due to the fact that these objects have smaller perihelion distances and are therefore more easily observable, as we will discuss later. Besides, since the resonance's strength is proportional to $e^{|n-m|}$, and given the very high resonance order of $|n-m|\ge10$, it is natural that the $e$ values of the relevant RKBOs are expected to be large in order to ensure that the resonance's strength is not too low.

It is well known that as the resonance order increases, the resonance's strength generally decreases. To provide a direct perception of resonance's strength using numerical values, we consider the 21 $m$:$n$ RKBOs associated with different resonance orders ($|n-m|$) of 10-20. Among them, we select objects with nearly constant perihelion distances $q$ and inclinations $i$, allowing for a more informative comparison regarding their $e$ values. Accordingly, three representative samples with $q\sim35.5$ AU and $i\sim24^{\circ}$ are found, located in the 6:17, 6:19, and 6:25 MMRs, respectively. We then calculate the resonance's strength parameter, $e^{|n-m|}$, where the averaged values of $e$, as presented in Fig. \ref{real}, are used. For these three resonances, the parametrised strength values are: $4.8\times10^{-5}$ for the 6:17 ($e=0.405$), $3.9\times10^{-5}$ for the 6:19 RKBO ($e=0.458$) and $0.9\times10^{-5}$ for the 6:25 RKBO ($e=0.543$). It is evident that the resonance's strength markedly decreases as the resonance order increases.

The number of RKBOs in the 10th- to 20th-order resonances is 21, which is much smaller than the 185 RKBOs associated with resonance orders lower than 10, as discussed just above. This significant decrease in the resonant population may result from the weakening of resonance's strength at higher resonance orders.
In a re-examination of the migration model from \citet{bras2013}, developed within the framework of the Nice model, \citet{Pike2017} conducted a detailed study of the resonant structures of test KBOs after the planet migration had ceased. They found dozens of Neptune's resonances beyond 50 AU that could potentially be occupied, with the highest-order one being the 3:23 MMR, which is a 20th-order resonance. This suggests that if the resonance order continues to increase to $>20$, the resonance's strength will weaken further, making it even more difficult for objects to be maintained within such resonances.

\begin{figure}
 \hspace{0cm}
  \centering
  \includegraphics[width=9cm]{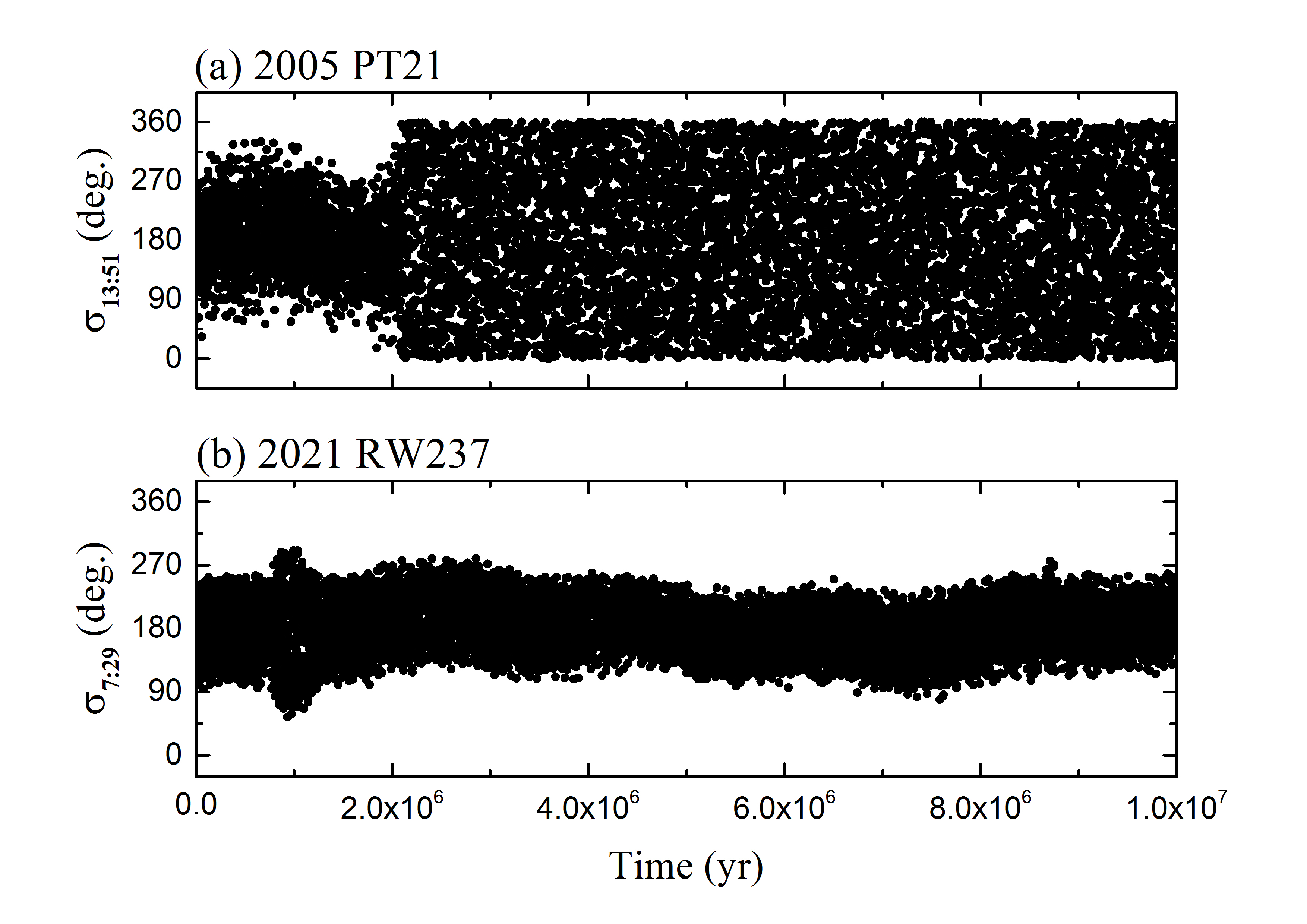}
   \caption{Two typical scenarios illustrating the time evolution of the resonant angle for two KBOs associated with resonances higher than the 20th-order. (Panel a) The object 2005 PT21 initially experiences the 13:51 resonance (38th-order)  but eventually transitions to a circulation state. (Panel b) The object 2021 RW237, trapped in the 7:29 resonance (22nd-order), exhibits stable libration over 10 Myr and is thus identified as an RKBO.}
  \label{RKBOs20}
\end{figure}

(3) \underline{Resonance order $>20$:} As we have just argued, the RKBOs associated with resonance order $>20$ could be rare. Regarding current observations, some candidate RKBOs, as found in Section 2.1, could inhabit resonances exceeding the 20th-order. 
For example, the candidate RKBO 2005 PT21 is associated with the 38th-order 13:51 MMR. This object has been observed at 6 oppositions, spanning an arc length of 17 years, and therefore has a very high orbital accuracy of $\Delta a / a \sim $0.046\%. The time evolution of its resonant angle, $\sigma_{13:51}$, is displayed in Fig. \ref{RKBOs20}(a).
We can observe that during the initial 1 Myr integration used for the selection of candidate RKBOs, $\sigma_{13:51}$ librates around the resonance centre at $180^{\circ}$, with a quite large resonant amplitude of about $150^{\circ}$. The 13:51 MMR maintains up to around 2 Myr, after which 2005 PT21 diffuses out of this resonance as $\sigma_{13:51}$ turns into a circulation state. In subsequent evolution, despite losing the phase protection provided by the resonance mechanism, 2005 PT21 can still survive for the remainder of the 10 Myr integration, aided by its high perihelion distance of $>35$ AU, which keeps it well beyond Neptune's orbit.

For the $m$:$n$ resonances of orders $>20$, the strongest ones--such as the 2:23, 3:25, and 4:25 MMRs--are outside the considered $a$-range of 50-100 AU. Accordingly, such extremely high-order resonances that could possibly be occupied are the 5:$n$ MMRs with $n\ge26$, the 6:$n$ MMRs with $n\ge29$, and the 7:$n$ MMRs with $n\ge29$, while the resonant populations with $m\ge8$ appear to be unstable, as discussed above.
Indeed, among the identified RKBOs, we find an object, 2021 RW237, residing in the 22nd-order 7:29 MMR. Remarkably, this is the only RKBO associated with the resonance order $>20$ within the $a$-range of 50-100 AU and is highlighted as a large black (open) circle in Fig. \ref{real}. As depicted in Fig. \ref{RKBOs20}(b), over the entire 10 Myr timescale used for the identification of the resonant population, 2021 RW237 can maintain a stable libration of its resonant angle $\sigma_{7:29}$. It should be noted that 2021 RW237 cannot be confidently classified as a secure RKBO. Of its 10 clones, each generated with a different semimajor axis, only one exhibits libration of $\sigma_{7:29}$ while the remaining nine display circulation. This dynamical state arises from the large orbital uncertainty of this object, as discussed in Appendix \ref{sec:uncommon RKBOs}.

Therefore, the two examples shown in Fig. \ref{RKBOs20}, namely 2005 PT21 and 2021 RW237 from observations, indicate that RKBOs in the highest-order resonances exceeding the 20th-order may exist, but the extremely weak strength of the resonance results in such RKBOs being rare. We will further discuss this in the subsequent stability analysis.\\

In the above discussion of RKBOs associated with resonance orders of 10-20, we argued that these objects should be on high-$e$ orbits in order to maintain sufficient resonance strength. In fact, a considerable fraction of RKBOs residing in relatively lower-order ($<10$) resonances are also found on very eccentric orbits. It is important to note that nearly all of the known MPC KBOs with $a>50$ AU--the range considered here--exhibit large values of $e$, as those objects have perihelia that are much closer and thus much, much more observable. Although a few less eccentric objects have been detected at greater distances, the overwhelming majority of large-$a$ KBOs are discovered near perihelion. As shown in Fig. \ref{real}(a), all of these large-$a$ objects have perihelion distances ($q$) smaller than 45 AU, close to the `kernel’ of the main classical Kuiper belt \citep{Li2020}. We additionally note that, at the time of this paper's data access, there were no submitted surveys with significant sensitivity to the low-$e$ 2:5 or more distant resonances. Indeed, among the 77 identified 2:5 RKBOs, all have $e>0.2$, and 74 of them (i.e. over 96\%) have $e>0.3$. From this perspective, since observations are biased toward detecting objects with large eccentricities, it is necessary to provide a plausible eccentricity distribution for the potential RKBOs beyond 50 AU, particularly regarding the lower limit of $e$, as we will discuss below in Sections 3 and 4.

As the fifth paper in our series on the high-inclination population in the Kuiper belt, this study also considers the inclination distribution of RKBOs as a key area of interest.
In Fig. \ref{real}(b), we focus on the 1:$n$ (red), 2:$n$ (blue), and 3:$n$ (green) resonances, which host the most numerous identified RKBOs. Furthermore, we limit our consideration to the RKBOs with semimajor axes $a<80$ AU, as these objects are relatively easier to detect. 
The plot in Fig. \ref{real}(b) shows that, whether for the 1:$n$ type resonances having both symmetric and asymmetric libration modes \citep{Beau94,Gall2006b,Li2014b}, or for the 2:$n$ and 3:$n$ type resonances exhibiting only symmetric libration mode, the inclination ranges of the corresponding RKBOs appear similar: the minimum inclination is close to $0^{\circ}$, while the maximum inclination reaches as high as $30^{\circ}$-$40^{\circ}$. 
Given their broad range of semimajor axes, spanning from 50 to 80 AU, it is likely that these RKBOs were not formed in situ. Instead, they were possibly gravitationally scattered by Neptune to such widely spread positions along with the orbital excitation, and subsequently captured by various resonances \citep{kaib16,nesv16b,Pike2017,yu2018}. This scenario helps to explain why they share similar inclination distributions.

Before concluding this subsection, we briefly discuss the candidate RKBOs that fail to be identified as RKBOs. There are 54 such objects, referred to as failed RKBOs, out of a total of 261 candidates. After the initial 1 Myr, during which their individual resonant angles $\sigma_{m:n}$ were librating, some of the failed RKBOs are completely released from resonances as $\sigma_{m:n}$ transitions into consistent circulation, similar to the evolution of 2005 PT21 displayed in Fig. \ref{RKBOs20}(a); while the remaining objects undergo frequent alternations between libration and circulation. In either case, the failed RKBOs go through periods of circulation, during which they are no longer under the phase protection of resonance. Nonetheless, nearly all of the 54 failed RKBOs survive around their original resonant locations. This is because, in fact, their perihelia are more than 3 AU beyond Neptune's orbit. For instance, the object 2005 PT21 can persist in the vicinity of the 13:51 MMR despite not being in this resonance, as evidenced by the consistent circulation of the resonant angle $\sigma_{13:51}$ in the latter part of its evolution. However, since the non-resonant population is not the subject of this study, we will not enter into more details.

\subsubsection{Asymmetric 1:$n$ resonances}

For the 1:2 MMR at 47.8 AU, observations over the past decade have shown that more RKBOs reside in the leading swarm than in the trailing one \citep{Li2014b, Lih23}. 
Previous studies expect that not only in the 1:2 MMR but also across all 1:$n$ resonances in the distant Kuiper belt beyond 50 AU, the leading swarm tends to be consistently more populated \citep{Chia2002, Murr05, PL2017, Lih24}.

Now, we turn our attention back to the 1:$n$ resonances beyond 50 AU, specifically those with $n\ge3$ which are the subject of this study. According to the libration states of the identified 1:$n$ RKBOs, Table \ref{1tonRKBO} details the number of objects classified into three swarms: leading, trailing, and symmetric. For the symmetric 1:$n$ RKBOs, these refer to objects whose resonant angles exhibit symmetric libration (i.e. around $180^{\circ}$) or, in rare cases, alternate between symmetric and asymmetric librations. Our analysis focusses primarily on secure RKBOs, but for reference, the number of probable RKBOs is also provided in brackets. Notably, all identified 1:$n$ RKBOs associated with both the leading and trailing asymmetric librations are secure RKBOs.

As shown in Table \ref{1tonRKBO}, the leading RKBOs outnumber the trailing ones in both the 1:3 and 1:4 MMRs. The comparison is not applicable for the higher-order 1:5 MMR, as no RKBOs are found in the asymmetric libration state. In the highest-order 1:6 MMR, there are two RKBOs, both of which reside in the leading swarm. Together with the 1:2 MMR within 50 AU, the number difference appears to be a distinctive feature across all 1:$n$ resonances, where the leading swarm generally hosts a larger population of RKBOs than the trailing swarm. This difference in number could be closely related to the origin of RKBOs in the planetary migration model \citep{Pike2017,PL2017}, with long-term evolution likely contributing further. For instance, a similar feature is observed among Jupiter Trojans, where more objects are in the L4 (leading) swarm than in the L5 (trailing) swarm. Under the current Solar system configuration, gigayear-scale evolution has been shown to produce approximately 10\% more Jupiter Trojans around L4 than around L5 \citep{Sisto14,Li2023b}. The long-term stability of the leading and trailing RKBOs in 1:$n$ resonances, including their dependence on eccentricity and inclination, will be discussed in detail in Section 4.


Nevertheless, it is necessary for us to recognise the limitations of the known MPC sample more clearly. It seems that the real 1:$n$ RKBOs in the MPC have been preferentially found in the leading islands rather than in the trailing islands. This apparent distribution may be biased by the specifics of the observational surveys in which objects were (and were not) discovered. \citet{Chen19}, based on OSSOS observations, suggests that there may be no significant difference in the number of 1:2 RKBOs between the leading and trailing swarms. That paper used a sample with known biases, accounted for those observing biases, and, after the proper analysis was done, the authors could neither prove that the intrinsic distributions of the leading and trailing islands are the same, nor prove that these two islands are different. However, a similar de-biasing process cannot be effectively applied to the full MPC sample. While it is possible that with the Legacy Survey of Space and Time (LSST), we will obtain a sample capable of testing whether there is an intrinsic difference between the leading and trailing populations.

\begin{table}
\hspace{-0.5cm}
\centering
\begin{minipage}{8.3cm}
\caption{Libration states of the observed RKBOs in the 1:$n$ resonances presented in Fig. \ref{real}. Regular digits refer to secure RKBOs, while bracketed digits refer to probable RKBOs.}      
\label{1tonRKBO}
\begin{tabular}{c c c c c c}        
\hline                 
Resonance    &    Leading    &    Trailing     &    Symmetric    &    Total         \\
 
\hline

   1:3       &       9       &        3        &      5+(3)      &    17+(3)       \\

   1:4       &       4       &        2        &      1+(1)      &    7+(1)       \\

   1:5       &       0       &        0        &      3+(1)      &    3+(1)       \\

   1:6       &       2       &        0        &      1+(1)      &    3+(1)       \\
    
\hline
\end{tabular}
\end{minipage}
\end{table}

Having discussed the general distribution of the real 1:$n$ RKBOs, we next examine the individual populations in the leading, trailing, and symmetric libration states.
We observe that if they experience one of these three states during the initial 1 Myr integration (i.e. for candidate RKBO selection), they tend to maintain their respective libration states throughout the entire 10 Myr integration (i.e. for resonant population identification). 
Although transitions between different libration states are theoretically possible and have been previously observed in 1:2 RKBOs near 47.8 AU \citep{Li2014b}, such transitions appear intrinsically rare for 1:$n$ RKBOs beyond 50 AU due to their larger $e$ values, which enhance stability. Among the 30 secure 1:$n$ RKBOs, only one object, 2011 WJ157, is detected to undergo transitions between different libration states, following a sequence of leading, trailing, symmetric and eventually returning to the leading libration by the end of the 10 Myr integration. 

The stability noted above may be attributed to the fact that the currently known 1:$n$ RKBOs have relatively high eccentricities of $e>0.3$. It should be kept in mind that the lack of lower-$e$ samples could result from observational bias favouring the detection of larger-$e$ objects with smaller perihelion distances, as we discussed in Section 2.2.1. If there are any low-$e$ 1:$n$ RKBOs to be found in the future, the stability associated with persistent leading, trailing, or symmetric libration state could be better understood.


\section{Resonant features and the effects of high inclinations}

For the RKBOs currently observed beyond 50 AU, as shown in Fig. \ref{real}(a), their eccentricities $e$ range from 0.18 to 0.65. Consequently, we will restrict our theoretical study to $e$ values within the range of 0.1-0.7. Adopting $e\ge0.1$ offers the advantage that, in this range, the libration centre in terms of the semimajor axis $a$ remains fixed at the nominal resonance location, $a_{res}$, as defined in equation (\ref{ares}). In contrast, if $e$ approaches zero, the libration centres of Neptune's resonances in the Kuiper belt diverge from $a_{res}$ and shift outward \citep{Lei2020, Malh2020}, making their precise locations difficult to determine. Here, for the theoretical analysis of the dynamical features of high-order resonances beyond 50 AU, we employ the circular restricted 3-body problem (CR3BP) model, which consists of the Sun, Neptune, and a test particle. In the following, Section 3.1 uses the planar CR3BP model, as the resonance width is calculated by examining the phase space on two-dimensional Poincar\'e surfaces of section. Since the resonance width generally decreases as the resonator’s inclination increases \citep{Gall2020,namo2020,Li2020,Li2023}, this approach allows us to estimate the maximum libration zone for a given resonance. Sections 3.2 and 3.3, on the other hand, incorporate the effects of high inclination on the resonant features by adopting the spatial CR3BP model.

\subsection{Resonance width and eccentricity range}

\begin{figure*}
  \centering
  \vspace{0 cm}
  \begin{minipage}[c]{1\textwidth}
  \hspace{-1 cm}
  \includegraphics[height=6cm]{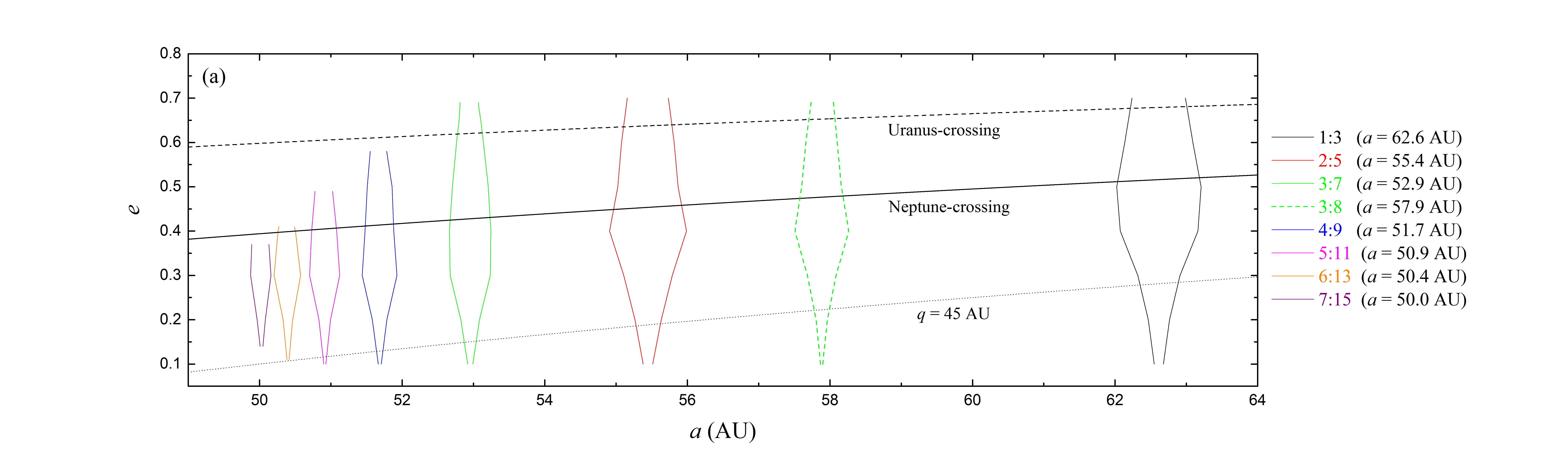}
  \end{minipage}
  \begin{minipage}[c]{1\textwidth}
  \centering
  \includegraphics[height=6cm]{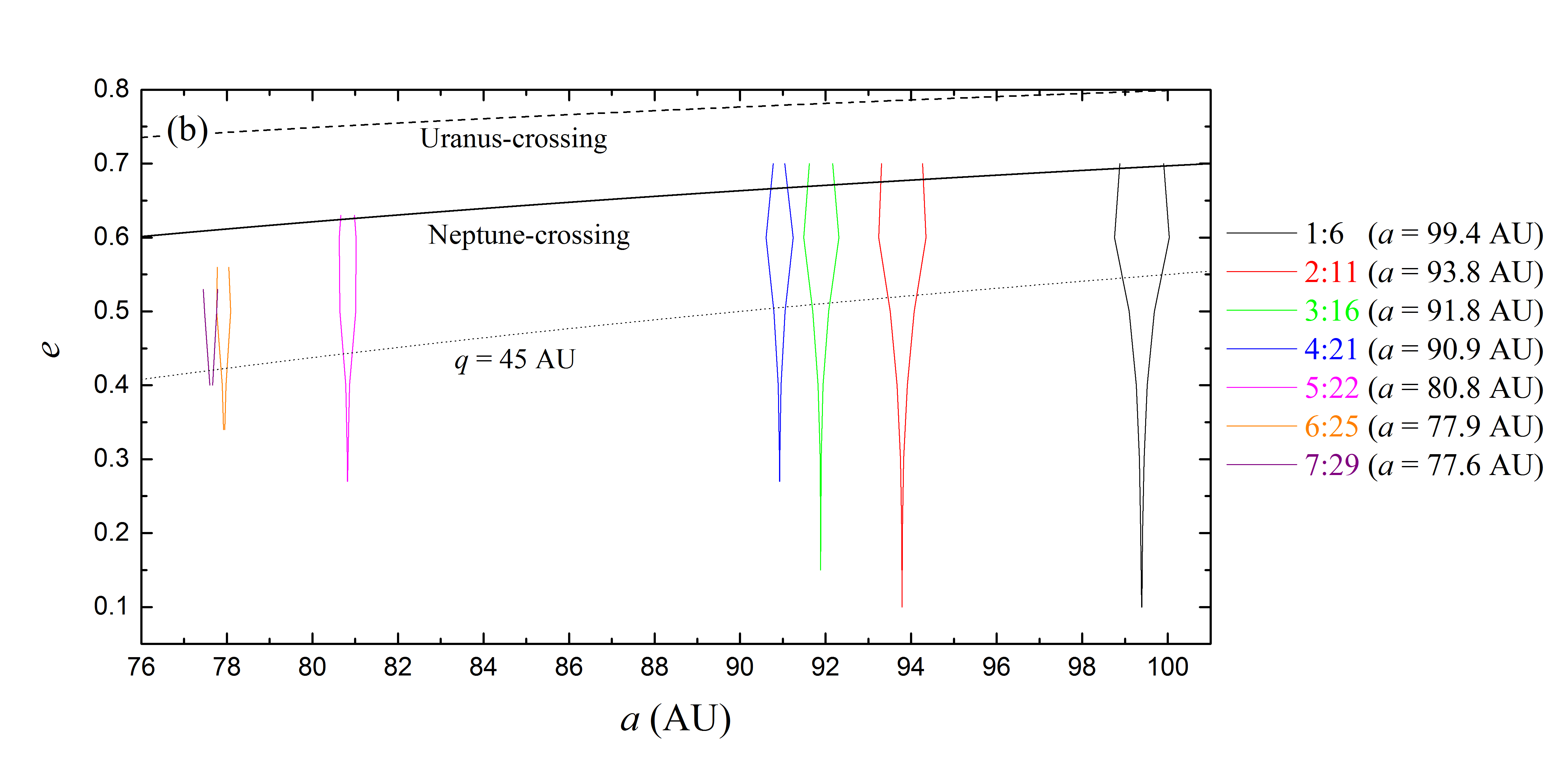}
  \end{minipage}
  \caption{Libration zones as a function of semimajor axis ($a$) and eccentricity ($e$) for a selection of Neptune’s MMRs in the distant Kuiper belt extending from 50 to 100 AU, based on the planar CR3BP model. For each MMR, the libration zone is bounded by two curves of a specific colour, which represent the sunward and anti-sunward resonance separatrices. The resonance width is defined as the range of $a$ variation between these separatrices. To better visualise the relative widths of the resonances, the aspect ratios of panels (a) and (b) are adjusted to ensure that both have the same scale of $a$. For reference, three curves of constant perihelion distance are plotted: the solid curve represents orbits with perihelion distances equal to Neptune's aphelion ($\sim30.3$ AU), the dashed curve represents orbits with perihelion distances equal to Uranus's aphelion ($\sim20.1$ AU), and the dotted curve represents orbits with perihelion distances of 45 AU.}
 \label{width}
\end{figure*}

In this subsection, we calculate the resonance widths in the planar CR3BP model, where the motions of all three objects are coplanar. For a detailed description of the methods used to determine the resonance width in terms of the semimajor axis, the readers are referred to \citet{Li2023}. Here, we merely reiterate one key point: the planar model yields the largest possible libration zone, which has the maximum resonance width, while the resonance width generally narrows for inclined orbits \citep{Gall2020, namo2020}.

For Neptune's $m$:$n$ resonances in the distant Kuiper belt beyond 50 AU, \citet{Malh2019} provided the resonance widths for the 1:$n$ and 2:$n$ resonances. Since a substantial fraction of the observed RKBOs are located in the resonances with $m\ge3$, as identified in Section 2, it is necessary to extend the previous theoretical study to larger values of $m$. However, the $m$:$n$ resonances with $m\ge 8$ are excluded because they cannot sustain RKBOs in stable libration mode. Then for a given $m$ between 1 and 7, we first examine the two most representative groups of $m$:$n$ resonances: the lowest-order resonance with the smallest $n$ (determined by the condition $a>50$ AU) and the highest-order resonance with the largest $n$ (constrained by the identified RKBOs). Specifically, these resonances include: Group \Rmnum1 of 1:3, 2:5, 3:7, 4:9, 5:11, 6:13, and 7:15 MMRs; and Group \Rmnum2 of 1:6, 2:11, 3:16, 4:21, 5:22, 6:25, and 7:29 MMRs. There are two points that we would like to note here: (1) among the Group \Rmnum1 resonances, the 7:15 MMR has not yet been detected to host RKBOs, probably due to current incomplete observations; (2) the Group \Rmnum 2 resonances represent the highest-order MMRs, which are expected to be the weakest ones. Thus, our identification of RKBOs associated with Group \Rmnum2 may provide additional constraints on the migration history of Neptune.

Fig. \ref{width} presents the resonance widths in the $(a, e)$ plane for Group \Rmnum1 (panel a) and Group \Rmnum2 (panel b) resonances. Each resonance is represented by a specific colour, with its libration zone bounded by two curves that correspond to the sunward and anti-sunward resonance separatrices. At a given $e$,  the resonance width is defined as the difference between the maximum and minimum values of $a$. Considering that the two groups of resonances, being closer to or farther from the Sun, span distinct ranges of $a$, we have adjusted the aspect ratios of the panels to ensure that the horizontal axes have the same scale of $a$ in both panel (a) and panel (b). This adjustment allows for a clearer and more direct comparison of the relative widths of the resonances.

For a given $m$:$n$ resonance, the resonance's strength, which is proportional to $e^{|n-m|}$, increases monotonically with $e$. However, the resonance width exhibits more complex behaviour. As shown in Fig. \ref{width}, an obvious feature of resonance widths is that they are quite narrow at low $e$. Then, as $e$ increases, the resonance widths expand, but after $e$ exceeds a critical value, they begin to diminish. Before proceeding with further analysis, we illustrate a usual concept: the resonance width is expressed as the maximum variation in $a$, which corresponds to the maximum variation of the resonant angle $\sigma_{m:n}$, i.e. the resonant amplitude $A$. Actually, in the context of the resonant Hamiltonian model, $a$ and $\sigma_{m:n}$ can be regarded as a pair of pseudo-canonical variables \citep{Lei2020}. Therefore, the resonant amplitude $A$ serves as an equivalent parameter for characterising the resonance width. Based on this concept, we now present the dependence of the resonance width on $e$:

\begin{table}
\vspace{-1 cm}
\centering
\begin{minipage}{7.5cm}
\caption{For each resonance, the maximum and minimum eccentricities $e$ that allow an object to exhibit libration behaviour. The full range of $e$ is set to 0.1-0.7. The last column gives the nominal values of the semi-major axes, $a$, for the resonances.}      
\label{erange}
\begin{tabular}{c c c c}        
\hline                 
Resonance    &    Minimum $e$    &    Maximum $e$  &  $a$ (AU)     \\
 
\hline

1:3	&	$\le0.1$	&	$\ge0.7$	&  62.6  \\
1:4	&	$\le0.1$	&	$\ge0.7$	&  75.8  \\
1:5	&	$\le0.1$	&	$\ge0.7$	&  88.0  \\
1:6	&	$\le0.1$	&	$\ge0.7$	&  99.4  \\

\hline

2:5	    &	$\le0.1$	&	$\ge0.7$	&  55.4  \\
2:7	    &	$\le0.1$	&	$\ge0.7$	&  69.4  \\
2:9	    &	$\le0.1$	&	$\ge0.7$	&  82.0  \\
2:11	&	$\le0.1$	&	$\ge0.7$	&  93.8  \\

\hline       

3:7	    &	$\le0.1$  &   0.69	    &  52.9  \\
3:8	    &	$\le0.1$  &	  $\ge0.7$	&  57.9  \\
3:10	&	$\le0.1$  &	  $\ge0.7$	&  67.2  \\
3:11	&	$\le0.1$  &	  $\ge0.7$	&  71.6  \\
3:13	&	0.11	  &	  $\ge0.7$	&  80.0  \\
3:14	&	0.13	  &	  $\ge0.7$	&  84.1  \\
3:16	&	0.15	  &	  $\ge0.7$	&  91.9  \\

\hline

4:9	    &	$\le0.1$  &   0.58	    &  51.7  \\
4:11	&	$\le0.1$  &   0.64	    &  59.1  \\
4:13	&	0.12	  &	  0.65	    &  66.0  \\
4:15	&	0.15	  &	  $\ge0.7$	&  72.7  \\
4:17	&	0.2	      &	  $\ge0.7$	&  79.0  \\
4:19	&	0.22	  &	  $\ge0.7$	&  85.1  \\
4:21	&	0.27	  &	  $\ge0.7$	&  90.9  \\

\hline

5:11	&	$\le0.1$  &	  0.49	&  50.9  \\
5:12	&	$\le0.1$  &	  0.54	&  54.0  \\
5:13	&	0.11	  &	  0.55	&  56.9  \\
5:14	&	0.12	  &	  0.59	&  59.8  \\
5:16	&	0.18	  &	  0.58	&  65.4  \\
5:17	&	0.19	  &	  0.6	&  68.1  \\
5:18	&	0.21	  &	  0.6	&  70.7  \\
5:19	&	0.21	  &	  0.64	&  73.3  \\
5:21	&	0.27	  &	  0.62	&  78.4  \\
5:22	&	0.27	  &	  0.63	&  80.8  \\

\hline

6:13	&	0.11	&	0.41	&  50.4  \\
6:17	&	0.16	&	0.52	&  60.3  \\
6:19	&	0.24	&	0.52	&  64.9  \\
6:23	&	0.26	&	0.55	&  73.7  \\
6:25	&	0.34	&	0.56	&  77.9  \\

\hline

7:15	&	0.14	&	0.37	&  50.0  \\
7:16	&	0.14	&	0.39	&  52.2  \\
7:17	&	0.14	&	0.41	&  54.4  \\
7:18	&	0.2	    &	0.41	&  56.5  \\
7:19	&	0.2	    &	0.43	&  58.6  \\
7:20	&	0.2	    &	0.44	&  60.6  \\
7:22	&	0.29	&	0.46	&  64.6  \\
7:23	&	0.29	&	0.47	&  66.5  \\
7:24	&	0.29	&	0.48	&  68.4  \\
7:25	&	0.33	&	0.49	&  70.3  \\
7:26	&	0.33	&	0.5	    &  72.2  \\
7:27	&	0.33	&	0.51	&  74.0  \\
7:29	&	0.4	    &	0.53	&  77.6  \\

\hline
\end{tabular}
\end{minipage}
\end{table}

(\rmnum1)  At low $e$, the resonance's strength is weak, and the resonance occurs only near the libration centre, where $A$ is small. Since the resonance's strength scales with $e^{|n-m|}$, this scaling effect becomes more pronounced for higher-order resonances with larger values of $|n-m|$. We illustrate this using the portraits in Fig. \ref{width}(b), which correspond to the highest-order $m$:$n$ resonance for each $m$, i.e. the Group \Rmnum2 resonances. For the smallest $e = 0.1$, although resonant particles can be found in the 2:11 MMR, they lead to a narrow libration zone at such a small $e$ (as shown by the red curves). Correspondingly, among these 2:11 resonant particles, we notice that they all exhibit medium amplitudes in the resonant angle, with $A \le 77^{\circ}$. Furthermore, for even higher-order resonances, such as the 3:16 MMR, the resonance's strength becomes so weak that at the libration centre itself, where $A=0^{\circ}$, resonant particles cannot exist. Consequently, the libration state does not occur in the 3:16 MMR at $e = 0.1$. 
For similar reasons, Fig. \ref{width}(b) shows that the libration zones of the 3:16 (green), 4:21 (blue), 5:22 (magenta), 6:25 (orange), and 7:29 (purple) MMRs are all confined to the regions where $e>0.1$.

(\rmnum2) As $e$ increases, the resonance's strength becomes stronger, leading to a gradual increase in the possible resonant amplitude $A$. In the most extreme case, $A$ can reach its upper limit of $179^{\circ}$ to sustain the libration state of the resonant particles, as $A=180^{\circ}$ indicates the circulation. Of course, this result is obtained within the framework of the restricted 3-body model. When additional perturbations from planets other than Neptune are included, the resonant particles can no longer possess such large values of $A$. For instance, our earlier work demonstrates that the maximum $A$ for stable 2:3 RKBOs over the age of the Solar system is approximately $120^{\circ}$ \citep{Li2014a}.

(\rmnum3) When $e$ continues to increase and exceeds the Neptune-crossing value, the perihelia of the resonant particles will be located inside the orbit of Neptune. At this stage, if the values of $A$ are too large, these particles may experience considerable perturbations during their perihelion passages, as they could be close to Neptune. As a result, the trends in the variation of $A$ and the resonance width observed in the cases (\rmnum1) and (\rmnum2) are reversed, both decreasing as $e$ increases. Such an outcome is particularly evident for resonances with relatively small $a$. Among the Group \Rmnum1 resonances shown in Fig. \ref{width}(a), this includes the 3:7 MMR (green curves) and other resonances interior to it. Similarly, among the Group \Rmnum2 resonances shown in Fig. \ref{width}(b), this includes the 5:22 MMR (magenta curves) and others within it. For these resonances, at the largest $e = 0.7$, stable resonant particles cannot be found even at the resonance centre with $A=0^{\circ}$. Consequently, their libration zones do not extend up to $e=0.7$. For example, the libration zones of the 7:15 and 7:29 MMRs, indicated by the purple curves in Fig. \ref{width}, are restricted to $e<0.37$ and $e<0.53$, respectively.

For the 1:$n$ and 2:$n$ resonances, \citet{Malh2019} conducted comprehensive studies, and our results in Fig. \ref{width} are in good agreement: their libration zones span the full tested eccentricity range of $e=0.1$-0.7. In contrast, for the newly examined $m$:$n$ resonances with $m\ge3$, this $e$ range is only partially covered. Case (\rmnum1) defines the lower eccentricity limit (e.g. the 3:16 MMR requires $e\ge0.15$), while Case (\rmnum3) defines the upper limit (e.g. the 3:7 MMR exists only for $e\le0.69$). Similarly, the resonances from 4:$n$ to 7:$n$ have individual $e$ ranges confined within 0.1-0.7. To obtain a comprehensive view of distant MMRs, we further predict the allowable $e$ ranges for $m$:$n$ resonances across all possible $n$ values, rather than considering only those with the minimum $n$ (Group \Rmnum1) and maximum $n$ (Group \Rmnum2) analysed above. The resulting $e$ limits are summarized in Table \ref{erange}. For the 1:$n$ and 2:$n$ resonances, $e$ can extend throughout the entire range of 0.1-0.7, whereas for the resonances from 3:$n$ to 7:$n$, the minimum $e$ may exceed 0.1 and the maximum $e$ fall below 0.7. These theoretical $e$ limits provide robust constraints on real RKBOs in resonances beyond 50 AU and are indeed consistent with the eccentricities of the currently known RKBOs.

Recall that in Section 2.2.1 we discussed the high-$e$ nature of the currently known RKBOs, as their smaller perihelion distances make them much more observable, whereas potential RKBOs beyond 50 AU could plausibly reside on low-$e$ orbits. For instance, in the range of $a>70$ AU, all identified RKBOs have $e>0.4$ (see Fig. \ref{real}(a)), while Fig. \ref{width}(b) shows that the libration zones can extend to $e$ values well below 0.4, even in the most distant region of $a=90$-100 AU. This suggests that the current lack of low-$e$ RKBOs may indeed be due to observational bias. We must note that the allowable $e$ ranges here are theoretically derived within the Sun+Neptune+particle restricted 3-body model. To better constrain the lower limit of $e$ for real distant RKBOs, in the next section we further consider a model of the actual Solar system, where additional perturbations from the other planets would narrow the $e$ ranges of these RKBOs.

\subsection{Resonance centre for 1:$n$ resonances}

To derive the resonance width from the phase space on the two-dimensional Poincar\'e surfaces of section, as conducted in \citet{Li2023}, we were limited to the planar model in the previous subsection. From now on, we will consider the influence of high inclinations on resonant features: the 1:$n$ resonances with both symmetric and asymmetric librations in this subsection, and the 2:$n$ to 7:$n$ resonances exhibiting only symmetric libration in the next. Based on the observed orbits of real RKBOs, as shown in Fig. \ref{real}(b), we adopt a similar inclination range, extending up to $i=40^{\circ}$. 

\begin{figure}
 \hspace{0cm}
  \centering
  \includegraphics[width=9cm]{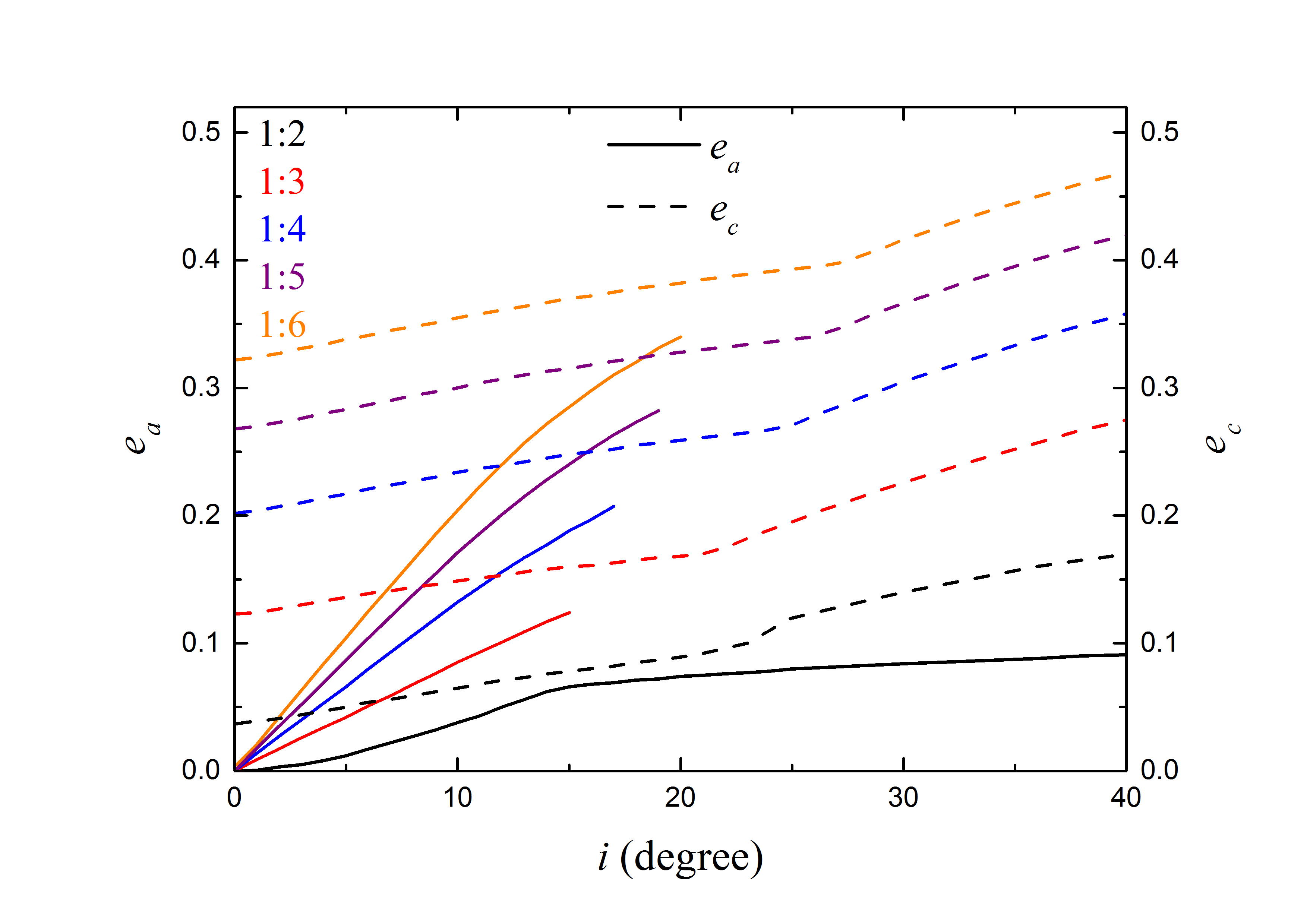}
   \caption{The critical eccentricities $e_a$ (solid line) and $e_c$ (dashed line) for the 1:$n$ MMRs as functions of the inclination $i$. This study focuses on the 1:3 (red), 1:4 (blue), 1:5 (purple), and 1:6 (orange) MMRs, while the $e_a$ and $e_c$ curves for the 1:2 MMR (black), adopted from Fig. 4 in \citet{Li2014b}, are also plotted for reference.}
  \label{eaec}
\end{figure}

In our previous study on the 1:2 MMR \citep{Li2014b}, we introduced two critical eccentricities, $e_a$ and $e_c$, to characterise the resonant states of inclined orbits. Specifically: (1) when the eccentricity satisfies $e < e_a$, the resonant angle consistently circulates, so $e_a$ marks the lower eccentricity threshold for libration; (2) when $e_a < e < e_c$, only symmetric libration of the resonant angle occurs, with the libration centre at $180^{\circ}$ (for $i<15^{\circ}$); (3) when $e > e_c$, asymmetric librations centred at $\sigma_L <180^{\circ}$ (leading) and $\sigma_T>180^{\circ}$ (trailing) become possible, while symmetric libration around $180^{\circ}$ may also co-occur. The phase space structure corresponding to this resonant state is sketched in Fig. 2 of \citet{Li2014b}. As $i$ increases, both $e_a$ and $e_c$ shift monotonically toward larger values (see Fig. \ref{eaec}, black lines), and for $e > e_c$, the asymmetric libration centres $\sigma_L$ and $\sigma_T$ also evolve in a mirror-like way. The calculation of all relevant parameters, e.g. $e_a$, $e_c$, $\sigma_L$, and $\sigma_T$, is carried out by analysing the variation of the SLCs at different values of the argument of perihelion $\omega$. For a detailed description of this method, the reader is referred to \citet{Li2014b}.

The resonant feature reviewed above for the 1:2 MMR can be generalised to the other 1:$n$ resonances between 50 and 100 AU, as indicated by Fig. \ref{eaec}. For the 1:3, 1:4, 1:5, and 1:6 MMRs, both $e_a$ (solid line) and $e_c$ (dashed line) generally increase with inclination $i$, similar to the pattern observed in the 1:2 MMR. Nevertheless, two notable features arise. First, for the 1:$n$ MMRs, both $e_a$ and $e_c$ increase with $n$. In other words, higher-order resonances require larger eccentricities for both symmetric and asymmetric librations to appear. This trend compensates for the weakening of the resonance's strength, which is proportional to $e^{|n-1|}$, as $n$ becomes larger. Second, unlike the 1:2 MMR, the $e_a(i)$ curves for the higher-order 1:3 to 1:6 MMRs terminate at relatively large inclinations of $i\sim15^{\circ}$-$20^{\circ}$. When $e_a$ is absent, the regime of exclusively symmetric libration under the condition $e_a < e < e_c$ does not exist. Despite this, the critical eccentricity $e_a$ may not be a key parameter in determining the dynamical behaviours of real 1:$n$ RKBOs beyond 50 AU. For example, for the 1:3 MMR, we have $e_a = 0.124$ at $i = 15^{\circ}$, but the observed 1:3 RKBOs all have substantially larger $e$ values, typically exceeding 0.34 (see Fig. \ref{real}).

\begin{figure*}
  \centering
  \begin{minipage}[c]{1\textwidth}
  \vspace{0 cm}
  \includegraphics[width=9cm]{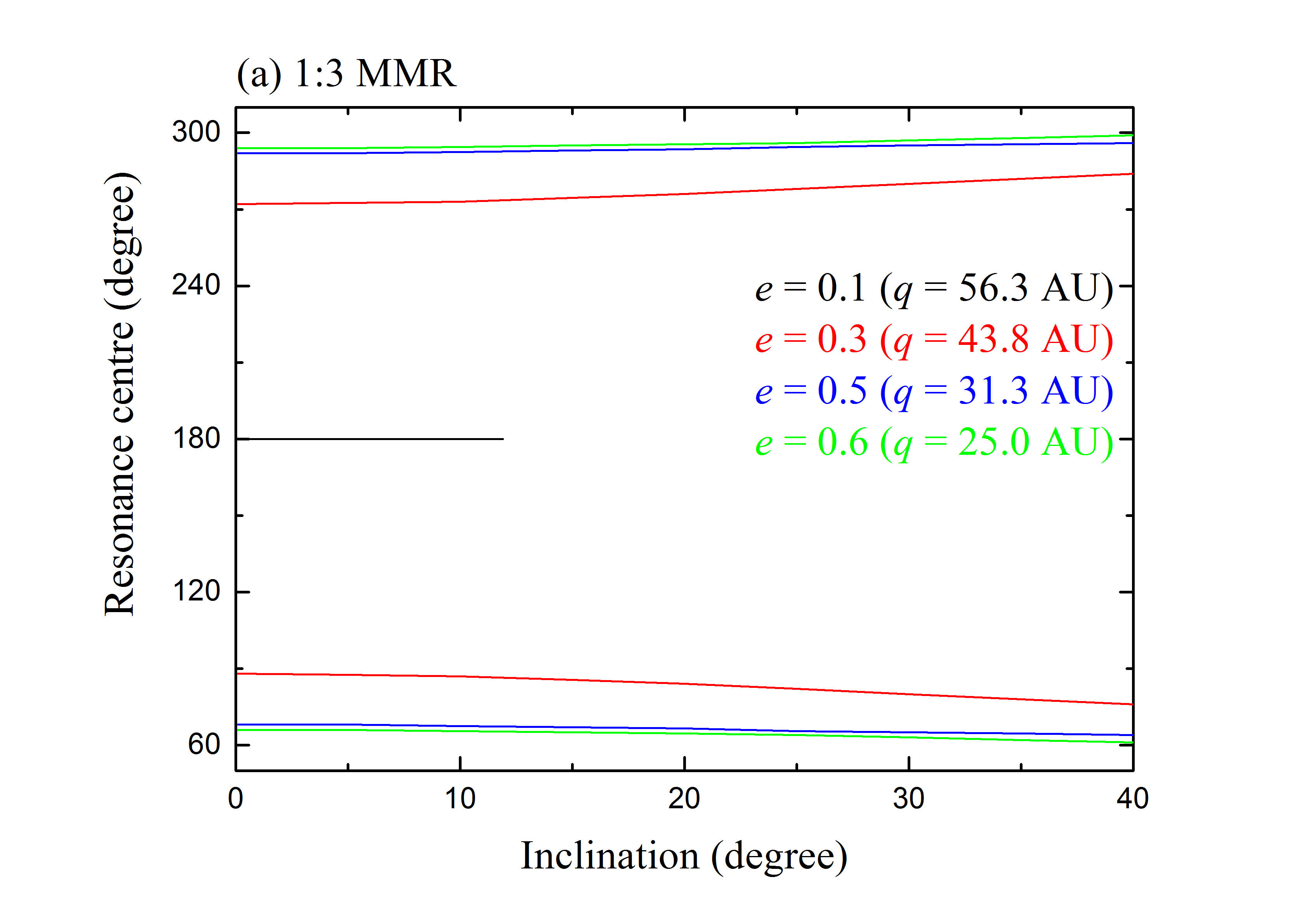}
  \includegraphics[width=9cm]{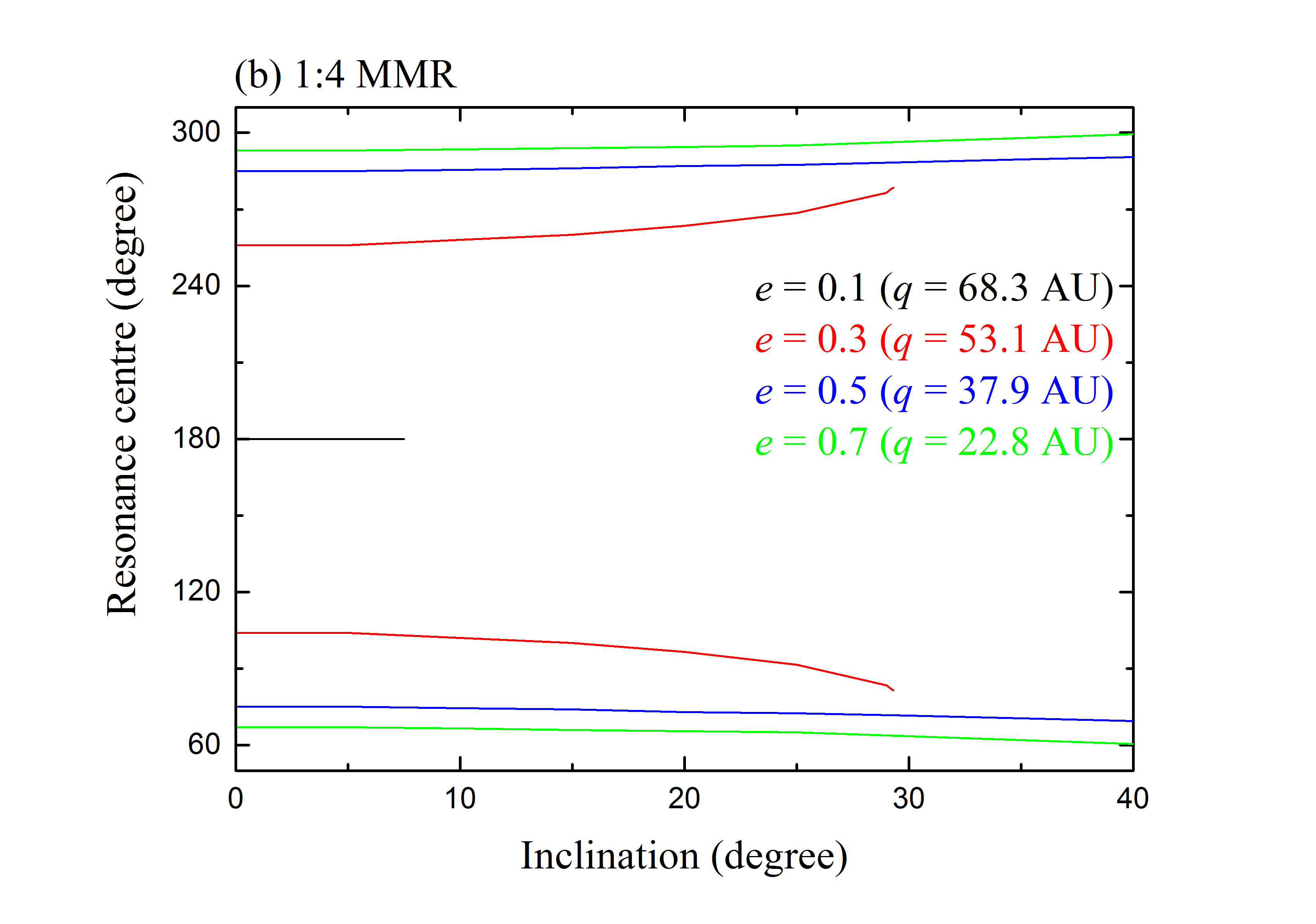}
  \end{minipage}
  \begin{minipage}[c]{1\textwidth}
  \vspace{0 cm}
  \includegraphics[width=9cm]{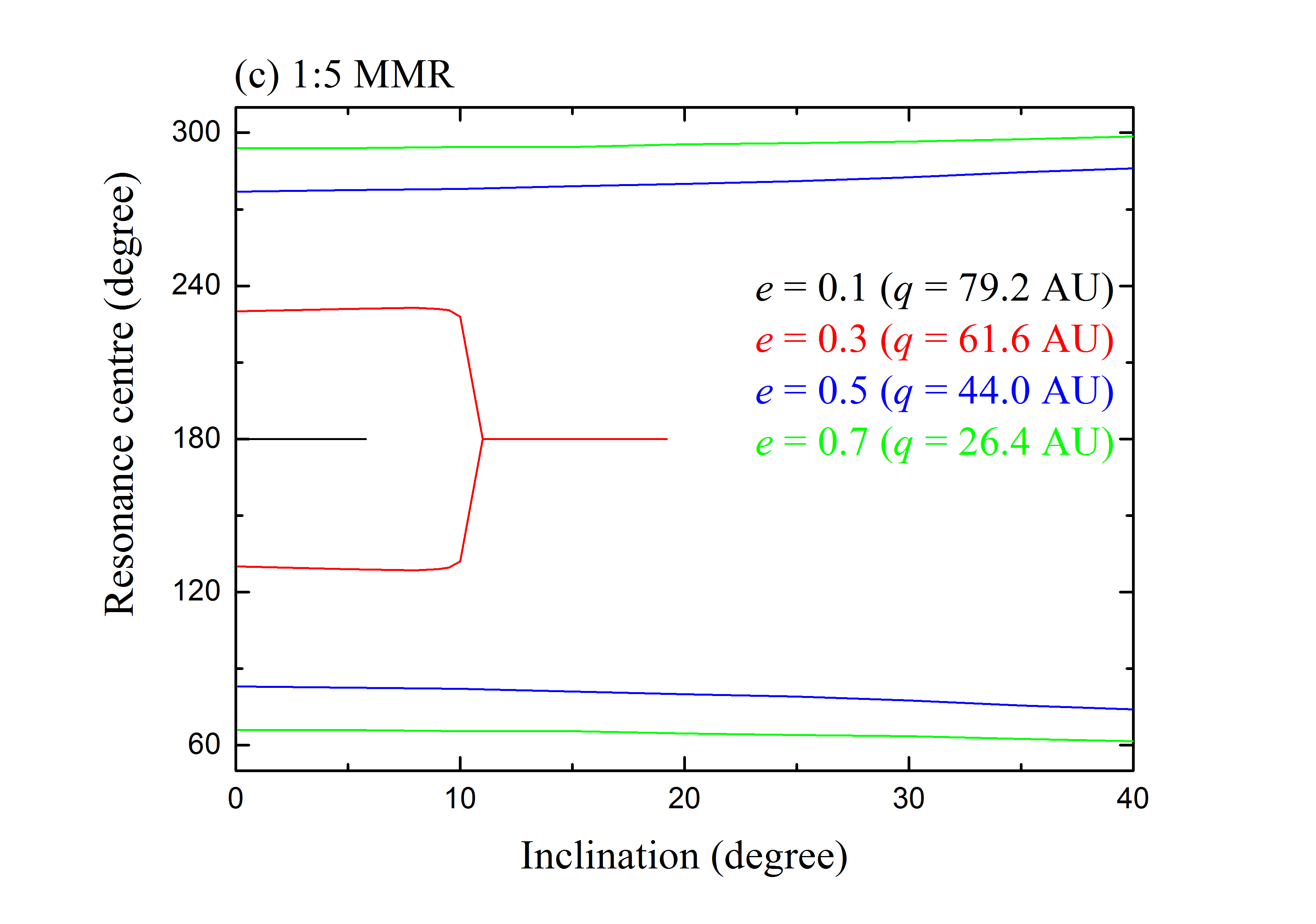}
  \includegraphics[width=9cm]{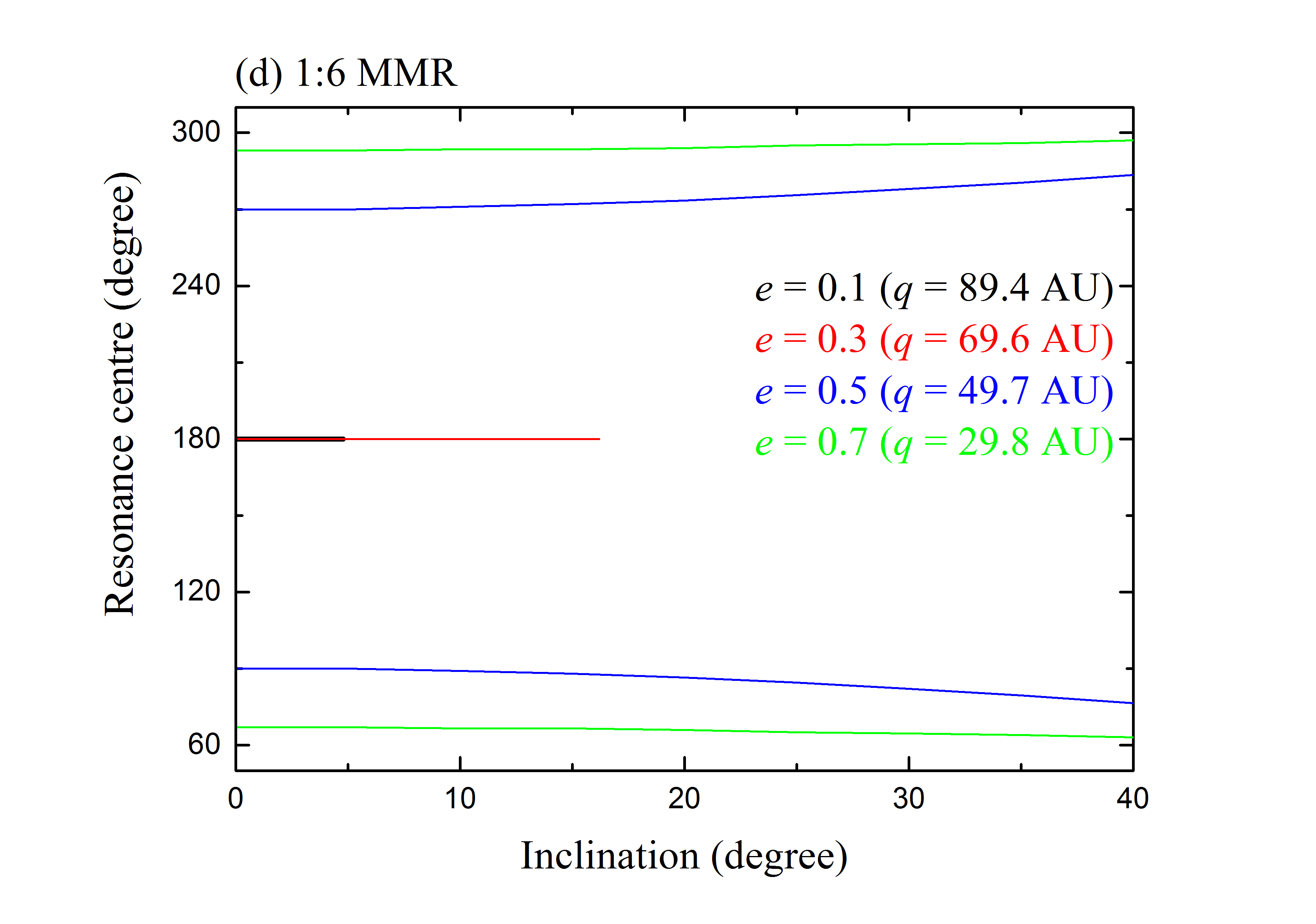}
  \end{minipage}
  \caption{Locations of the resonance centres with respect to the inclination for the 1:3 (panel a), 1:4 (panel b), 1:5 (panel c), and 1:6 (panel d) MMRs. For each MMR, four representative values of the eccentricity $e$ are adopted, as indicated by different colours. If continuous $e$-sampling is considered, one can read the critical eccentricities $e_a$ and $e_c$ from Fig. \ref{eaec}, and then expect the different types of libration behaviours as: (1) when $e_a < e < e_c$, only symmetric libration around $180^{\circ}$ occurs; (2) when $e > e_c$, asymmetric libration arises and co-occurs with the symmetric libration.}
 \label{GLC}
\end{figure*}

\begin{table}
\hspace{-1 cm}
\centering
\begin{minipage}{8.8cm}
\caption{Drawn from Fig. \ref{GLC},  the eccentricity ($e$) and inclination ($i$) values corresponding to specific resonant states for 1:$n$ resonances. The letters `S', `A', and `C' denote symmetric libration, asymmetric libration, and circulation, respectively. The last column indicates the symmetric resonance centre at $180^{\circ}$, the leading centre at angles $<180^{\circ}$, and the trailing centre at angles $>180^{\circ}$.}      
\label{ResState}
\begin{tabular}{c c c c}        

\hline\hline                 

1:3 MMR      \\

\hline
$e$       &      $i$         & Resonant state &             Resonance centre                \\
\hline

0.1       &  $<12^{\circ}$    &     S       &              $180^{\circ}$                    \\

0.1       &  $>12^{\circ}$    &     C       &                   --                            \\

0.3       &  $<40^{\circ}$    &    A\&S      &   $76^{\circ}$-$88^{\circ}$, $284^{\circ}$-$272^{\circ}$, $180^{\circ}$  \\

0.5       &  $<40^{\circ}$    &    A\&S      &   $64^{\circ}$-$68^{\circ}$, $292^{\circ}$-$296^{\circ}$, $180^{\circ}$  \\

0.6       &  $<40^{\circ}$    &    A\&S      &   $61^{\circ}$-$66^{\circ}$, $294^{\circ}$-$299^{\circ}$, $180^{\circ}$  \\

\hline 
1:4 MMR      \\
\hline                 
$e$       &      $i$         & Resonant state &             Resonance centre                \\

\hline

0.1       &  $<8^{\circ}$    &     S       &              $180^{\circ}$                    \\

0.1       &  $>8^{\circ}$    &     C       &                   --                            \\

0.3       &  $<30^{\circ}$   &    A\&S      &   $82^{\circ}$-$104^{\circ}$, $256^{\circ}$-$278^{\circ}$, $180^{\circ}$  \\

0.3       &  $>30^{\circ}$    &     C       &                   --                            \\

0.5       &  $<40^{\circ}$    &    A\&S      &   $70^{\circ}$-$75^{\circ}$, $285^{\circ}$-$290^{\circ}$, $180^{\circ}$  \\

0.7       &  $<40^{\circ}$    &    A\&S      &   $60^{\circ}$-$67^{\circ}$, $293^{\circ}$-$300^{\circ}$, $180^{\circ}$  \\

\hline 
1:5 MMR      \\
\hline                 
$e$       &      $i$                    & Resonant state &             Resonance centre                \\

\hline

0.1       &  $<6^{\circ}$                &     S       &              $180^{\circ}$                    \\

0.1       &  $>6^{\circ}$                &     C       &                   --                            \\

0.3       &  $<11^{\circ}$               &    A\&S      &   $129^{\circ}$-$132^{\circ}$, $228^{\circ}$-$231^{\circ}$, $180^{\circ}$  \\

0.3       &  $11^{\circ}$-$19^{\circ}$   &     S       &                   $180^{\circ}$                            \\

0.3       &  $>19^{\circ}$               &     C       &                   --                            \\

0.5       &  $<40^{\circ}$               &    A\&S      &   $74^{\circ}$-$83^{\circ}$, $277^{\circ}$-$286^{\circ}$, $180^{\circ}$  \\

0.7       &  $<40^{\circ}$               &    A\&S      &   $62^{\circ}$-$66^{\circ}$, $294^{\circ}$-$298^{\circ}$, $180^{\circ}$  \\

\hline
1:6 MMR      \\
\hline                 

$e$       &      $i$                    & Resonant state &             Resonance centre                \\

\hline

0.1       &  $<5^{\circ}$                &     S       &              $180^{\circ}$                    \\

0.1       &  $>5^{\circ}$                &     C       &                   --                            \\

0.3       &  $<16^{\circ}$               &    S      &                   $180^{\circ}$      \\

0.3       &  $>16^{\circ}$               &     C       &                   --                            \\

0.5       &  $<40^{\circ}$               &    A\&S      &   $77^{\circ}$-$90^{\circ}$, $270^{\circ}$-$283^{\circ}$, $180^{\circ}$  \\

0.7       &  $<40^{\circ}$               &    A\&S      &   $63^{\circ}$-$67^{\circ}$, $293^{\circ}$-$297^{\circ}$, $180^{\circ}$  \\
\hline

\end{tabular}
\end{minipage}
\end{table}

As briefly introduced in Section 1, for any given ($e$, $i$) pair, the variation of the SLCs can be computed as the argument of perihelion $\omega$ changes from $0^{\circ}$ to $360^{\circ}$. The average of these SLCs defines the GLC, also referred to as the resonance centre \citep{Li2014a}. Fig. \ref{GLC} presents the locations of the resonance centres for the 1:3, 1:4, 1:5, and 1:6 MMRs with $i$ ranging from $0^{\circ}$ to $40^{\circ}$, given a few representative $e$ values. Note that for cases where $e > e_c$, asymmetric libration is always accompanied by symmetric libration. To keep the figures clear, we therefore plot only the asymmetric resonance centres, $\sigma_L$ ($<180^{\circ}$) and $\sigma_T$ ($>180^{\circ}$), omitting the symmetric centre at $180^{\circ}$. To directly illustrate the resonant behaviours produced by the various $(e,i)$ combinations, we compiled Table \ref{ResState} using the information from Fig. \ref{GLC}. The letters `S', `A', and `C' denote symmetric libration, asymmetric libration, and circulation, respectively, and the notation ‘A\&S’ indicates that both asymmetric and symmetric librations are allowed (i.e. when $e > e_c$). The corresponding libration centres are listed in the last column.

Regarding the 1:3 MMR (see Fig. \ref{GLC}(a)), for a small eccentricity of $e=0.1$ (black curve), only symmetric libration around $180^{\circ}$ is possible. However, when $i>12^{\circ}$, the symmetric libration disappears, corresponding to the absence of $e_a$ at such large inclinations, as seen in Fig. \ref{eaec} (red solid line) and discussed earlier. For moderate to large eccentricities of $e \ge 0.3$ (indicated by the red, blue, and green curves), which exceed the maximum critical value $e_c^{1:3}(i=40^{\circ}) = 0.275$, asymmetric libration is always allowed for any $i\le40^{\circ}$. As $i$ increases from $0^{\circ}$ to $40^{\circ}$, the leading centre $\sigma_L$ ($<180^{\circ}$) decreases monotonically, while the trailing centre $\sigma_T$ ($>180^{\circ}$) increases in a mirror image of the leading centre. These variations of resonant states and the ranges of $\sigma_L$ and $\sigma_T$ can also be found in Table \ref{ResState}. We note that, for the 1:3 MMR, we adopt a maximum $e$ of 0.6 rather than 0.7 as previously considered. This is because the nominal location of this resonance is at $a_{res} = 62.6$ AU, and a resonator with $e=0.7$ would have a perihelion distance of $q=18.8$ AU, which is inside Uranus’s orbit at 19.2 AU, resulting in an inevitable orbital instability due to close encounters with Uranus.

 For the other three 1:$n$ MMRs (see Fig. \ref{GLC}(b)-(d)), the variation of the resonance centres generally follows the pattern observed for the 1:3 MMR, while notable differences arise at moderate $e = 0.3$ (all these differences can be seen in Table \ref{ResState}):

(1) 1:4 MMR: Asymmetric libration is only allowed for $i<30^{\circ}$, as shown by the red curve in Fig. \ref{GLC}(b). This arises because $e_c$ increases with resonance order. Specifically, $e_c^{1:4}$ reaches 0.305 at $i=30^{\circ}$; since $e=0.3$ is below this critical value, asymmetric libration would not occur.

(2) 1:5 MMR: The critical eccentricity $e_c^{1:5}$ exceeds 0.3 at an even lower inclination of $i\sim10^{\circ}$, where asymmetric libration disappears (red curve in Fig. \ref{GLC}(c)). Beyond this inclination, the two asymmetric centres merge into a single centre at $180^{\circ}$, meaning that only symmetric libration is possible. Moreover, if $i$ continues to increase beyond $19^{\circ}$, objects would be certain to move outside the 1:5 MMR, as any kind of libration of the resonant angle $\sigma_{1:5}$ becomes completely prohibited.

(3) 1:6 MMR: For $e=0.3$, asymmetric libration has entirely vanished, leaving only symmetric libration (red curve in Fig. \ref{GLC}(d)). This is because, at $i=0^{\circ}$, the smallest critical eccentricity for this resonance is $e_c^{1:6}=0.322$ (orange dashed line in Fig. \ref{eaec}), which is greater than $e=0.3$, meaning asymmetric libration cannot occur at any inclination. In addition, symmetric libration is limited to $i<17^{\circ}$ due to the increase of another critical eccentricity $e_a$ (orange solid line in Fig. \ref{eaec}).

In fact, these peculiar libration behaviours at $e=0.3$ in the 1:4, 1:5, and 1:6 MMRs have counterparts in the lower-order 1:3 MMR at smaller $e$ values. For instance, Fig. \ref{eaec} (red dashed line) shows that $e_c^{1:3}(i=26^{\circ})=0.202$, so for $e=0.2$, asymmetric libration disappears at $i\gtrsim26^{\circ}$, analogous to its disappearance at $i>30^{\circ}$ for $e=0.3$ in the 1:4 MMR.

\subsection{Permission region for all other resonances}

\begin{figure*}
  \centering
  \begin{minipage}[c]{1\textwidth}
  \vspace{0 cm}
  \includegraphics[width=9cm]{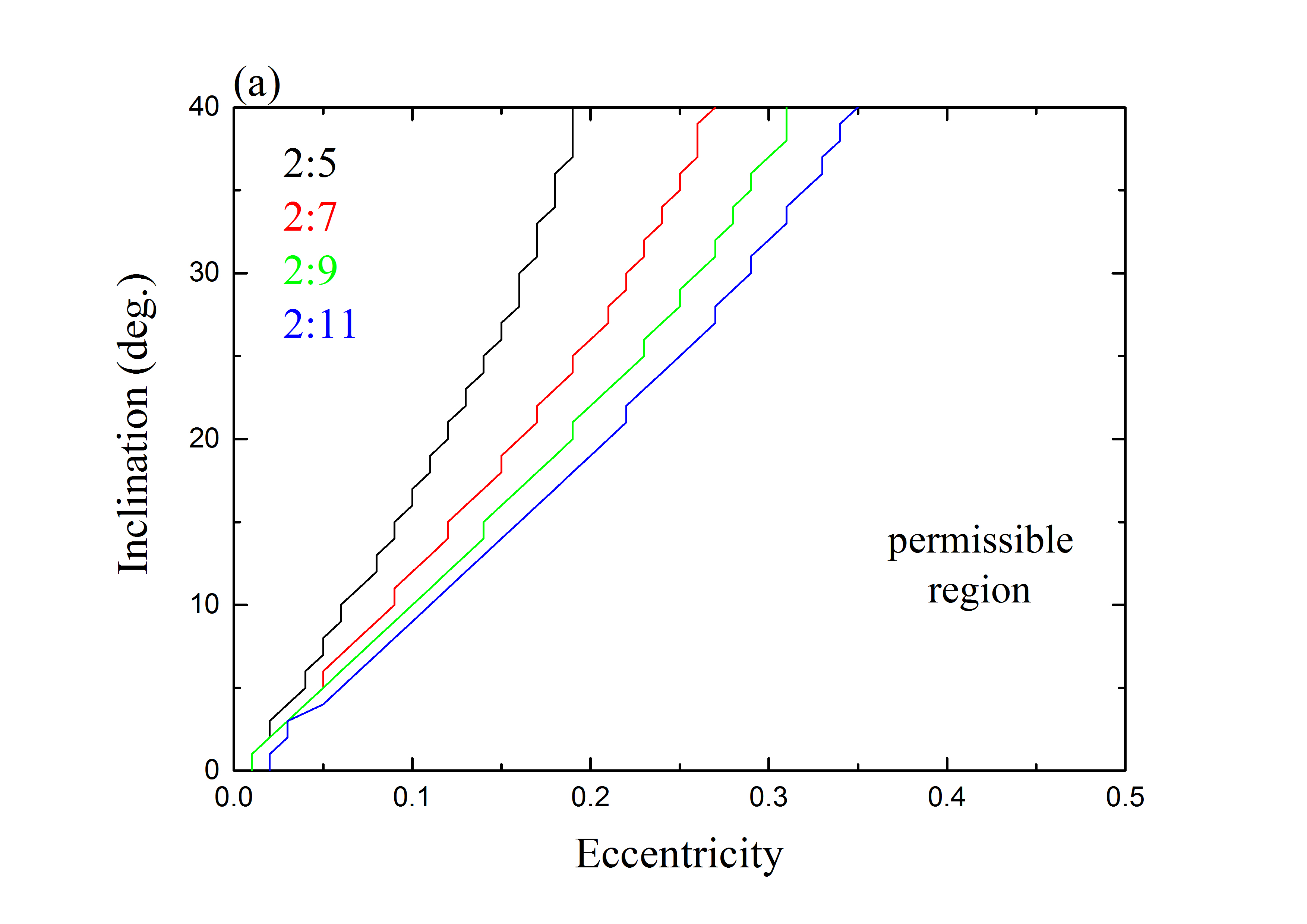}
  \includegraphics[width=9cm]{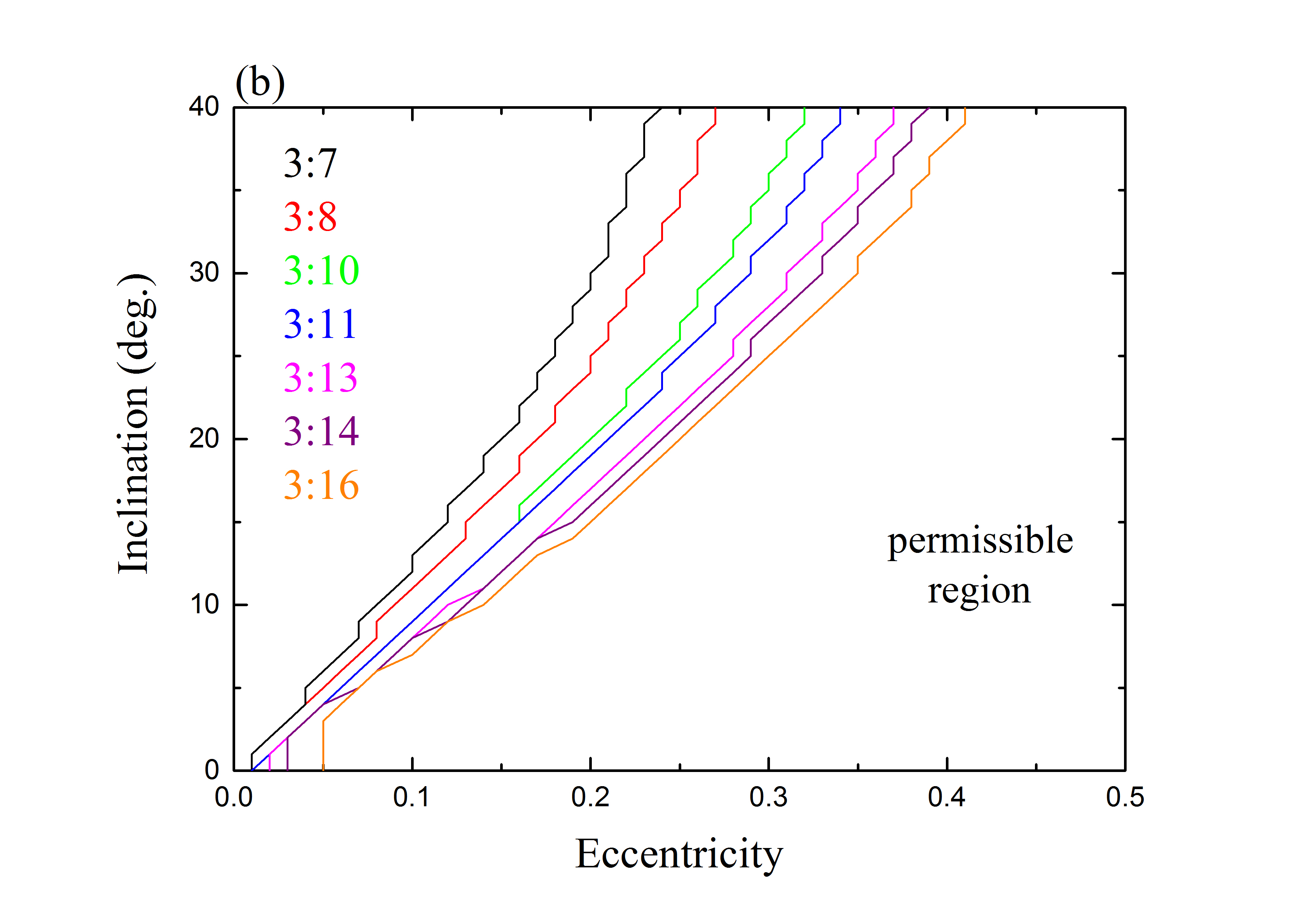}
  \end{minipage}
  \begin{minipage}[c]{1\textwidth}
  \vspace{0 cm}
  \includegraphics[width=9cm]{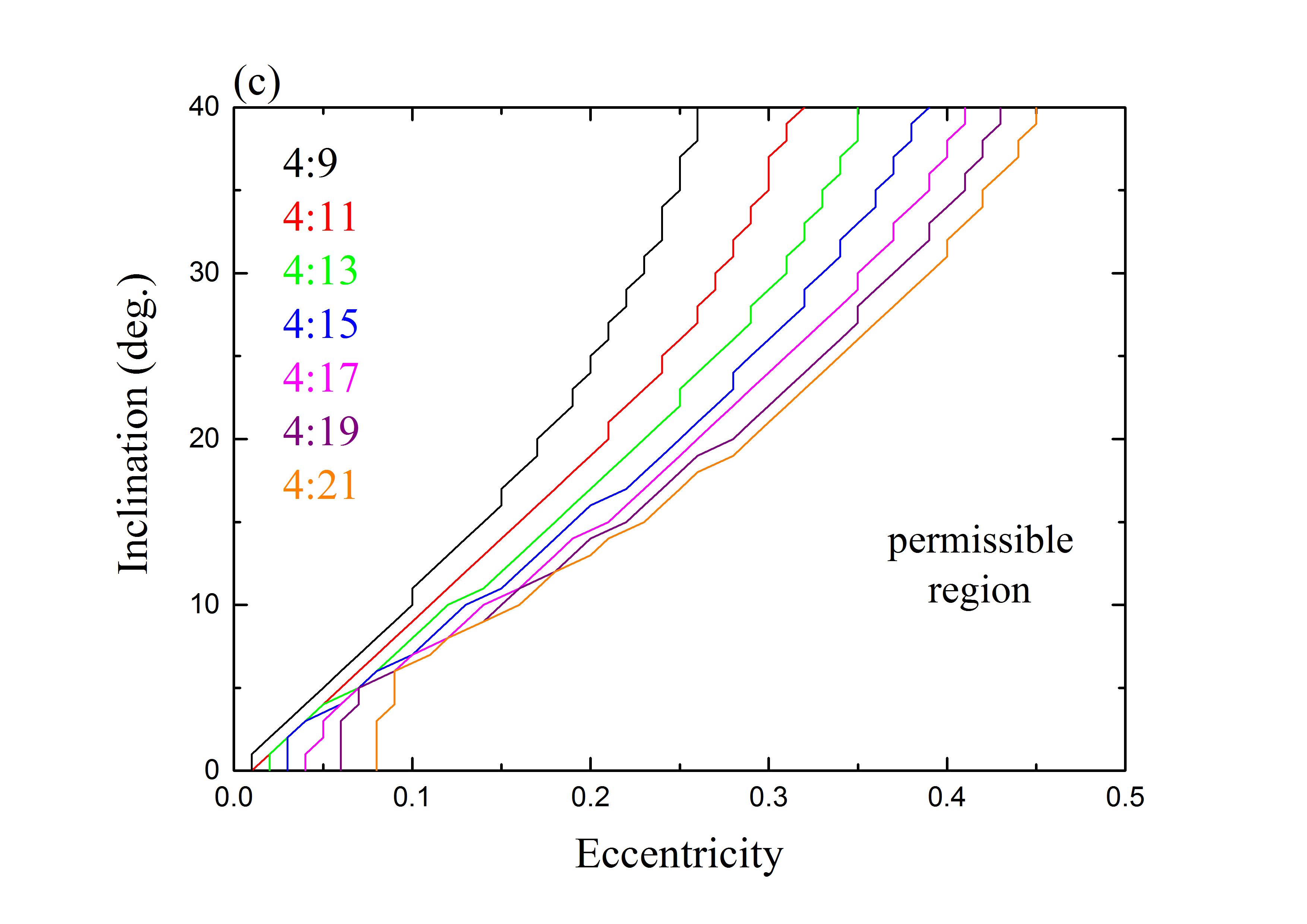}
  \includegraphics[width=9cm]{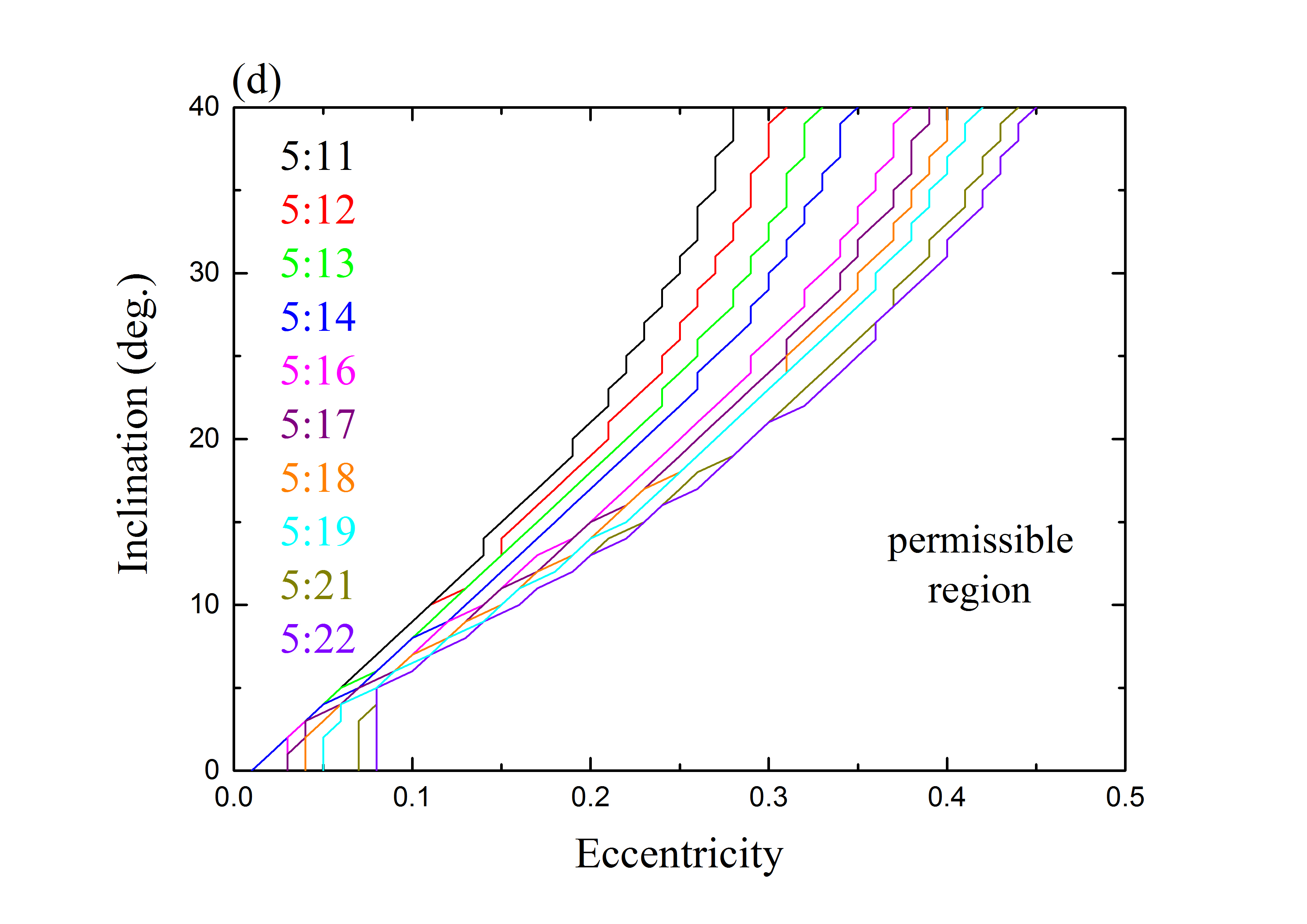}
  \end{minipage}
  \begin{minipage}[c]{1\textwidth}
  \vspace{0 cm}
  \includegraphics[width=9cm]{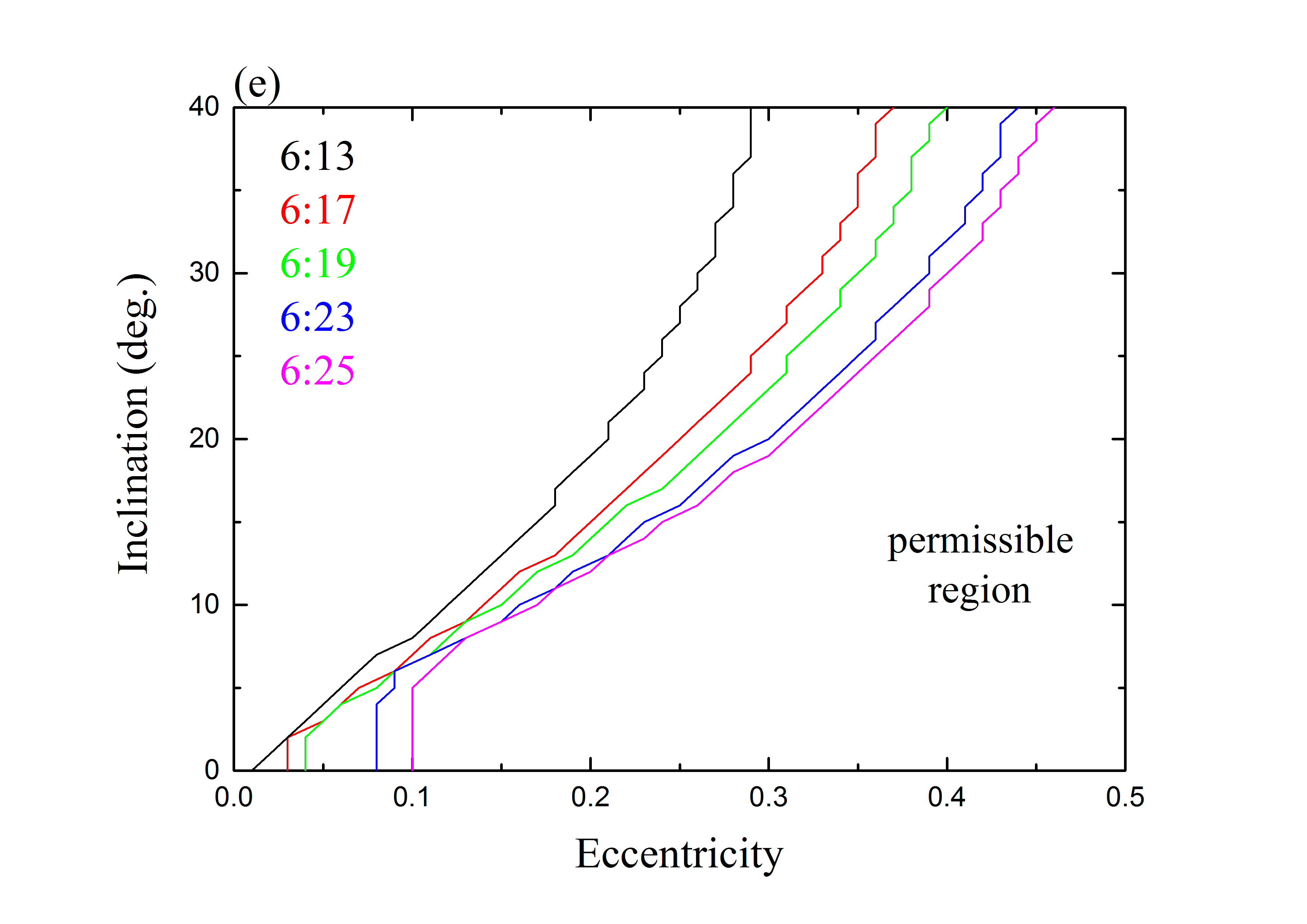}
  \includegraphics[width=9cm]{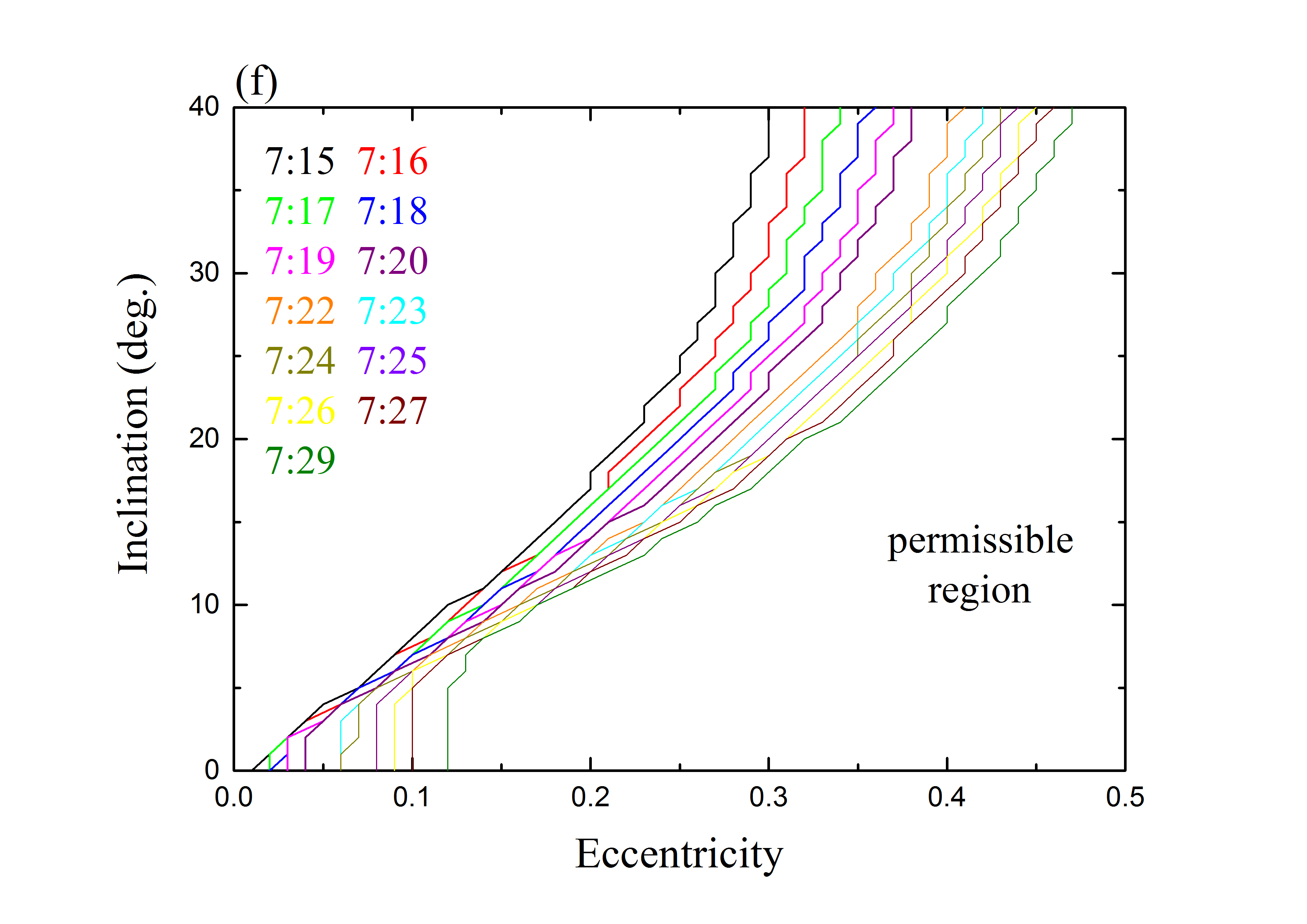}
  \end{minipage}
  \caption{Limiting curves on the eccentricity-inclination plane for a comprehensive set of $m:n$ resonances ($m\ge 2$) in the 50–100 AU range: (a) 2:$n$ resonances; (b) 3:$n$ resonances; (c) 4:$n$ resonances; (d) 5:$n$ resonances; (e) 6:$n$ resonances; (f) 7:$n$ resonances. In every panel, the limiting curve for each resonance is represented by a specific colour, and the permissible region of this resonance is on the right side.} 
 \label{PerReg}
\end{figure*}

For the $m$:$n$ resonances with $m\ge2$, unlike the 1:$n$ resonances discussed above, they only exhibit symmetric libration with a centre at $180^{\circ}$. 
By libration, we mean that the resonant angle oscillates with an amplitude $A$ less than $180^{\circ}$. \citet{Li2014a} proposed that for high-inclination orbits, the lower limit of $A$, denoted as $A_{\min}$, is determined by the maximum variation of the SLCs when the argument of perihelion $\omega$ changes between $0^{\circ}$ and $360^{\circ}$. In that paper, we applied this theory to the 2:3 MMR, finding that $A_{\min}$ reaches a largest value of $\sim75^{\circ}$ for inclinations up to $90^{\circ}$. Therefore, the libration motion is always possible for this 1st-order resonance. However, when the resonance order increases, such as for the 2nd-order 3:5 MMR \citep{Li2023} and the 3rd-order 4:7 MMR \citep{Li2020}, $A_{\min}$ will only be limited to values of $<180^{\circ}$ for orbits with $e$ exceeding a critical value, $e_{crit}$, which increases with $i$. Accordingly, for the 2nd- and higher-order MMRs, $e_{crit} = e_{crit}(i)$ defines a `limiting curve' in the $(e, i)$ plane. To the right of the limiting curve, there is the region where the condition $e \ge e_{crit}$, or equivalently $A_{\min} < 180^{\circ}$, is satisfied. This region is referred to as the `permissible region' of an MMR.

In short, the permissible region can place strong constraints on the eccentricity and inclination distribution of potential RKBOs in the 2nd- and higher-order MMRs. We now generalise this theory to a broader class of MMRs in the distant Kuiper belt beyond 50 AU. It is worth noting that, as the order of the resonance increases, such constraints would become even more pronounced. Considering the MMRs listed in Table \ref{erange}, which could reach resonance orders greater than 20, the permissible regions would be confined to much smaller areas in the $(e, i)$ plane, thereby imposing more stringent constraints on the orbital distribution of the associated resonators.

In Fig. \ref{PerReg}, we present the limiting curves $e_{crit}(i)$ for the resonances of 2:$n$ to 7:$n$, considering inclinations up to $i=40^{\circ}$. For each resonance, the permissible region lies to the right of its corresponding limiting curve. This figure can be considered as a supplement to Table \ref{erange}, where only the planar case of $i=0^{\circ}$ is considered. Taking the highest-order 7:29 MMR as an example, Table \ref{erange} gives a minimum $e$ of 0.4 for this resonance to occur. However, when the high inclination is taken into account, such as for $i=40^{\circ}$, the minimum $e$ increases to 0.47. Similarly, in Fig. \ref{PerReg}, one can easily read, for any given $i$, the minimum $e$ required for the libration motion to emerge for different $m$:$n$ resonances with $m\ge2$ between 50 and 100 AU. We believe that Fig. \ref{PerReg} can serve as an informative database for predicting the orbital distributions of potential RKBOs to be discovered in the future throughout the distant Kuiper belt.


\section{Long-term stability and possible occupancy}

Having theoretically presented the complex features of Neptune's MMRs beyond 50 AU, this section performs long-term numerical simulations to explore the stable regions within these resonances. Our simulations are essential for predicting the intrinsic distribution of potential RKBOs. We will examine the orbits of the simulated resonators and compare them with observational data, such as eccentricity and inclination distributions, the leading-to-trailing number asymmetry for the 1:$n$ MMRs, and the (relative) numbers of resonators associated with different MMRs. Particularly, since some of the considered distant MMRs could be extremely weak (e.g. with resonance orders higher than 20), we are interested in determining whether resonators can possibly exist. We recall that among the 207 observed RKBOs between 50 and 100 AU, only one is found in such weak MMRs.

As the order of the resonance increases, the resonance width narrows, leading to a decreased probability of occupancy \citep{Li2023}. For the MMRs we consider between 50 and 100 AU, Fig. \ref{width} shows that the libration zone of the lowest-order 1:3 MMR is significantly broader than that of the highest-order 7:29 MMR. Therefore, to simulate resonators that may be distributed across these MMRs, which span a large range of resonance orders, if a certain number of resonators are to be generated for the highest-order MMRs, many more will be produced accordingly in the lowest-order MMRs. Considering that we need to examine up to 50 MMRs, as listed in Table \ref{erange}, we will ultimately have to include a huge number of test particles in the simulations.

\subsection{Generation of test particles}

The classical method for generating test particles is that, given a specific resonance, the particles are assigned semimajor axes equal to the nominal resonant value $a_{res}$ (see Eq. (\ref{ares})), while eccentricities, inclinations, and resonant angles are chosen from appropriate ranges \citep{Lyka2007, volk2016}. Actually, the resonance width in terms of semimajor axis is also typically used to cover the entire resonant zone, as we will consider later.
Another method involves referencing the distribution of observed objects. For example, \citet{Lyka2023} conducted simulations to investigate the long-term stability of resonant particles generated from the L7 synthetic model, which was derived from the Canada-France Ecliptic Plane Survey (CFEPS)\footnote{https://www.cfeps.net/?page\underline{~}id=105} \citep{L7Kavelaars2009, Peti2011, Glad2012}. However, since the L7 model is limited to resonances for which objects were found in the CFEPS survey, beyond 50 AU only objects associated with the 1:3, 1:5, and 2:5 MMRs are included, resulting in too few sampled resonances.

Given the large number of resonances spanning from 50 to 100 AU -- 50 in total that we have considered -- if each $m$:$n$ resonance is treated individually, such as by using the classical method with different $a_{res}$ values for different combinations of $m$ and $n$, the process of assigning initial $a$ values would be exceedingly complex. In addition, a challenge lies in choosing the initial values of $e$ and $i$. For the 1:$n$ type resonances, different $(e, i)$ pairs correspond to different resonance centres, making the selection of initial resonant angles exceptionally complicated. And for the remaining 46 $m$:$n$ resonances with $m\ge 2$, each has a specific permissible region for the libration motion, which imposes constraints on the initial $(e, i)$ space. 
Moreover, for each of these 50 resonances, it is not easy to determine how many test particles should be simulated. Consequently, treating numerous resonances one by one is impractical for conducting the long-term stability study. 

To address the aforementioned issues, we adopt the most straightforward approach: uniformly covering the entire semimajor axis space when generating test particles. 
Therefore, the particles are assumed to have initial semimajor axes $a_0$ randomly distributed within the range of 50-100 AU, representing the distant Kuiper belt considered in this study. The initial eccentricities and inclinations are assigned representative values of $e_0=0.1$, 0.3, 0.5, 0.6, and 0.7, and $i_0=0^{\circ}$, $20^{\circ}$, and $40^{\circ}$. These $e_0$ and $i_0$ values are chosen to cover the extremal values of currently observed RKBOs beyond 50 AU (i.e. $e=0.1$-0.7 and $i=0^{\circ}$-$40^{\circ}$, as shown in Fig. \ref{real}), as well as to align with the theoretical analysis presented in Section 3.

For the other three orbital elements, namely the ascending node $\Omega$, the longitude of perihelion $\varpi$, and the mean longitude $\lambda$, their initial values are randomly selected within the range of $0^{\circ}$-$360^{\circ}$. In fact, these three angles determine the resonant angles $\sigma_{m:n}$ of the test particles, as well as their resonant amplitudes $A$. As illustrated in Section 3, for any MMR beyond 50 AU, the occurrence of libration motion is constrained by the values of $e$ and $i$. For instance, the resonance centre of a 1:$n$ MMR changes with $e$ and $i$. Since the variation trends of the resonance centres differ for distinct values of $n$, if the initial resonant angle needs to be within a specific range near the resonance centre, e.g. requiring $A<100^{\circ}$, the assignment of initial $\Omega$, $\varpi$, and $\lambda$ could become quite complex. The advantage of randomly selecting these three angles is that it includes all possible $\sigma_{m:n}$, which range from $0^{\circ}$ to $360^{\circ}$, thus covering the full range of $A$ up to $180^{\circ}$. It is important to note that our simulations involve a large number of resonances, so, for the sake of simplicity, we adopt initial conditions of test particles that uniformly cover the entire orbital parameter space. However, when focussing on one or a few specific resonances, the resonant features outlined in Section 3 will be valuable for generating more appropriate initial conditions for test particles. More importantly, these resonant features will also aid in analysing the stability of certain resonances in the following subsections.

According to the approach for generating test particles described above, for each $(e_0, i_0)$ pair, 30,000 particles are produced with semimajor axes of $a_0=50$-100 AU and resonant angles of $\sigma_{m:n}=0^{\circ}$-$360^{\circ}$. Note that the full range of $\sigma_{m:n}$ is consistently covered for all the $m$:$n$ resonances. For the highest-order 7:29 MMR, which has the narrowest libration zone, the maximum resonance width in terms of $a$ is about 0.3 AU (see Fig. \ref{width}). Then the number of test particles generated for this resonance can be roughly estimated as $0.3~\mbox{AU} / (100~\mbox{AU} - 50~\mbox{AU}) \times 30,000 = 180$. The adoption of the maximum resonance width, rather than the area of the libration zone, to estimate the relative numbers of resonant populations for different MMRs has been validated in \citet{Li2023}. As a result, the resolution of the resonant angle for the 7:29 MMR is calculated to be $360^{\circ} / 180 = 2^{\circ}$. With such a high resolution, test particles can be assigned initial $A$ values of $<2^{\circ}$, indicating that they are sufficiently close to the resonance centre and, theoretically, should be the most stable resonators. Furthermore, considering the chosen values of $e_0$ and $i_0$, there are $5 \times 3 = 15$ pairs of $(e_0, i_0)$, so the total number of test particles is $30,000 \times 15 = 450,000$. This leads to even higher resolution in the coverage of both the initial $a$ and $A$.

\subsection{Numerical simulations}

To explore the dynamical stability of resonators populating different MMRs beyond 50 AU, we numerically integrate the orbits of test particles, as generated in Section 4.1, within the framework of the current outer Solar system. In the numerical integrations performed here, we also employ the SWIFT\_RMVS3 integrator with a time-step of 0.5 yr \citep{Levi1994}. Due to the considerably large number of test particles, we first carry out 10 Myr pre-runs to filter out non-resonant particles. This 10 Myr integration timescale is the same as that adopted to identify the observed RKBOs, as it is long enough for the preliminary selection of the candidate resonators. We have to point out that among the 50 $m$:$n$ resonances associated with the observed RKBOs (see Table \ref{erange}), those with $m \geq 4$ actually do not include the individual highest-order resonances for each $m$, i.e. the farthest ones interior to 100 AU. For completeness, our numerical studies will also take into account all such highest-order resonances, specifically up to the 4:23, 5:29, 6:35, and 7:41 MMRs.

In order to demonstrate that a sufficient number of candidate resonators are available for subsequent long-term simulations, we use the planar case of $i_0=0^{\circ}$ as an example. Given $e_0=0.5$, from a sample of 30,000 test particles, the number of candidate resonators found in the 1:$n$, 2:$n$, 3:$n$, and (4-7):$n$ resonances are 1458, 1078, 958, and 1057, respectively. In fact, for the selected $e_0$ values ranging from 0.1 to 0.7 and $i_0$ values between $0^{\circ}$ and $40^{\circ}$, a total of about 10,000 candidate resonators are produced in the pre-runs. Even though a significant proportion may escape the resonances during the long-term evolution, a considerable number of resonators can still remain afterwards for statistical analysis.

\begin{figure*}
  \centering
  \hspace{-1 cm}
  \begin{minipage}[c]{1\textwidth}
  \centering
  \includegraphics[height=12cm]{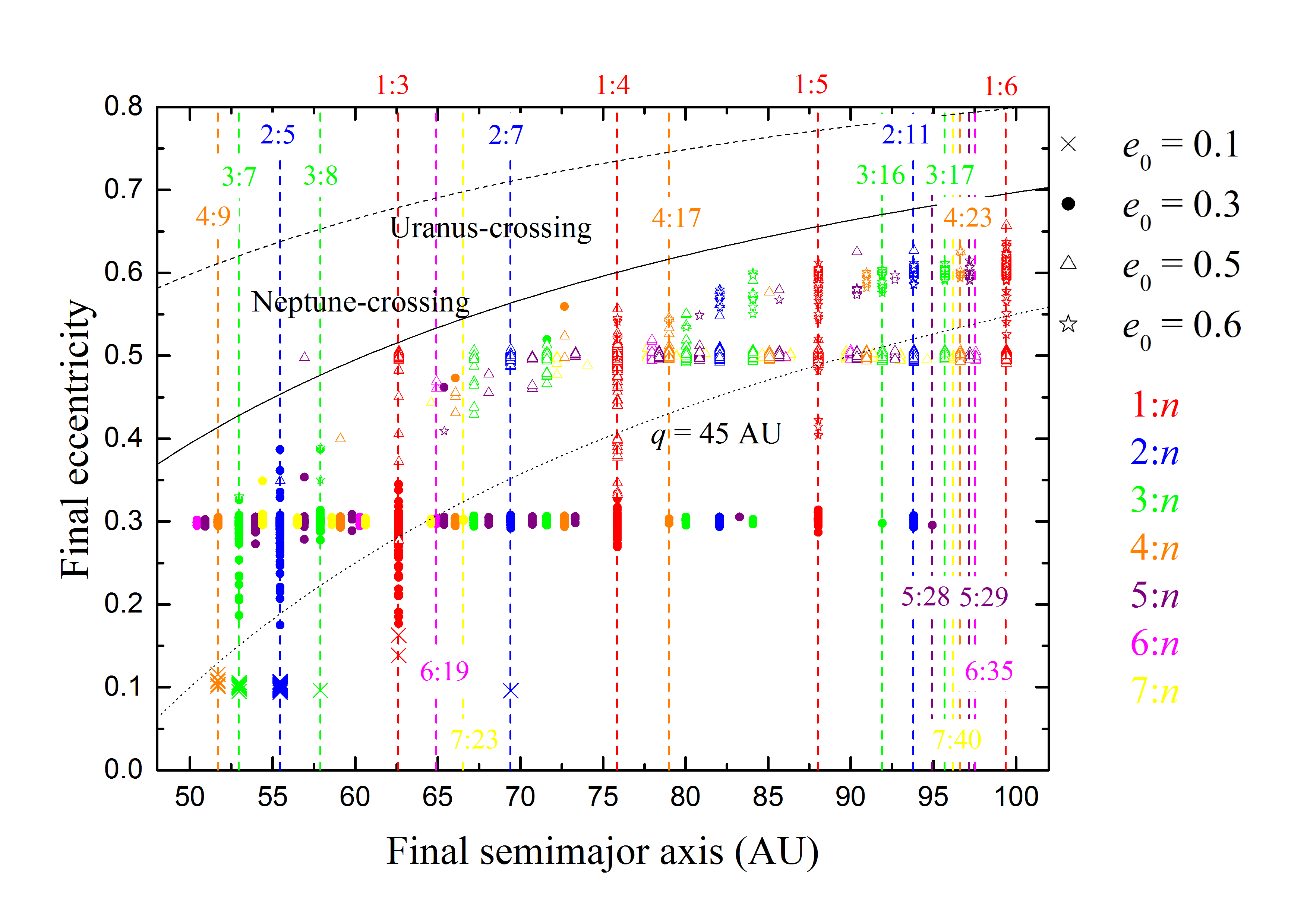}
  \end{minipage}
  \caption{Final semimajor axes and eccentricities of the stable resonators in the planar case at the end of the 4 Gyr simulations, for initial eccentricities $e_0=0.1, 0.3, 0.5$, and 0.6 (indicated by different symbols). In each panel, different colours represent the stable resonators associated with the $m$:$n$ resonances for varying values of $m$ from 1 to a maximum of 7. The vertical dashed lines indicate, among the occupied $m$:$n$ resonances, some interesting ones for each $m$. Similar to Fig. \ref{width}, three curves of constant perihelion distance (Uranus crossing, Neptune crossing, and $q=45$ AU) have been plotted for reference.}
 \label{casei0}
\end{figure*}

Regarding the candidate resonators obtained from the pre-runs, we further examine their long-term evolution over the age of the Solar system, of 4 Gyr. After completing the 4 Gyr simulations, we extend the integrations for an additional 10 Myr to analyse the resonant angle behaviours of the surviving objects. Those survivors who consistently librate within their respective MMRs during the 10 Myr extension are classified as `stable resonators'. In this time window, we also calculate their average semimajor axes, eccentricities, and inclinations, which are recorded as the final orbital elements $a_f$, $e_f$, and $i_f$, respectively. In the following, we will present the results in detail for stable resonators, separately for the planar and high-inclination cases, across a variety of MMRs.

\subsection{Results for the planar case ($i_0=0^{\circ}$)}

\subsubsection{$e_0=0.1$}

As shown in Fig. \ref{real}, for the RKBOs currently observed beyond 50 AU, all have eccentricities $e>0.1$. Accordingly, in our numerical simulations, we set the minimum initial eccentricity of the test particles to be $e_0=0.1$. This eccentricity value corresponds to the narrowest libration zone (see Fig. \ref{width}), where the number of resonators is theoretically expected to be the lowest. In the planar case of $i_0=0^{\circ}$, there are 30,000 test particles initialised with $e_0=0.1$. After 4 Gyr of evolution, only 48 stable resonators are found, leading to an extremely small production fraction of only $\sim0.16$\%. Their final orbital distribution is presented in Fig. \ref{casei0} (crosses). It shows that the minimum final eccentricity of these stable resonators reaches $e_f\sim0.1$, whereas no MPC RKBOS have $e < 0.18$. Although RKBOs with such small $e$ values may potentially exist, their proportion could be much smaller compared to those with larger $e$ values. This would significantly reduce the probability of detection. It should be noted that the low-e orbit also reduces the brightness at perihelion, which further decreases the likelihood of detection.

For the 1:3 MMR, in the planar case, the critical eccentricities marking the onset of symmetric and asymmetric librations are $e_a(i=0^{\circ})\sim0$ and $e_c(i=0^{\circ})=0.12$, respectively (see Fig. \ref{eaec}). Theoretically, at a low eccentricity of $e=0.1$, where $e < e_c(i=0^{\circ})$, stable asymmetric libration should not appear. As indicated by the red crosses in Fig. \ref{casei0}, among the 48 stable resonators, only two reside in the 1:3 MMR, with their final eccentricities excited to $e_f \sim 0.14$-0.16. Although $e_f$ exceeds $e_c(i=0^{\circ}) = 0.12$, which allows both symmetric and asymmetric librations, these two 1:3 resonators actually exhibit symmetric libration and an alternating state between symmetric and asymmetric librations, respectively. For the higher-order 1:4, 1:5, and 1:6 MMRs, since the critical eccentricity $e_c$ increases with the order of the resonance, their $e_c(i=0^{\circ})$ values remain consistently larger than 0.1, while their $e_a(i=0^{\circ})$ values are still as small as $\sim0$. Consequently, similar to the 1:3 MMR, only symmetric libration is expected. However, no stable resonators are found in the 1:4, 1:5, or 1:6 MMRs. We note that although the resonance widths of these three 1:$n$ MMRs are very narrow at $e=0.1$, a total of 52 test particles with $e_0 = 0.1$ fall within the corresponding $a$-ranges. Among them, the 10 Myr pre-runs identify 15 candidate resonators, but none maintain libration after the 4 Gyr evolution. Therefore, the absence of stable resonators in the 1:4, 1:5, and 1:6 MMRs may be attributed to the fact that these resonances are of higher order than the 1:3 MMR, which makes them too weak to sustain resonators at $e$ values as small as $\sim 0.1$.

Next, we examine the other $m$:$n$ resonances with $m \geq 2$. As indicated by the non-red crosses in Fig. \ref{casei0}, for a given $m$, the stable resonators tend to cluster in the lowest-order resonances (i.e. those with the smallest $|n-m|$ values), such as the 2:5 and 2:7 MMRs or the 3:7 and 3:8 MMRs. For $m=4$, only the lowest-order 4:9 MMR hosts stable resonators. Like the 3:8 MMR, it is a 5th-order resonance. Bearing in mind that the resonance's strength weakens as the resonance order increases, the number of stable resonators could consequently decrease. This leads to no stable resonators being found in the 6th- or higher-order resonances, including all $m$:$n$ resonances with $m \geq 5$, e.g. the empty 5:11 MMR.

\subsubsection{$e_0=0.3$}

The stable resonators produced in the case of $e_0=0.3$ are indicated by dots in Fig. \ref{casei0}. It is evident that these more eccentric resonators are significantly more numerous than those from the case of $e_0=0.1$ (denoted by crosses). This is a natural consequence of the substantial increase in resonance's strength with eccentricity. At $i_0=0^{\circ}$, out of the 30,000 test particles initialised with $e_0=0.3$, 1,464 stable resonators are produced, yielding a fraction of 4.88\%, which is more than 30 times higher than the production fraction in the case of $e_0=0.1$. Note that this production rate is derived under the assumption that the test particles initially cover the entire orbital space, including both the resonant and non-resonant regions. When considering the survival rate, $R_S$, of resonators, i.e. the fraction of the 1,464 stable resonators relative to the 2,487 candidate resonators generated in the pre-runs, $R_S$ reaches a remarkably high value of approximately 59\%.

For the 1:$n$ resonances, a careful examination of the associated stable resonators, indicated by the red dots in Fig. \ref{casei0}, reveals the emergence of asymmetric resonators in the 1:3 and 1:4 MMRs. This is attributed to the fact that, in the planar case, the critical eccentricities associated with the asymmetric libration of these two MMRs are $e_c^{1:3}(i=0^{\circ}) = 0.12$ and $e_c^{1:4}(i=0^{\circ}) = 0.20$, both of which are significantly lower than the initial eccentricity of $e_0 = 0.3$. In contrast, for the higher-order 1:5 MMR, the critical eccentricity $e_c^{1:5}(i=0^{\circ}) = 0.27$ is very close to $e_0 = 0.3$. Similarly to the behaviours of the 1:3 stable resonators found earlier in the case of $e_0=0.1$, here all 1:5 stable resonators exhibit either symmetric libration or an alternating state between symmetric and asymmetric librations. However, for the highest-order 1:6 MMR, no stable resonators are found. With the emergence of 1:$n$ asymmetric resonators, an important subject to investigate is the number difference between the leading and trailing resonators. This will be analysed in the following sections, incorporating results from other cases with even larger $e_0$.

Regarding the $m$:$n$ resonances with $m \geq 2$, as indicated by the non-red dots in Fig. \ref{casei0}, we observe that the stable resonators can occupy very high-order resonances. Given different values of $m$, the highest-order resonances found to be occupied are the 2:11, 3:16, 4:17, 5:28, 6:19, and 7:23 MMRs. Note that within the considered region interior to 100 AU, only the 2:11 MMR serves as the outermost 2:$n$ resonance. However, for each of the remaining resonances with $m=3$-7, there exist more distant resonance(s) with larger $n$ value(s). These more distant, higher-order resonances are expected to be populated by stable resonators, as some of them, such as the 4:21, 6:25, and 7:29 MMRs, have been identified as being inhabited by observed RKBOs. This potential occupancy will be confirmed later in cases with $e_0>0.3$.

\subsubsection{$e_0=0.5$ and $0.6$}

Since the initial eccentricity $e_0$ in the range of 0.5–0.6 is quite large, significantly exceeding the critical eccentricities of $e_{c}^{1:5} (i=0^{\circ}) = 0.27$ and $e_{c}^{1:6} (i=0^{\circ}) = 0.32$ for the 1:5 and 1:6 MMRs, respectively, asymmetric stable resonators begin to appear in these two highest-order 1:$n$ resonances (indicated by the red triangles and stars in Fig. \ref{casei0}). Furthermore, compared to the case with a smaller $e_0 = 0.3$ (red dots), the proportion of asymmetric stable resonators in the lower-order 1:3 and 1:4 MMRs has increased substantially, while only a small fraction of symmetric stable resonators remain. In particular, nearly all of the 1:3 stable resonators now exhibit asymmetric librations.

For the $m$:$n$ resonances with $m \geq 2$, as indicated by the non-red triangles and stars in Fig. \ref{casei0}, a major result of increasing $e_0$ to values as large as 0.5–0.6 is that the stable resonators can exist in even higher-order resonances. Regarding $m=2$-6, the highest-order resonances occupied by stable resonators are 2:11, 3:17, 4:23, 5:29, and 6:35 MMRs, which are the farthest resonances inside 100 AU for their respective values of $m$. When $m = 7$, stable resonators have been found in the 7:40 MMR. We note that among the 7:$n$ resonances interior to 100 AU, the 7:40 MMR is the second-highest-order resonance. 
As for the highest-order 7:41 MMR, we deem its occupancy quite plausible if the number of test particles is increased, while additional simulations would not be performed, given that the overall distribution of stable resonators is already well established. 
In fact, the only observed 7:$n$ RKBO is identified to be located in the 7:29 MMR (at $a \sim 77.6$ AU), which is much lower in resonance order than the 7:40 MMR. Therefore, the presence of stable resonators up to the 7:40 MMR (at $a \sim 96.2$ AU) is sufficient to imply that the 7:$n$ RKBOs could possibly spread across the entire 50-100 AU region.


\subsubsection{$e_0=0.7$}

Given the largest $e_0$ of 0.7, the perihelia ($q$) of the test particles can be very small. For particles trapped in Neptune's MMRs, even if they begin on Neptune-crossing orbits (i.e. $q<30.1$ AU), they may still remain stable due to the resonant phase-protection mechanism \citep{nesv01}, while such resonant particles must avoid crossing Uranus’s orbit at 19.2 AU, which requires $a_0>64.97$ AU. From this perspective, a considerable fraction of the 30,000 test particles could, in principle, remain stable for a long time. However, none of them survive the 4 Gyr integration, and the longest lifetime is only on the order of $10^8$ yr (specifically, $\sim 0.82 \times 10^8$ yr). Since no long-term survivors exist, stable resonators on such extremely eccentric orbits are naturally absent. This numerical result is consistent with the eccentricity distribution of the observed RKBOs beyond 50 AU, as all of them have $e \leq 0.65$ (see Fig. \ref{real}(a)).

Here, we considered the planar case, where the inclinations of the test particles are initially set to $0^\circ$. If the test particles move on high-inclination orbits, geometrically, they would spend only a small fraction of their time near the ecliptic plane, thereby reducing the likelihood of experiencing strong gravitational perturbations from the planets. Consequently, high-inclination particles starting with $e_0$ as large as 0.7 may survive for up to 4 Gyr of evolution. This scenario will be explored in the inclined case below, as detailed in Section 4.4.

\subsubsection{Summary of all $e_0$-cases}

\begin{figure*}
  \centering
  \hspace{-1 cm}
  \begin{minipage}[c]{1\textwidth}
  \centering
  \includegraphics[height=12cm]{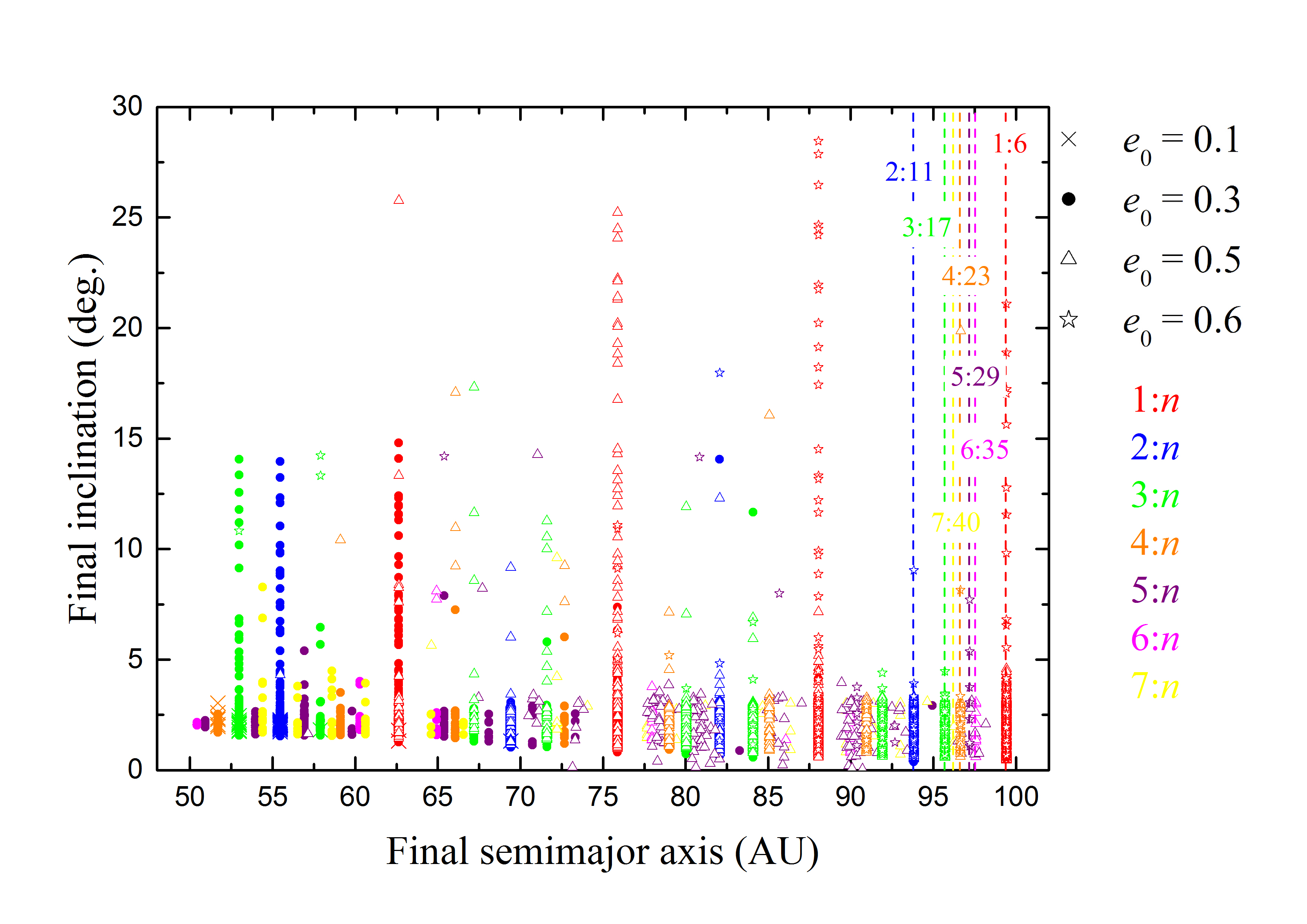}
  \end{minipage}
    \caption{Similar to Fig. \ref{casei0}, but showing the distribution in final semimajor axes and inclinations for stable resonators. For reference, the highest-order $m$:$n$ resonances for each $m$, which are equivalently the farthest, are marked by vertical dashed lines.}
  \label{casei0all}
\end{figure*}

In this subsection, we summarise the planar case results by combining the outcomes for different $e_0$ values discussed above. For the stable resonators obtained from simulations with $e_0$ ranging from 0.1 to 0.7, Fig. \ref{casei0} presents their distributions in final semimajor axes and eccentricities, while Fig. \ref{casei0all} further shows their final semimajor axes and inclinations. For clarity, in Fig. \ref{casei0all}, only the highest-order (equivalent to the farthest) $m$:$n$ resonances for each $m$ are highlighted with vertical dashed lines.

First, we start by examining the occupancy of Neptune's MMRs in the distant Kuiper belt beyond 50 AU. As shown in Fig. \ref{casei0}, the stable resonators populate all the 1:$n$ and 2:$n$ resonances, with the 1:6 and 2:11 MMRs being the respective highest-order resonances, consistent with the current observations. For the 3:$n$, 4:$n$, 5:$n$, 6:$n$, and 7:$n$ resonances, the highest-order resonances occupied by stable resonators are the 3:17, 4:23, 5:29, 6:35, and 7:40 MMRs, respectively. This suggests that RKBOs can potentially exist in nearly all $m$:$n$ resonances with $m=3$–7 within the 50–100 AU region. The only exception is the 7:41 MMR; however, as discussed in Section 4.4.3, not only is its occupancy quite plausible, but its absence does not affect the overall resonance occupancy in the distant Kuiper belt. In contrast, MPC RKBOs are found only in resonances up to 3:16, 4:21, 5:22, 6:25, and 7:29 MMRs, and all resonant objects have perihelion distances of $q<45$ AU (see Fig. \ref{real}(a)). Even higher-order MMRs at $a\approx80$-100 AU remain unoccupied, but potential RKBOs there would still be detectable at perihelion if $e\gtrsim 0.44$-0.55 (i.e. $q<45$ AU).

In Section 2.2.1, we noted that the currently known RKBOs from the MPC tend to have high-$e$ orbits. However, according to the evolutionary simulations of the full outer Solar system performed here, we find that potential RKBOs can remain stable on lower-$e$ orbits. For example, Fig. \ref{casei0} shows that simulated RKBOs with $e\sim0.3$ are distributed throughout the region between $a=50$ and 100 AU, whereas all MPC RKBOs with $a>70$ AU have $e>0.4$ (see Fig. \ref{real}(a)). Although even less eccentric objects ($e<0.3$) can be expected in the simulations if an initial eccentricity of $e_0 = 0.2$ is adopted, the difference in the lower limit of $e$ between the simulated and MPC RKBOs illustrated here is sufficient to demonstrate the observational bias toward detecting high-$e$ objects with large $a$, i.e. those with relatively smaller perihelion distances.

Regarding resonance occupancy, another important aspect to investigate is the number asymmetry of 1:$n$ resonators between the leading and trailing swarms. Current MPC RKBOs in the 1:3, 1:4, 1:5, and 1:6 MMRs beyond 50 AU consistently show that the leading swarm hosts at least as many objects as the trailing swarm (see Table \ref{1tonRKBO} and the discussion in Section 2.2.2). This number asymmetry is in agreement with the distribution of 1:2 RKBOs interior to 50 AU. The number ratio of leading to trailing 1:$n$ RKBOs, denoted as $R_{LT}$, is almost certainly linked to Neptune's outward migration and resonance capture in the early Solar system \citep{Li2014b, Lih23}. After Neptune and its resonances reached their current locations, the subsequent long-term evolution could have further altered the number ratio $R_{LT}$. In our numerical simulations, the test particles are initially distributed uniformly across all resonance regions. This means that, at the beginning of long-term evolution, there is little to no difference in the number of particles placed in the leading and trailing swarms for any 1:$n$ MMR. After 4 Gyr of evolution, the leading-to-trailing number ratios are calculated to be $R_{LT}=63/85$, 90/99, 89/74, and 109/107 for the 1:3, 1:4, 1:5, and 1:6 MMRs, respectively. These ratios indicate that, except for the 1:5 MMR, the long-term stability of leading 1:$n$ MMRs is weaker than that of trailing ones, which appears to contradict the current observations.

It should be noted that the current MPC RKBOs in the 1:3, 1:4, 1:5, and 1:6 MMRs are quite scarce. As listed in Table \ref{1tonRKBO}, their total number is only 30, comparable to the 27 reported by \citet{Lyka2023}. Thus, for each of these four 1:$n$ MMRs, whether the leading swarm is truly more populated than the trailing swarm is still an open question. The apparent population difference between these two swarms may be due to unknown MPC observational biases, as recognised by \citet{Chen19} (see the discussion in Section 2.2.2). Despite observational biases, \citet{PL2017} showed that, in the early stages of formation under the framework of planetary migration, the leading 1:$n$ resonators consistently outnumber the trailing ones. Moreover, the leading-to-trailing number ratios $R_{LT}$ exceed 2 in all cases, except for the 1:5 MMR, which has a relatively smaller $R_{LT} = 1.2$. This suggests that when the 1:$n$ RKBOs originally arrived at their current locations, in general, $R_{LT}$ may be significantly greater than 1. Even if more leading resonators became unstable and escaped during the later long-term evolution, the value of $R_{LT}$ would only have decreased while still remaining $>1$. Based on the above discussion, it is likely that all 1:$n$ RKBOs beyond 50 AU are not formed in situ. The intrinsic number asymmetry of these distant 1:$n$ RKBOs between the leading and trailing swarms could provide further constraints on the planetary migration models.


Second, we estimate the survival rates of various resonant populations in the range of 50–100 AU. In the planar case, we combine the results of all simulations with five different values of $e_0$, each of which contains 30,000 test particles, resulting in a total of 150,000. As described in Section 4.2, we begin by identifying objects associated with the $m$:$n$ resonances ($m = 1$–7) from the 10 Myr pre-runs, resulting in 10,424 candidate resonators. These candidates are then included in the 4 Gyr long-term runs to determine the stable resonators. By comparing the numbers of candidate and stable resonators, we can measure the survival rate for each set of $m$:$n$ resonances with the same $m$-value.

Under the representative starting distribution that is assumed, the resonant population occupies the entire dynamically allowed phase space of each $m$:$n$ resonance. Then for the 1:$n$, 2:$n$, and 3:$n$ resonances, the survival rates are 51.7\%, 52.7\%, and 48.5\%, respectively. This indicates that these three sets of resonances can preserve around half of their resonators initially captured from the primordial Kuiper belt. However, for the 4:$n$, 5:$n$, 6:$n$, and 7:$n$ resonances, the survival rates drop sharply to 34.9\%, 34.7\%, 28.7\%, and 19.3\%, respectively. These much lower survival rates of only $\sim20$–30\% are evidently due to the higher-order nature of these four sets of resonances, which makes them too weak to effectively sustain their resonant populations. Furthermore, we point out that among all $m$:$n$ resonances embedded in 50-100 AU, the four sets with $m = 4$–7 contain a total of 54 resonances, whereas the other three sets with $m = 1$–3 contain only 16. Even though the former has the advantage of a larger total of resonances, this cannot compensate for the disadvantage of weaker strengths associated with higher resonance orders. In fact, among the 207 observed RKBOs between 50 and 100 AU, as identified in Section 2, only about 17.9\% reside in the 4:$n$, 5:$n$, 6:$n$, and 7:$n$ resonances. Therefore, our long-term stability study suggests that the small number of RKBOs located in these resonances is likely a genuine feature rather than a consequence of observational bias.

Third, we briefly examine the distribution of final inclinations for stable resonators. In the planar case, although these objects are set to have initial inclinations of $i_0=0^\circ$, referenced to the J2000.0 ecliptic plane, they are still slightly inclined by $1^\circ$-$2^\circ$ relative to the orbits of the giant planets. Therefore, in general, their inclinations can be excited, but only to a small extent. As shown in Fig. \ref{casei0all}, most stable resonators from our simulations indeed remain on nearly planar orbits, with final inclinations $i_f$ of just a few degrees. However, we observe that some stable resonators can achieve highly inclined orbits, with $i_f$ up to $30^\circ$. A careful check of their 4 Gyr evolution reveals that the major factor is the temporary Kozai mechanism, as characterised by the short-term libration of the argument of perihelion. Since inclination can be traded for eccentricity due to this mechanism, an object’s orbit may become more inclined. If the Kozai mechanism happens to cease when the object's inclination reaches a high value, it can remain on such a high-inclination orbit, similar to the inclined stable resonators we observe in Fig. \ref{casei0all}.

The appearance of the Kozai mechanism typically requires high inclinations, with a minimum value of $39.2^\circ$ \citep{Naoz16}. However, when an object lies within the exterior resonance of a planet, the minimum inclination for the Kozai mechanism can be significantly reduced, e.g. to about $10^\circ$ for Neptune’s 2:3 MMR at 39.4 AU \citep{Wan07}. Furthermore, within Neptune’s higher-order MMRs beyond 50 AU, the Kozai mechanism could even act on nearly planar orbits. In our planar case simulations, a small fraction of stable resonators are indeed noticeably excited in their inclinations by this mechanism. However, we find that only about 0.4\% reside on orbits with $i_f > 20^\circ$, and all of them are located in the 1:$n$ type resonances, as indicated by the red symbols in Fig. \ref{casei0all}. In contrast, the observed RKBOs with such high inclinations are much more numerous (see Fig. \ref{real}(b)); moreover, they spread across various resonances rather than just the 1:$n$ type resonances. These discrepancies suggest that the RKBOs should have already acquired high inclinations before reaching their current locations, that is, prior to the starting point of the long-term evolution considered in this paper. Therefore, we will next analyse the evolution and distribution of stable resonators in the inclined cases with initial inclinations of $i_0 \geq 20^\circ$.

\subsubsection{Number reversal}

For the $m$:$n$ resonances with a fixed $m$, the resonance order increases with $n$, leading to a narrower libration zone. Consequently, a common expectation is that higher-order resonances host fewer resonators. This expectation is supported by our simulations at small $e_0=0.1$: after 4 Gyr of evolution, 7 stable resonators are detected in the lower-order 3:7 MMR, but only 1 in the higher-order 3:8 MMR. However, an unexpected exception arises at the larger $e_0 = 0.3$, where the 3:7 MMR hosts fewer stable resonators (119) than the 3:8 MMR (140). This unusual phenomenon, which we refer to as `number reversal', is a key highlight of this study. In the following, we will perform detailed analyses to reveal the underlying mechanism.

A well-known property of the resonance phase space is that the maximum deviation of the semimajor axis $a$ from the nominal resonance location $a_{res}$, as defined in Eq. (\ref{ares}),  characterises the maximum deviation of the resonant angle from the libration centre, i.e. the resonant amplitude $A$ \citep{LiX2022}. In the case of $e_0 = 0.3$, we observe that fewer than 7\% of the 3:7 stable resonators occupy orbits with $A > 120^\circ$, and none have $A > 140^\circ$. In contrast, a significantly larger fraction, about 40\%, of 3:8 stable resonators possess $A$ values in the range $120^\circ$-$180^\circ$. Therefore, we argue that the number difference of stable resonators between the 3:7 and 3:8 MMRs is primarily governed by two factors: (1) The 3:7 MMR, being a lower-order resonance, has a broader libration zone, which generally allows for a larger capacity to accommodate resonators. (2) The 3:8 MMR, despite being a higher-order resonance, is located at a greater heliocentric distance. Thus, for a given eccentricity, the 3:8 resonators have larger perihelion distances than the 3:7 resonators. This enhances their stability at larger $A$, leading to a higher proportion of the libration zone being occupied by stable resonators, i.e. a wider stable region. The straightforward explanation for this $A$-dependent stability can be understood from the resonant configuration in physical space. The value of $A$ directly measures the longitude separation of Neptune from the resonator at perihelion \citep{Malh1995}. Larger $A$ corresponds to a smaller separation between these two objects, which may increase the likelihood of stronger gravitational encounters and consequently weaken stability. As a result, the greater heliocentric distance of the 3:8 resonators helps mitigate this destabilising effect, thereby supporting their survivability at larger $A$.

\begin{figure*}
  \centering
  \begin{minipage}[c]{1\textwidth}
  \vspace{0 cm}
  \includegraphics[width=9cm]{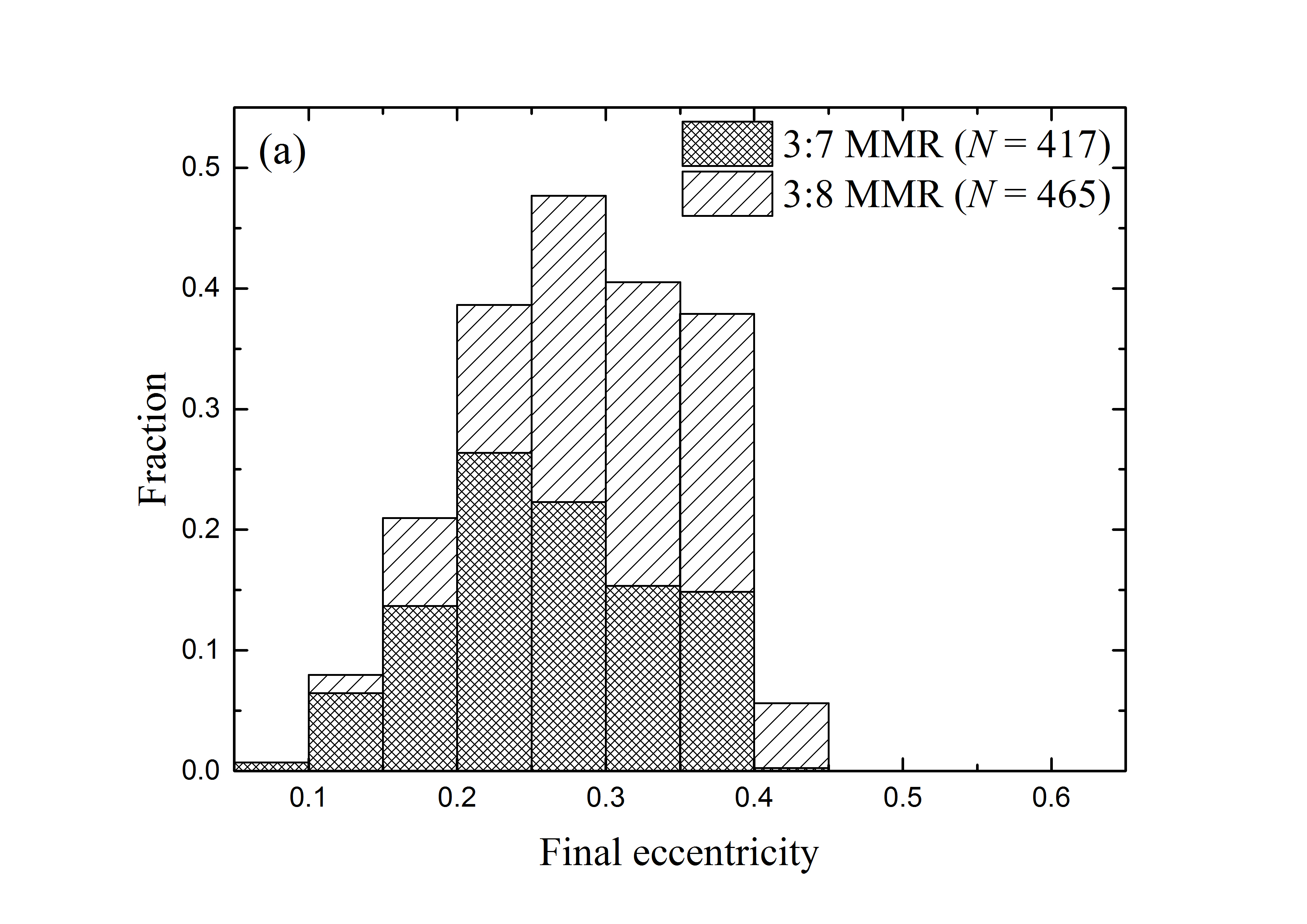}
  \includegraphics[width=9cm]{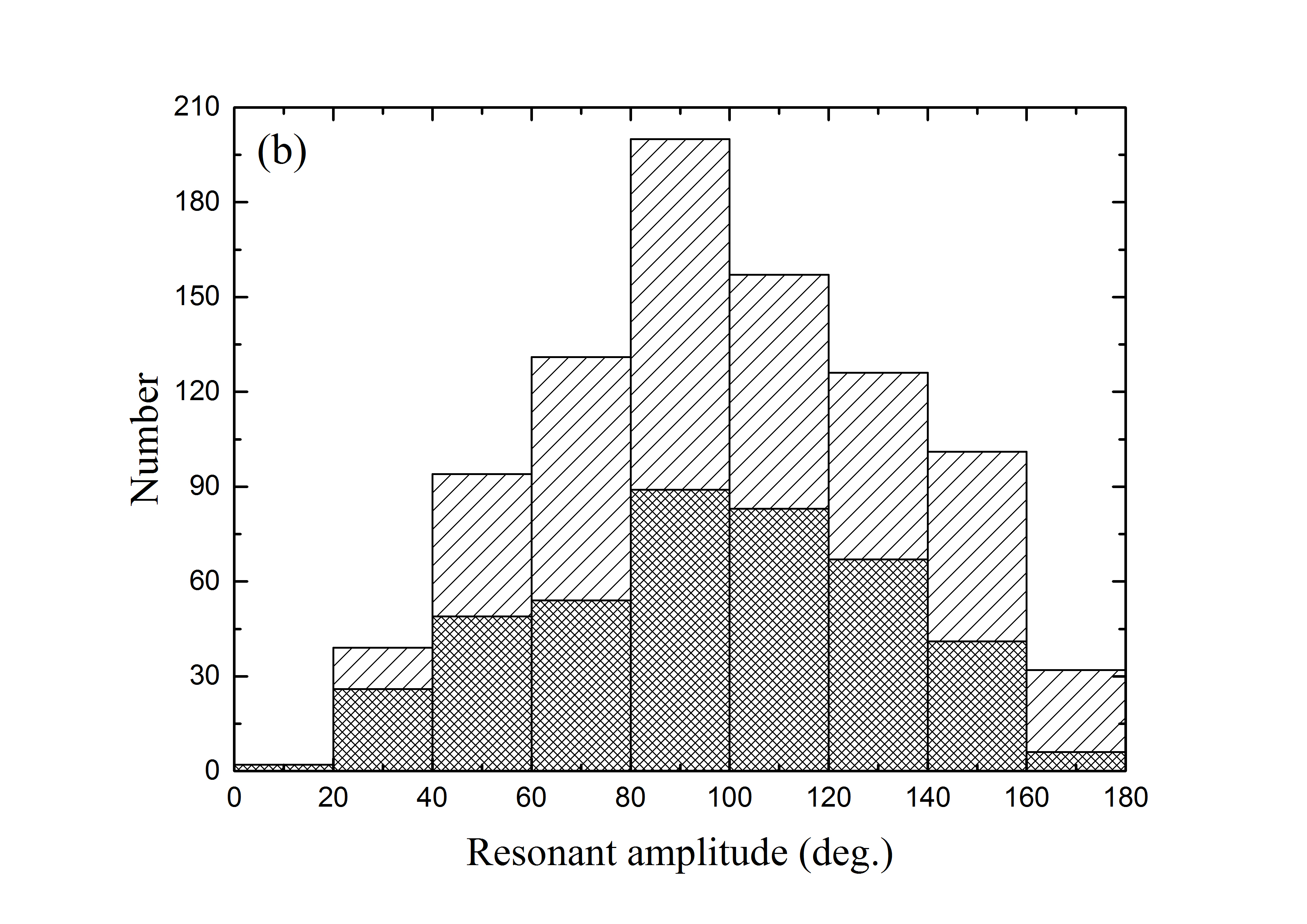}
  \end{minipage}
  \caption{Histogram of the distribution of the 3:7 and 3:8 stable resonators from an additional simulation, where test particles are initially spaced between 50 and 60 AU (i.e. in the neighbourhoods of the 3:7 and 3:8 MMRs), with initial eccentricities continuously distributed in the region of 0.1-0.7 and initial inclinations of $0^\circ$. (Panel a) Final eccentricity distribution. (Panel b) Resonant amplitude distribution.}
  \label{res3n}
\end{figure*}


Obviously, the competition between resonance order and heliocentric distance, i.e. the two factors discussed above, depends on the eccentricity of the resonators. At lower eccentricities (e.g. $e = 0.1$), the perihelion distances of resonators with $a>50$ AU are uniformly large. In this case, nearly the entire libration zones of both the 3:7 and 3:8 MMRs are stable. Since the lower-order 3:7 MMR has a broader libration zone, it can accommodate more objects. However, at higher eccentricities (e.g. $e = 0.3$), the perihelion distances of the resonators decrease significantly. Although the libration zone of the higher-order 3:8 MMR is narrower, the stable region within it is relatively wider, allowing more objects with large $A$ to survive. Unfortunately, the phenomenon of number reversal cannot be further examined at $e_0\ge0.5$ due to the scarcity of 3:7 or 3:8 stable resonators produced under these initial conditions.


The hints of a possible unexpected number distribution discovered above inspire further investigation. Since the occurrence of number reversal depends on eccentricity, it is more appropriate to adopt a continuous distribution of $e_0$ over a range instead of a few discrete values. Focusing on the 3:7 and 3:8 MMRs, located at 52.9 AU and 57.9 AU respectively, we carry out an additional simulation. The initial semimajor axes and eccentricities of the test particles are randomly selected within the ranges $a_0=50$–60 AU and $e_0=0.1$–0.7, while still considering the planar case with initial inclinations $i_0 = 0$. Under these initial conditions, we reassess the long-term stability of the 3:7 and 3:8 MMRs using 90,000 test particles. This setup achieves a high spatial resolution of $\Delta a_0\sim 1.1 \times 10^{-4}$ AU, increasing the initial number density of test particles in $a_0$ within these two resonances by a factor of three compared to the main simulations conducted in this study.


After 4 Gyr of evolution, 417 stable resonators survive in the 3:7 MMR and 465 in the 3:8 MMR. This further confirms that the higher-order 3:8 MMR indeed hosts more stable resonators, resulting in the phenomenon of number reversal. We then wondered whether this unexpected number distribution was shaped during formation or resulted from later long-term evolution. Considering that the initial distribution of test particles covers the entire libration zones for both the 3:7 and 3:8 MMRs, we examined the candidate resonators obtained from the 10 Myr pre-runs. There are 993 candidate resonators in the 3:7 MMR and 921 in the 3:8 MMR. This aligns with the common expectation that the lower-order 3:7 MMR should host more objects, i.e. no reversal in number. Comparing the number of stable resonators to that of candidate resonators, we find that after 4 Gyr of evolution, the survival rate of 3:7 resonators is $R_S \approx 42.0\%$ (417/993), which is obviously lower than $R_S\approx 50.2\%$ (465/921) for 3:8 resonators. The underlying reason is the same as we argued above for the case of $e_0=0.3$: at larger eccentricities $e$, the resonance amplitudes $A$ of the stable resonators tend to be smaller, reducing the size of the stable region within the libration zone. Compared to the 3:7 MMR, the higher-order 3:8 MMR is located at a greater heliocentric distance. Consequently, for the same $e$, the 3:8 resonators have larger perihelion distances, which can partially compensate for the shrinking effect of large $e$ on their stable region.


Fig. \ref{res3n}(a) presents the distribution of final eccentricities for stable resonators in the 3:7 and 3:8 MMRs from the additional simulation conducted here. For the 3:7 stable resonators, the largest fraction is concentrated around the final eccentricities of $e_f=0.2$–0.25, while their number begins to decrease at larger $e_f$, with almost none found at $e_f > 0.4$. In contrast, for the 3:8 stable resonators, the fraction remains consistently high within the range of $e_f=0.25$–0.4; in addition, a notable proportion is still present even at $e_f > 0.4$. This difference in eccentricity distribution indicates that the higher-order 3:8 MMR allows more resonators to occupy highly eccentric orbits compared to the 3:7 MMR. In other words, for the same range of large $e_f=0.25$–0.4, the greater heliocentric distance of the 3:8 MMR enhances the stability of its resonant population, as previously reasoned for the occurrence of number reversal between the 3:7 and 3:8 stable resonators. To further support our argument regarding the possible cause of this phenomenon, Fig. \ref{res3n}(b) provides the resonant amplitude ($A$) distribution of the 3:7 and 3:8 stable resonators. It is visually apparent that the number of 3:7 resonators drops sharply for $A > 140^\circ$, while many more 3:8 resonators with similarly large $A$ values can survive. The predominance of 3:8 stable resonators at such large $A$, which indicates a higher resonance capacity, substantially contributes to the overall size of this population.


So far, our results on the phenomenon of number reversal have been achieved under the initial conditions where the regions of all $m$:$n$ resonances are entirely covered by test particles. We believe these results are robust for the long-term stability of RKBOs up to the age of the Solar system. However, still using the pair of 3:7 and 3:8 MMRs as an example, if future observations reveal that more objects reside in the former resonance but not in the latter, i.e. contrary to our stability analysis, then the formation process of the RKBOs should be thoroughly reconsidered. This could potentially offer valuable insights into the eccentricity distribution of primordial KBOs beyond 50 AU, particularly regarding how their number decreases with increasing eccentricity. Additionally, if these distant primordial KBOs were formed in the planetary scattering scenario, further constraints could be placed on the dynamical models of the Solar system, e.g. the giant planet instability and migration model \citep{nesv12, PL2017}. 


Finally, we note that this newly discovered phenomenon of number reversal may extend to other $m$:$n$ resonances with $m$ values different from 3. Specifically, when combining all cases with $e_0=0.1$, 0.3, 0.5, 0.6, and 0.7, number reversal emerges in the following resonance pairs: 1:3 and 1:4 MMRs, 2:5 and 2:7 MMRs, 4:9 and 4:11 MMRs, 5:11 and 5:12 MMRs, 6:13 and 6:17 MMRs, as well as 7:15 and 7:17 MMRs. In each pair, the lower-order resonance consistently hosts fewer resonators, while the higher-order resonance accommodates more. Collectively, these results suggest that the phenomenon of number reversal is likely an intrinsic feature of such adjacent $m$:$n$ resonance pairs.

\subsection{Results for inclined cases ($i_0=20^{\circ}$, $40^{\circ}$)}

\begin{figure*}
  \centering
  \begin{minipage}[c]{1\textwidth}
  \vspace{0 cm}
  \includegraphics[width=9cm]{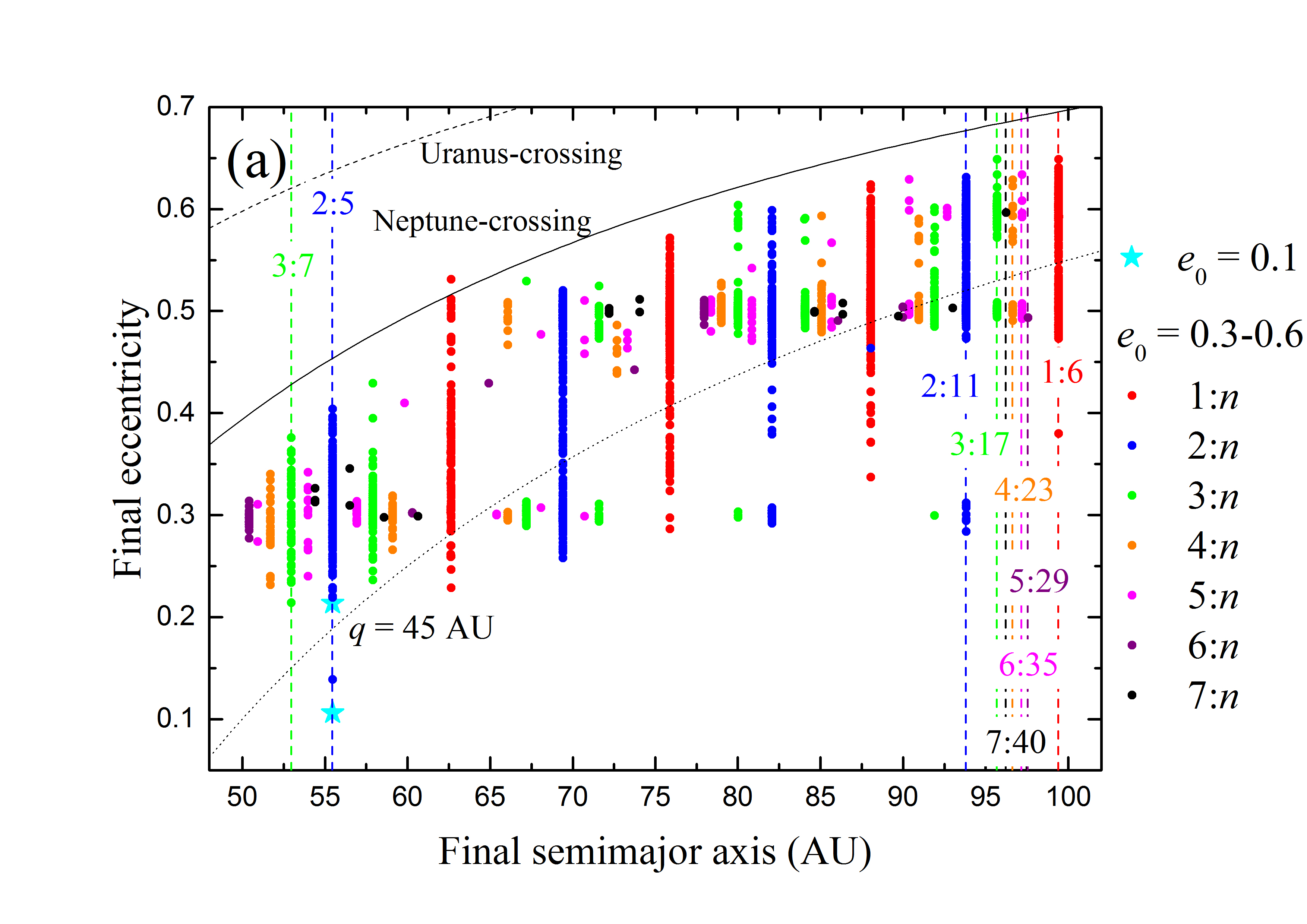}
  \includegraphics[width=9cm]{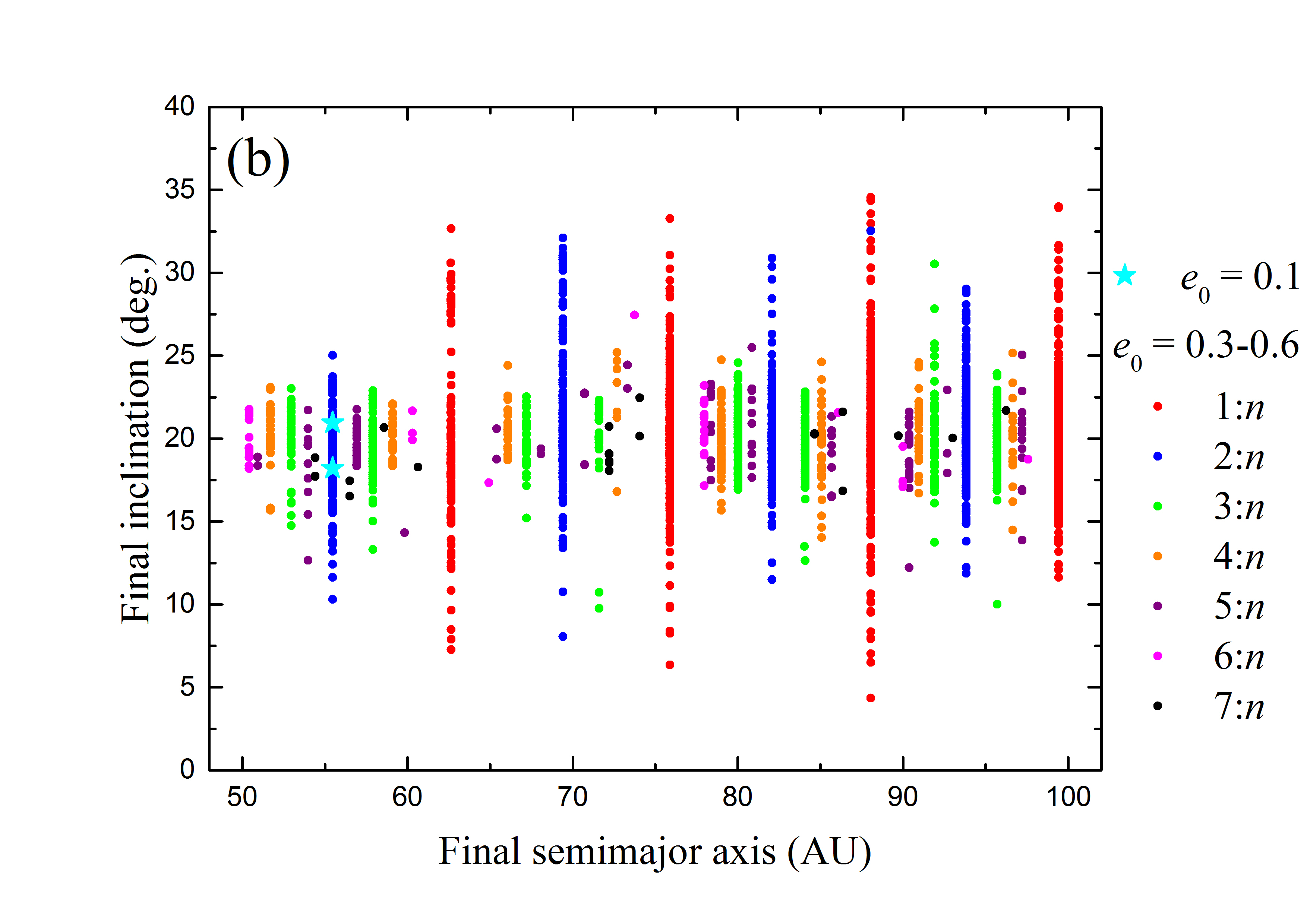}
  \end{minipage}
    \caption{Similar to Figs. \ref{casei0} and \ref{casei0all}, but for stable resonators in the inclined case with initial inclinations of $i_0=20^\circ$. The cyan stars highlight the objects associated with the lowest initial eccentricity of $e_0=0.1$, while the dots indicate those with $e_0=0.3$-0.6. For reference, three curves of constant perihelion distance (Uranus crossing, Neptune crossing, and $q=45$ AU) have been plotted.}
  \label{casei20all}
\end{figure*}

\begin{figure*}
  \centering
  \begin{minipage}[c]{1\textwidth}
  \vspace{0 cm}
  \includegraphics[width=9cm]{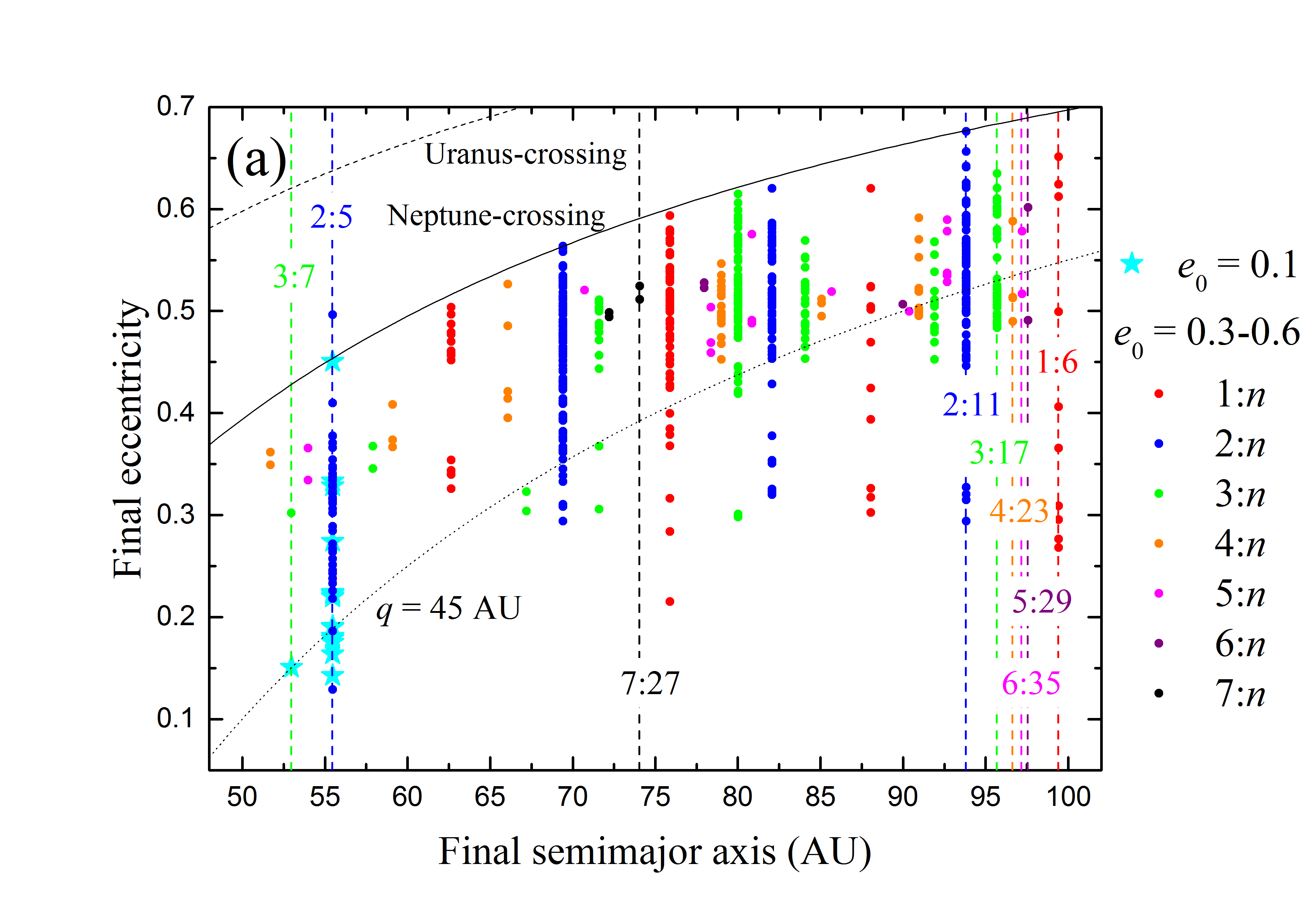}
  \includegraphics[width=9cm]{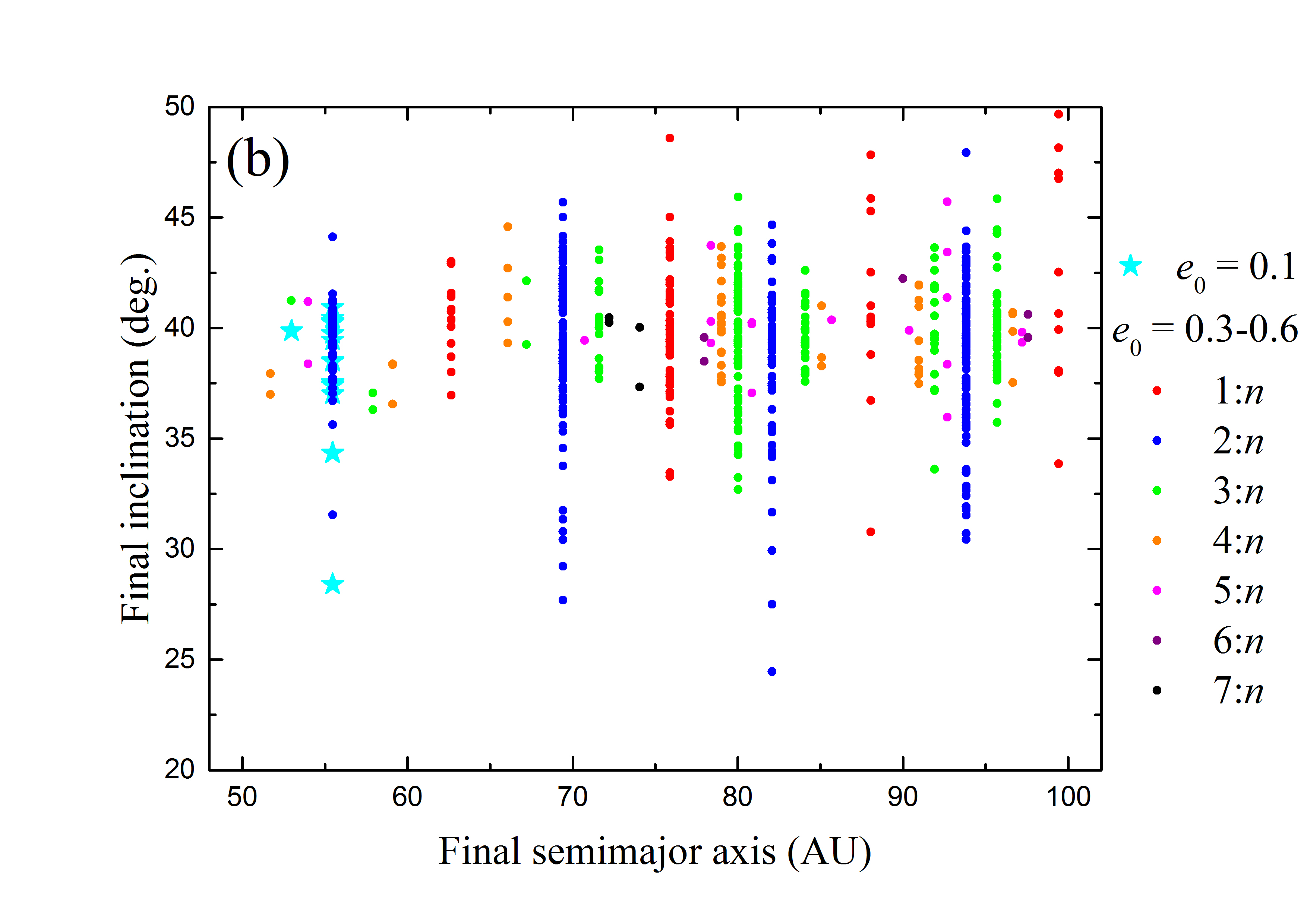}
  \end{minipage}
    \caption{Similar to Fig. \ref{casei20all}, but for even higher initial inclinations of $i_0=40^\circ$.}
  \label{casei40all}
\end{figure*} 

For the high-inclination cases, in general, the effect of eccentricity on the stability of Neptune's MMRs beyond 50 AU is similar to that in the planar case discussed before. Furthermore, since our main focus here is on how inclined orbits influence the distribution of stable resonators, we present in Figs. \ref{casei20all} and \ref{casei40all} the overall distribution of final orbits for the stable resonators from our simulations, with initial inclinations of $i_0 = 20^\circ$ and $40^\circ$, respectively. This is achieved by combining the results from all cases with different $e_0$ values ranging from 0.1 to 0.7. Similar to Fig. \ref{casei0all}, in these two figures, vertical dashed lines are plotted to mark the highest-order $m$:$n$ resonances occupied by stable resonators for each value of $m$.

Taking into account the influence of high inclinations, only an extremely small fraction of test particles with the smallest initial eccentricity of $e_0 = 0.1$ could evolve into stable resonators after 4 Gyr of evolution. These objects are particularly highlighted by the cyan stars in Figs. \ref{casei20all} and \ref{casei40all}. As for test particles initially on the most eccentric orbits with $e_0 = 0.7$, there are a few survivors, but none are trapped in any resonance. These two specific cases of $e_0 = 0.1$ and $0.7$ will first be discussed separately below, followed by the overall results for $e_0 = 0.3$–0.6.

\subsubsection{$e_0=0.1$}

For the 1:$n$ resonances, as in the planar case, neither asymmetric nor symmetric stable resonators appear. This is because both critical eccentricities, $e_a$ (for symmetric libration) and $e_c$ (for asymmetric libration), increase with inclination. As shown in Fig. \ref{eaec}, for the 1:$n$ resonances with $n \ge 3$ beyond 50 AU, when $i \gtrsim 20^\circ$, $e_a$ disappears and thus symmetric libration is no longer possible; meanwhile, $e_c$ far exceeds 0.1 and asymmetric libration cannot occur either. Consistently, in Figs. \ref{casei20all}(a) and \ref{casei40all}(a), none of the stable resonators with $e_0=0.1$ (cyan stars), nor those produced from more eccentric initial conditions ($e_0=0.3$–0.6; red dots), end with eccentricities near 0.1 in any 1:$n$ resonance.

For the other $m$:$n$ resonances with $m \geq 2$, only 18 stable resonators form out of 60,000 test particles with $e_0=0.1$ at $i_0=20^\circ$ and $40^\circ$. This number is much smaller than in the planar case, and all of these stable resonators are found in the lowest-order 2:5 and 3:7 MMRs. The critical eccentricities (i.e. $e_{crit}$) of these two resonances are always $>0.12$ at $i=20^\circ$ and $40^\circ$. Thus, particles starting with eccentricities as small as $e_0 = 0.1$ would not satisfy $e \ge e_{crit}$, and these two lowest-order resonances should not be populated. However, first, as indicated by the cyan stars in Fig. \ref{casei20all}(a) (for $i_0 = 20^\circ$) and Fig. \ref{casei40all}(a) (for $i_0 = 40^\circ$), some 2:5 stable resonators reach $e_f > 0.2$, which is consistently larger than $e_{crit}^{2:5}$, so libration is allowed. Second, a few stable resonators retain very small eccentricities. For example, at $i_0=40^\circ$, some 2:5 and 3:7 resonators have final eccentricities down to $e_f \sim 0.15$, which is below the corresponding $e_{crit}$. These exceptions can be explained by invoking the Kozai mechanism within an MMR. With the Kozai mechanism in effect, the argument of perihelion of an object no longer traverses the full range $0^\circ$–$360^\circ$. This dynamical factor allows for the violation of the usual libration condition \citep{Li2020}. Nevertheless, such cases remain extremely rare.

Finally, although a few slightly eccentric, highly inclined stable resonators appear in the 2:5 MMR, only one occurs in the higher-order 3:7 MMR (see the cyan stars in Fig. \ref{casei40all}(a)), which corresponds to a larger $e_{crit}$. This may account for why stable resonators of this kind have not been found in even higher-order resonances. Hence, for the high-inclination population beyond 50 AU, we propose that RKBOs with small eccentricities of $e \sim 0.1$ are expected to be intrinsically scarce in both the 1:$n$ resonances and all other resonances.

\subsubsection{$e_0=0.7$}

For test particles starting with $e_0 = 0.7$, since their semimajor axes range from 50 to 100 AU, the perihelia actually lie inside the orbit of Neptune. In the absence of the resonant phase-protection mechanism, these particles will eventually experience close encounters with Neptune and be scattered away after a sufficiently long period of evolution. As a result, in the planar case, no test particles on such most eccentric orbits can survive up to 4 Gyr, whereas in the inclined case, a very small number of survivors are observed.

This is because for test particles on inclined orbits, the larger their inclinations, the shorter the time they spend near the ecliptic plane, which is proportional to $1/\sin(i)$ \citep{Brow01}. Consequently, even without the resonant phase-protection mechanism, it is still possible that they will not approach the vicinity of Neptune and thus avoid strong perturbations within the 4 Gyr timescale. We find that, given initial inclinations of $i_0 = 20^\circ$, about 0.02\% of the test particles (6 out of 30,000) can survive for such a long period. For much higher $i_0 = 40^\circ$, the time that the test particles spend near the ecliptic plane becomes even shorter, and the survival rate increases to 0.11\% (32 out of 30,000 particles). However, none of the survivors from the inclined cases with $i_0=20^{\circ}$ and $40^{\circ}$ are located in Neptune’s MMRs. Upon examining the orbits of these non-resonant survivors, we observe that their eccentricities decreased somewhat during the evolution, which results in final perihelion distances of $\gtrsim 32$ AU, i.e. at least 2 AU beyond the orbit of Neptune. This provides moderate protection against severe close encounters with Neptune in our simulations, though not permanently. Further analysis of these survivors is out of the scope of this work, as they belong to the non-resonant population.

Building on the results from both the planar and inclined cases that have been discussed, we propose that the resonant objects with $e\ge0.7$ in the distant Kuiper belt between $a=50$ and 100 AU should be extremely rare.

\subsubsection{$e_0=0.3$-$0.6$}

Figs. \ref{casei20all} and \ref{casei40all} display the distribution of final orbits for stable resonators starting on high-inclination orbits with $i_0 = 20^\circ$ and $i_0 = 40^\circ$, respectively. The most notable feature is that nearly all of them originate from the cases with $e_0=0.3$-0.6, as indicated by the dots in the figures. In the following, we provide a detailed analysis of these highly inclined resonators.

We find that even at high inclinations up to $40^\circ$, the stable resonators can occupy all 1:$n$ to 6:$n$ resonances within the 50-100 AU range, while for the 7:$n$ resonances, only a few of the highest-order ones are empty. In the case of $i_0 = 20^\circ$, as shown in Fig. \ref{casei20all}, the 7:$n$ stable resonators (black dots) extend up to the 7:40 MMR, except for the highest-order 7:41 MMR inside 100 AU. This pattern of resonance occupancy is exactly the same as in the planar case. For even higher $i_0 = 40^\circ$, the stability of the 7:$n$ resonances appears to weaken substantially. As seen in Fig. \ref{casei40all}, among the 7:$n$ stable resonators (black dots), the highest-order one they can occupy is the 7:27 MMR. It is very interesting to note that, as highlighted by the large black circle in Fig. \ref{real}, the only observed 7:$n$ RKBO, 2021 RW237, is located in the 7:29 MMR and moves on the orbit with $i \approx 20^\circ$. This observational evidence seems to support the resonance occupancy predicted by our long-term simulations for the high-inclination resonant populations.

\begin{table}
\hspace{0 cm}
\centering
\begin{minipage}{8cm}
\caption{Statistics of stable resonators in the 1:$n$ resonances for the inclined cases of $i_0=20^\circ$ and $40^\circ$ from the 4 Gyr evolution simulations. The numbers of leading and trailing objects are listed in the second and third columns, and their number ratio ($R_{LT}$) is given in the last column.}      
\label{1tonInc}
\begin{tabular}{c c c c}        

\hline\hline                 

$i_0=20^{\circ}$      \\

\hline
Resonance    &    Leading    &    Trailing     &    $R_{LT}$             \\
\hline

1:3       &          30       &       31       &      0.97             \\

1:4       &          62       &       86       &      0.72              \\

1:5       &          74       &       63       &      1.17             \\

1:6       &          85        &      61        &     1.39                \\

\hline
$i_0=40^{\circ}$      \\

\hline
Resonance    &    Leading    &    Trailing     &    $R_{LT}$             \\
\hline

1:3       &          9       &       7       &      1.28             \\

1:4       &          26       &      29       &     0.90              \\

1:5       &          6       &       6       &      1.00             \\

1:6       &          7        &      2        &     3.50                \\

\hline

\end{tabular}
\end{minipage}
\end{table}

In the planar case, as discussed in Section 4.3, the stable resonators in the 1:3, 1:4, and 1:6 MMRs consistently show fewer objects in the leading swarm than in the trailing swarm, whereas only the 1:5 MMRs host more leading objects. However, for the 1:$n$ stable resonators on highly inclined orbits, as $n$ varies from 3 to 6, the number asymmetry becomes even more uncertain. The specific numbers are listed in Table \ref{1tonInc}. For $i_0 = 20^\circ$, the number ratio of leading to trailing stable resonators in the 1:3 MMR is $R_{LT} = 0.97$, indicating that the two populations are nearly equal. In contrast, the 1:4 MMR hosts noticeably fewer leading resonators ($R_{LT} = 0.72$), while the 1:5 and 1:6 MMRs host more ($R_{LT}=1.17$ and 1.39, respectively). As $i_0$ increases further to $40^\circ$, a similar variability in number asymmetry is likewise observed across the 1:$n$ resonances. The number ratios $R_{LT}$ are 1.28 for the 1:3 MMR, 0.90 for the 1:4 MMR, 1.00 for the 1:5 MMR, and 3.50 for the 1:6 MMR. It should be noted, however, that except for the 1:4 MMR, the leading and trailing swarms each contain fewer than 10 stable resonators in the other three resonances, so the statistical significance of these number differences is limited.

In addition to the effects of long-term evolution, the leading-to-trailing number ratio of 1:$n$ RKBOs may also depend on their formation processes, such as planetary migration and resonance sweeping/capture. In our previous study of the high-inclination population in the 1:2 MMR \citep{Li2014b}, we found that during the outward migration of Neptune, particles are preferentially captured into the leading rather than the trailing libration zone at $i \lesssim 10^\circ$. However, this number asymmetry reverses at $i = 20^\circ$ and continues to vary at even larger $i$. Accordingly, to better constrain the intrinsic number asymmetry of 1:$n$ RKBOs beyond 50 AU, further detailed investigations are needed to assess the influence of inclination across both their formation stage and subsequent long-term evolution.

By combining the planar and inclined cases, we observe a clear trend: the number of stable resonators decreases as the inclination increases. For the same total of 90,000 test particles with initial eccentricities $e_0=0.3$-0.6, the number of candidate resonators detected during the first 10 Myr of evolution, as simulated in the pre-runs, is 10,221 for $i_0 = 0^\circ$, 9,134 for $i_0 = 20^\circ$, and 6,125 for $i_0 = 40^\circ$. After 4 Gyr of long-term evolution, the number of stable resonators corresponding to these three $i_0$ values is reduced to 4,739, 3,004, and 637, respectively. These results not only indicate that the number of stable resonators decreases with increasing inclination, but also reveal a simultaneous decline in the survival rate, defined as the number ratio of stable resonators to candidate resonators. Specifically, the survival rate decreases from 46.3\% at $i_0 = 0^\circ$ to 32.8\% at $i_0 = 20^\circ$, and it further drops to just 10.4\% at $i_0 = 40^\circ$.

We suppose that the decrease in the number of stable resonators with increasing inclination can be attributed to two possible factors: (1) As the inclination increases, the critical eccentricity, which is equivalent to the minimum eccentricity required for stable libration, also increases. In other words, on high-inclination orbits, test particles with relatively small eccentricities may no longer be allowed to librate in the resonances. This could result in a decrease in the total number of stable resonators. (2) In this work, the libration zones shown in Fig. \ref{width} are determined using the planar model, while they tend to shrink as the inclination of the resonant particle increases \citep{Gall2020, Lei2020, namo2020, Li2023}. Due to this libration zone contraction, the number of stable resonators may become smaller on higher-inclination orbits.

\section{Conclusions and discussion}

Our previous studies have systematically investigated the dynamics of high-inclination objects in Neptune's MMRs within the Kuiper belt, up to 50 AU, including Plutinos \citep{Li2014a}, Twotinos \citep{Li2014b}, and other RKBO populations in the main classical Kuiper belt \citep{Li2020, Li2023}. In this work, we extend this series by providing a comprehensive overview of the dynamical structure of the more distant Kuiper belt beyond 50 AU, where a broad space exists and could be substantially populated. The outer boundary is set at 100 AU to minimise the possible influence of Planet 9.

We first identified the observed RKBOs that occupy Neptune's $m$:$n$ MMRs between 50 and 100 AU. For these resonances, we considered the most complete combinations of $m$ and $n$, with $m \leq 20$ and all possible $n$. From the multi-oppositional KBOs registered in the MPC database as of April 29, 2024, we identified 207 RKBOs within the considered region. Among them, 177 are classified as secure RKBOs, since for each of them, all 10 clones can persist in the same resonance; 29 are categorised as probable RKBOs, each of which has at least one resonant clone; and 2021 LS43 is identified as an insecure RKBO in the 3:16 resonance. The distribution of these 207 observed RKBOs reveals several notable properties: (1) they are distributed across a large number of resonances, ranging from the 2nd-order 1:3 MMR to the 22nd-order 7:29 MMR. 
However, for the $m$:$n$ resonances with $m \geq 8$, RKBOs cannot be sustained stably. (2) The observed RKBOs exhibit a broad inclination distribution, reaching up to $i = 40^\circ$. This $i$-distribution appears similar in different sets of $m$:$n$ resonances with varying values of $m$ ($\leq 7$). 
In addition, their eccentricities range from $e=0.18$ to 0.65, with all objects having $a>70$ AU exhibiting $e>0.4$. This prevalence of high eccentricities likely results from observational bias, as objects with smaller perihelion distances are much more easily detected.
(3) Among the 1:$n$ RKBOs, as $n$ increases from 3 to 6, the leading swarm consistently hosts at least as many objects as the trailing swarm.

According to the distribution of observed RKBOs, such as resonance occupancy and ranges of $e$ and $i$, we then theoretically investigated the dynamical features of numerous Neptune’s MMRs beyond 50 AU. The planar CR3BP model was used to derive the libration zones of these MMRs on the plane of semimajor axis and eccentricity. From this, we obtained the $e$ range that allows an object to remain in each MMR. Furthermore, we examined the influence of high inclination on the structures of two representative sets of MMRs: (1) for the 1:$n$ resonances with $n=3$-6, we determined the conditions required for the symmetric and asymmetric librations, represented by $e \geq e_a$ and $e \geq e_c$, respectively. Here, both the critical eccentricities $e_a$ and $e_c$ vary with $i$. We also calculated the symmetric and asymmetric libration centres for these 1:$n$ resonances, which are strongly dependent on $i$. This dependence in the high-$i$ case can differ significantly from that in the low-$i$ case. Particularly, in some high-$i$ scenarios, even the symmetric resonance centre will disappear, indicating the non-existence of the associated 1:$n$ resonators. (2) For the other $m$:$n$ resonances with $m \geq 2$, where only the symmetric libration motion exists, we similarly determined the critical eccentricity $e_{crit}$, which represents the lower limit of $e$ for the occurrence of the libration. Since $e_{crit}$ increases as a function of increasing $i$, the libration condition of $e \geq e_{crit}(i)$ defines a permissible region on the $(e, i)$ plane, as we first introduced for the 4:7 MMR in \citet{Li2020}.

After theoretically studying the complex dynamical features of Neptune’s MMRs beyond 50 AU, we performed numerical simulations to explore the long-term stability of these resonances. For this purpose, we considered test particles with initial semimajor axes uniformly distributed from 50 to 100 AU, initial eccentricities of $e_0=0.1$, 0.3, 0.5, 0.6, and 0.7, and initial inclinations of $i_0=0^{\circ}$, $20^{\circ}$, and $40^{\circ}$. The remaining three orbital elements ($\lambda$, $\varpi$, and $\Omega$) were randomly assigned between $0^\circ$ and $360^\circ$. This setup ensures that the test particles initially cover the entire libration zones for all resonances within the 50-100 AU range, regardless of the resonance order. To achieve a high resolution in the coverage of these resonances, we generated a total of 450,000 test particles. Within the framework of the current outer Solar system, we numerically integrated the orbits of the test particles over 4 Gyr. Particles exhibiting libration behaviours within the first 10 Myr are classified as candidate resonators, and those surviving in resonances after 4 Gyr of evolution are classified as stable resonators. For the classification of these resonators, according to the resonance occupancy of the observed RKBOs, we focused on the $m$:$n$ resonances with $m$ values ranging from 1 to 7. For each $m$, the maximum $n$ was chosen to correspond to the farthest resonance interior to 100 AU. It should be noted that, while some stable resonators may experience inclination variations, most of them maintain orbits with final inclinations close to their initial inclinations $i_0$.


Based on our numerical results, we may address the three issues raised in Section 1. The first two are: (1) the potential resonance occupancy, and (2) the possible ranges of $e$ and $i$ for the associated resonators. 

For initial inclinations up to $i_0=40^\circ$, the stable resonators can occupy all resonances of the type 1:$n$ through 6:$n$, with varying $n$ corresponding to the 50-100 AU region. The individual highest-order resonances within this region are MMRs of 1:6 (99.4 AU), 2:11 (93.8 AU), 3:17 (95.7 AU), 4:23 (96.6 AU), 5:29 (97.2 AU), and 6:35 (97.5 AU). In contrast, for the observed RKBOs, apart from the 1:$n$ and 2:$n$ type resonances, they are found only in MMRs up to 3:16 (91.9 AU), 4:21 (90.9 AU), 5:22 (80.8 AU), and 6:25 (80.0 AU). This suggests that in the distant Kuiper belt, the RKBOs could potentially occupy the farthest-out resonances located beyond 80-90 AU.


Regarding the 7:$n$ type resonances, only one associated RKBO, 2021 RW237, has been identified. It is located in the 7:29 MMR at 77.6 AU and has an inclination of about $20^\circ$. For this object of particular interest, we also considered its actual orbital uncertainty in $a$ from the MPC Explorer. By simulating 10 clones with different $a$ values within the uncertainty range, we find that one clone remains in the 7:29 MMR while the other nine escape. Therefore, 2021 RW237 is classified as probable RKBO.
Nevertheless, the existence of this MPC RKBO is supported by our long-term simulations, which show that for $i_0=0^\circ$-$20^\circ$, the stable resonators can be spread up to the 7:40 MMR at 96.2 AU. The 7:40 MMR is much higher in resonance order than the 7:29 MMR, where 2021 RW237 resides, suggesting that the potential 7:$n$ RKBOs could span the entire 50-100 AU region. We also find that at even higher $i_0 = 40^\circ$, the highest-order 7:$n$ resonance occupied by stable resonators is the 7:27 MMR. However, the overall occupancy of the 7:$n$ resonances does not change.

As illustrated above, across the resonances from 1:$n$ to 7:$n$ between $a=50$ and 100 AU, the inclinations of potential RKBOs can consistently extend to as high as $i \sim 40^\circ$. For eccentricities, there is an upper limit at $e=0.7$, corresponding to perihelion distances of $q=15$-30 AU. In theory, RKBOs located on Neptune-crossing orbits (i.e. with $q < 30.1$ AU), but exterior to Uranus’s orbit (i.e. with $q > 19.2$ AU), may remain long-term stable due to the resonant phase-protection mechanism \citep{nesv01}.
However, in practice, once this $e$-limit is exceeded, no objects can remain stably in any of these resonances over the age of the Solar system. The detailed influence of $i$ and $e$ on the distribution of final orbits for stable resonators is presented in the main text.

The issue (3) listed in Section 1 concerns the number asymmetry of the 1:$n$ resonators between the leading and trailing swarms. For $i_0$ up to $20^\circ$, we found that the stable resonators in the lower-order 1:3 and 1:4 MMRs consistently exhibit fewer objects in the leading swarm compared to the trailing swarm. This number asymmetry appears to contradict current observations. Nevertheless, the formation model of the Kuiper belt may induce an opposite trend in the number distribution of objects in the 1:3 and 1:4 MMRs. In the context of planetary migration, \citet{PL2017} demonstrated that for these resonances, the leading swarm could capture twice as many objects as the trailing swarm. 
On the other hand, for the higher-order 1:5 and 1:6 MMRs with $i_0=0^\circ$-$20^\circ$, or for all 1:$n$ MMRs with an even higher $i_0=40^\circ$, the number asymmetry of the 1:$n$ stable resonators shows considerable variability across different resonances. 
As a matter of fact, we must acknowledge the limitations of the known MPC sample: (1) severe observational incompleteness, with a total of 18 RKBOs identified in the 1:3 and 1:4 MMRs and only 2 in the higher-order 1:5 and 1:6 MMRs; and (2) an apparent distribution that may be biased by the specifics of the observational surveys, while an effective de-biasing process cannot yet be applied to the current MPC sample. Therefore, it remains inconclusive whether the leading population outnumbers the trailing one. It is possible that, in the near future, the LSST will discover tens of thousands of KBOs and enable us to assess the intrinsic difference between the leading and trailing populations.

In addition to the number asymmetry problem that arises only in the 1:$n$ resonances, issue (3) also concerns the relative number of RKBOs in all $m$:$n$ resonances with different values of $m$. We estimated the survival rates of various resonance sets with $m=1$-7 by comparing the number of candidate and stable resonators. In the planar case of $i_0 = 0^\circ$, we found that the survival rates for the 1:$n$, 2:$n$, and 3:$n$ resonances are all around 50\%. However, for the 4:$n$, 5:$n$, 6:$n$, and 7:$n$ resonances, the survival rates drop sharply to $\sim20$–30\%, indicating that these resonances likely host a much smaller fraction of the resonant population due to their higher-order nature. This result is consistent with current observations, as among the 207 observed RKBOs between 50 and 100 AU, only about 17.9\% belong to the four resonance sets with $m=4$–7. Taking into account the effect of high inclination, we further found that the survival rate of the resonant populations declines significantly with increasing inclination. For all $m$:$n$ resonances combined, the survival rate decreases from 46.3\% at $i_0 = 0^\circ$ to 32.8\% at $i_0 = 20^\circ$, and further drops to just 10.4\% at $i_0 = 40^\circ$. 

We finish this paper by discussing the newly discovered phenomenon of number reversal, which may place additional constraints on the eccentricity and inclination distributions of primordial KBOs. This phenomenon is characterised by an unexpected trend: as the order of a resonance increases, the number of stable resonators increases rather than decreases. It is first recognised in the 3:7 and 3:8 MMRs, where the former is lower-order and the latter is higher-order. To investigate this phenomenon more thoroughly, we conducted an additional simulation adopting a continuous distribution of $e_0=0.1$–0.7, instead of a few discrete values used in our main simulations, while still focussing on the planar case with initial inclinations $i_0 = 0$. Considering test particles with initial semimajor axes $a_0=50$-60 AU (i.e. in the neighbourhoods of the 3:7 and 3:8 MMRs), we find that, after 4 Gyr of evolution, the higher-order 3:8 MMR unexpectedly hosts more stable resonators than the lower-order 3:7 MMR, despite the latter possessing a wider libration zone. The underlying reason for this number reversal is that as eccentricities $e$ increase, stable resonators tend to have smaller resonant amplitudes $A$, leading to a reduction in the size of the stable region within the libration zone. Because the 3:8 MMR lies at a greater heliocentric distance than the 3:7 MMR, its resonators have larger perihelion distances, which may partially compensate for the shrinking effect of the larger $e$ on their stable region and enhance long-term stability. From these results, it is evident that the appearance of the number reversal for the 3:7 and 3:8 MMRs is closely related to the $e$ values of the resonators. If future observations were to show, contrary to our stability analysis, that more objects occupy the 3:7 MMR than the 3:8 MMR, this could offer valuable insight into the $e$ distribution of primordial KBOs beyond 50 AU, e.g. how their numbers decrease with increasing $e$. We note that, although the phenomenon is derived from in situ evolution, it is not incompatible with the classical formation model of RKBOs in the context of planetary migration and resonance capture. This is so because, in this model, the 3:8 MMR sweeps through the primordial Kuiper belt ahead of the 3:7 MMR, and as a result, it should preferentially capture more objects.

The above analysis is based mainly on simulations in the planar case ($i_0 = 0^\circ$). Furthermore, the phenomenon of number reversal may also appear, and even become more pronounced, in inclined cases with high $i_0$. By combining all main simulations with discrete initial eccentricities of $e_0=0.1$, 0.3, 0.5, 0.6, and 0.7, we find that the number ratio of stable resonators in the 3:8 to the 3:7 MMR is approximately 1.1 at $i_0 = 0^\circ$, but increases to 1.7 at $i_0 = 20^\circ$. This can be regarded as an enhanced manifestation of number reversal, which may give us new insight into the inclination distribution of primordial KBOs. We expect that future investigations will explore this aspect in more detail.

Interestingly, the phenomenon of number reversal is not restricted to the 3:7 and 3:8 resonance pair. We have also observed it in several other adjacent resonance pairs, including 1:3 and 1:4 MMRs, 2:5 and 2:7 MMRs, 4:9 and 4:11 MMRs, 5:11 and 5:12 MMRs, 6:13 and 6:17 MMRs, as well as 7:15 and 7:17 MMRs. For each of these resonance pairs, our simulations show that the lower-order MMR hosts fewer stable resonators than the neighbouring higher-order resonance. These findings collectively suggest that number reversal may be an intrinsic characteristic of such adjacent $m$:$n$ resonance pairs in the distant Kuiper belt.


\section*{Acknowledgments}

This work was supported by the National Natural Science Foundation of China (Nos. 12473061, 11973027, 12150009), and the China Manned Space Program (CMS-CSST-2025-A16). The author would like to express his thanks to the anonymous referee for the valuable comments that helped to considerably improve the manuscript.

\section*{Data Availability}

The data underlying this article are available in the article and in its online supplementary material.


\appendix
\section{Classification of RKBOs within sparsely populated resonances}
\label{sec:uncommon RKBOs}

\begin{table*}
\hspace{0 cm}
\centering
\begin{minipage}{13.8 cm}
\caption{List of particular RKBOs of interest, each being the sole object identified in its resonance. Orbital information is obtained from the MPC Explorer. ‘$\Delta a / a$’ denotes the orbital uncertainty in the semimajor axis. ‘Opps.’ indicates the number of oppositions at which the object has been observed. For reference, the last column provides the observational arc length.}
\label{uncommonRKBOs}
\begin{tabular}{c c c c c c c}        
\hline                 
Resonance  &  Total  &  MPC designation   &    Class          & $\Delta a / a$ (per cent) &  Opps. &  Arc length \\
 
\hline

  3:16    &    1     &     2021 LS43      &    insecure RKBO   &       59.96              &    2   &   305 days   \\

  4:21    &    1     &     2009 KX36      &    probable RKBO   &       0.27               &   6   &    2009-2020    \\

  5:19    &    1     &     2015 RK258     &    probable RKBO   &        8.73              &    2   &   2015-2016   \\
   
  7:29    &    1     &     2021 RW237     &    probable RKBO   &        1.39              &    3   &   2019-2021    \\
    
\hline
\end{tabular}
\end{minipage}
\end{table*}



Having confirmed the RKBOs with best-fit orbits, we found several sparsely populated resonances. For instance, only a single object, 2021 RW237, resides in the 7:$n$ resonances. For such particular RKBOs of interest---those located in uncommon resonances hosting no more than three samples---we further examined their actual orbital uncertainties, $\Delta a / a$, from the MPC Explorer. Some of these RKBOs have $\Delta a / a$ values significantly larger than 0.05\%, which is the standard uncertainty adopted in the main text. Table \ref{uncommonRKBOs} lists the four particular RKBOs of interest: 2021 LS43 (3:16 MMR), 2009 KX36 (4:21 MMR), 2015 RK258 (5:19 MMR), and 2021 RW237 (7:29 MMR).


For each of these four RKBOs, we regenerated 10 clones by varying the semimajor axis within the actual uncertainty range given in Table \ref{uncommonRKBOs}, instead of the standard uncertainty of 0.05\%, and integrated their orbits over 10 Myr. For 2009 KX36, 2015 RK258, and 2021 RW237, two, one, and one of the clones, respectively, remain in the same resonance; thus, these three objects are classified as probable RKBOs, consistent with the inference based on the standard uncertainty of $\Delta a / a=0.05\%$. In contrast, for 2021 LS43, none of the clones are retained in the 3:16 resonance, and its classification is therefore updated as insecure. This outcome is unsurprising, as the actual $\Delta a / a$ of 2021 LS43 is as large as 60\%.

In addition, by examining the orbital information provided in Table \ref{uncommonRKBOs}, a reference can be made to assess the accuracy of the orbital determination. In general, this accuracy is primarily governed by the number of oppositions. For example, the object 2009 KX36, which has the smallest value of $\Delta a / a$, also has the largest number of oppositions (6). By contrast, 2021 RW237 and 2015 RK258 have fewer oppositions--3 and 2, respectively--and the uncertainty $\Delta a / a$ increases monotonically when 2009 KX36 is taken as a reference. Nevertheless, 2021 LS43 has the same number of oppositions as 2015 RK258, but a much larger $\Delta a / a$ ($\approx60\%$). This is likely due to its very short arc length of less than one year.



\label{lastpage}

\end{document}